\documentclass[fleqn,usenatbib]{mnras}

\usepackage{newtxtext,newtxmath}

\newcommand{\VardT}{\texttt{Thermal}\-\texttt{Kine}\-\texttt{tic}\_\texttt{var}$\mathtt{\Delta T}_{\mathtt{SN}}$}
\newcommand{\colibrefixedagntemp}
{\texttt{Thermal}\-\texttt{Kine}\-\texttt{tic}\_\texttt{var}$\mathtt{\Delta T}_{\mathtt{SN}}$\texttt{var}$\mathtt{f}_{\mathtt{E}}$}
\newcommand{\colibre}{\textsc{colibre}}
\newcommand{\gsmf}{GSMF}

\newcommand{\ssm}{SSMR}
\newcommand{\bhmsm}{BSMR}
\newcommand{\bhmsms}{BSMRs}
\newcommand{\shmr}{SHMR}
\newcommand{\shmrs}{SHMRs}
\newcommand{\modelbasiccolor}{green}
\newcommand{\modelthermalkineticcolor}{yellow}
\newcommand{\modelvardtcolor}{navy-blue}
\newcommand{\colibrefixedagndtcolor}{light-blue}
\newcommand{\colibrefiducialcolor}{dark-red}

\usepackage[T1]{fontenc}

\DeclareRobustCommand{\VAN}[3]{#2}
\let\VANthebibliography\thebibliography
\def\thebibliography{\DeclareRobustCommand{\VAN}[3]{##3}\VANthebibliography}

\usepackage{booktabs}
\usepackage{graphicx}	
\usepackage{amsmath}	

\title[Calibration of subgrid feedback in COLIBRE]{COLIBRE: calibrating subgrid feedback in cosmological simulations that include a cold gas phase}

\author[E. Chaikin et al.]{Evgenii Chaikin,$^{1}$\thanks{E-mail: chaikin@strw.leidenuniv.nl} 
Joop Schaye,$^{1}$ 
Matthieu Schaller,$^{2,1}$ 
Sylvia Ploeckinger,$^{3}$ 
Yannick M. Bah\'{e},$^{4,5}$ \newauthor 
Alejandro Ben\'{i}tez-Llambay,$^{6}$ 
Camila Correa,$^{1}$ 
Victor J. Forouhar Moreno,$^{1}$ 
Carlos S. Frenk,$^{7}$ 
Filip Hu\v{s}ko,$^{1}$ \newauthor 
Roi Kugel,$^{1}$ 
Robert McGibbon,$^{1}$ 
Alexander J. Richings,$^{8,9}$ 
James W. Trayford,$^{10}$ 
Josh Borrow,$^{11}$ \newauthor 
Robert A. Crain,$^{12}$ 
John C. Helly,$^{7}$ 
Cedric G. Lacey,$^{7}$ 
Aaron Ludlow,$^{13}$ and
Folkert S. J. Nobels$^{1,14}$ \\
$^{1}$Leiden Observatory, Leiden University, PO Box 9513, 2300 RA Leiden, the Netherlands \\
$^{2}$Lorentz Institute for Theoretical Physics, Leiden University, PO Box 9506, 2300 RA Leiden, the Netherlands \\
$^{3}$Department of Astrophysics, University of Vienna, T\"{u}rkenschanzstrasse 17, 1180 Vienna, Austria \\
$^{4}$School of Physics and Astronomy, University of Nottingham, University Park, Nottingham NG7 2RD, UK \\
$^{5}$Institute of Physics, Ecole Polytechnique F\'{e}d\'{e}rale de Lausanne (EPFL), Observatoire de Sauverny, 1290 Versoix, Switzerland \\
$^{6}$Dipartimento di Fisica G. Occhialini, Universit\`{a} degli Studi di Milano Bicocca, Piazza della Scienza, 3 I-20126 Milano MI, Italy \\
$^{7}$Institute for Computational Cosmology, Department of Physics, University of Durham, South Road, Durham, DH1 3LE, UK \\ 
$^{8}$Centre for Data Science, Artificial Intelligence and Modelling, University of Hull, Cottingham Road, Hull, HU6 7RX, UK \\
$^{9}$E. A. Milne Centre for Astrophysics, University of Hull, Cottingham Road, Hull, HU6 7RX, UK \\
$^{10}$Institute of Cosmology and Gravitation, University of Portsmouth, Dennis Sciama Building, Burnaby Road, Portsmouth PO1 3FX, UK \\
$^{11}$Department of Physics and Astronomy, University of Pennsylvania, 209 South 33rd Street, Philadelphia, PA 19104, USA \\
$^{12}$Astrophysics Research Institute, Liverpool John Moores University, 146 Brownlow Hill, Liverpool L3 5RF, UK \\
$^{13}$International Centre for Radio Astronomy Research, University of Western Australia, 35 Stirling Highway, Crawley, Western Australia, 6009, Australia \\ 
$^{14}$Netherlands Organisation for Applied Scientific Research (TNO), Molengraaffsingel 8, 2629 JD Delft, the Netherlands
}
\date{Accepted XXX. Received YYY; in original form ZZZ}

\pubyear{2015}

\begin{document}
\label{firstpage}
\pagerange{\pageref{firstpage}--\pageref{lastpage}}
\maketitle

\begin{abstract}
We present the calibration of stellar and active galactic nucleus (AGN) feedback in the subgrid model for the new \colibre{} hydrodynamical simulations of galaxy formation. \colibre{} directly simulates the multi-phase interstellar medium and the evolution of dust grains, which is coupled to the chemistry. \colibre{} is calibrated at three resolutions: particle masses of $m_{\rm gas} \approx m_{\rm dm} \sim 10^7$ (m7), $10^6$ (m6), and $10^5~\mathrm{M_\odot}$ (m5). To calibrate the \colibre{} feedback at m7 resolution, we run Latin hypercubes of $\approx 200$ simulations that vary up to four subgrid parameters in cosmological volumes of ($50~\mathrm{cMpc}$)$^{3}$. We train Gaussian process emulators on these simulations to predict the $z=0$ galaxy stellar mass function (\gsmf) and size -- stellar mass relation (\ssm) as functions of the model parameters, which we then fit to observations. The trained emulators not only provide the best-fitting parameter values but also enable us to investigate how different aspects of the prescriptions for supernova and AGN feedback affect the predictions. In particular, we demonstrate that while the observed $z=0$ \gsmf{} and \ssm{} can be matched individually with a relatively simple supernova feedback model, simultaneously reproducing both necessitates a more sophisticated prescription. We show that the calibrated m7 \colibre{} model not only reproduces the calibration target observables, but also matches various other galaxy properties to which the model was not calibrated. Finally, we apply the calibrated m7 model to the m6 and m5 resolutions and, after slight manual adjustments of the subgrid parameters, achieve a similar level of agreement with the observed $z=0$ \gsmf{} and \ssm.
\end{abstract}

\begin{keywords}
methods: numerical -- galaxies: general -- galaxies: formation -- galaxies: evolution
\end{keywords}



\section{Introduction}

In the last few decades, numerical simulations of galaxy formation have become an indispensable tool for advancing our understanding of the physics of galaxy formation \citep[see][for a recent review]{2023ARA&A..61..473C}. The rapid growth of computational facilities has opened up the possibility of simulating large cosmological volumes ($\gtrsim 100^3~\mathrm{comoving \, Mpc}^3$; hereafter $\mathrm{cMpc}^3$) with self-consistent modelling of baryonic processes \citep[e.g.][]{2010MNRAS.402.1536S,2014MNRAS.444.1453D,2015MNRAS.446..521S,2017MNRAS.465.2936M,2017MNRAS.470.1121T,2018MNRAS.473.4077P,2019MNRAS.486.2827D,2022MNRAS.512.3703B,2023MNRAS.524.2539P,2023MNRAS.526.4978S,2025arXiv250401061D}, as well as studying the properties of the multi-phase interstellar medium (ISM) of galaxies in smaller volumes situated in a cosmological environment \citep[e.g.][]{2021A&A...651A.109D,2023MNRAS.522.3831F}. A major part of this success stems from the progress in computational methods, which has greatly increased the efficiency with which large computational machines can be exploited. In particular, modern astrophysical codes demonstrate impressive performance in standard scaling tests extending up to $\sim 10^5$ compute cores \citep[e.g.][]{2021MNRAS.506.2871S,2024MNRAS.530.2378S}.

At the same time, the advent of advanced observational facilities such as the Atacama Large Millimeter/submillimeter Array \citep[ALMA;][]{ALMA5136193} and \textit{JWST} \citep[][]{JWST63183b28f893464d8ea2c9680f466d66}, has allowed us to study spatially resolved properties of galaxies with unprecedented sensitivity and accuracy, both in the local and high-redshift Universe. Dense, cold interstellar gas can be probed by ALMA at (sub-)kpc resolution either through CO rotation-line emission at $z\lesssim 2$ \citep[e.g.][]{2022ApJ...933...11I,2022A&A...658A.155R} or via [C~\textsc{ii}] 158~$\mu$m line at higher redshifts \citep[e.g.][]{2020A&A...643A...1L,2023A&A...680L...8B}. The properties of the warmer gas of high$-z$ objects can be studied at comparable resolution with \textit{JWST}, using emission lines such as [O~\textsc{iii}] or $\mathrm{H}~\upbeta$ \citep[e.g.][]{2023MNRAS.518.5607C, 2023ApJ...948..126G}. Clearly, for a fair comparison between theory and observations, these cutting-edge observational data demand numerical simulations that reach comparable or higher spatial resolutions, and that self-consistently model the multi-phase interstellar gas.

Simulations of galaxy formation from the past 10 years have achieved remarkable success in matching observational data and producing galaxies with realistic properties \citep[see e.g.][and references therein]{2023ARA&A..61..473C}. Various observed relations are broadly reproduced by the simulations, including the observed galaxy stellar mass functions (\gsmf) and luminosity functions at different redshifts, the galaxy size -- stellar mass relation (\ssm), the galaxy star-forming main sequence, the galaxy stellar mass -- metallicity relation and many others \citep[e.g.][]{2015MNRAS.446..521S,2018MNRAS.473.4077P,2019MNRAS.486.2827D}. However, those successful simulations largely neglected the modelling of the cold neutral gas, which is believed to play a key role in the cosmic baryon cycle \citep[e.g.][]{2020ARA&A..58..363P}. 

In fact, the large, high-resolution cosmological-volume simulations from the past decade such as \textsc{HorizonAGN} \citep{2014MNRAS.444.1453D}, \textsc{eagle} \citep{2015MNRAS.446..521S}, \textsc{IllustrisTNG} \citep{2018MNRAS.473.4077P}, and \textsc{Simba} \citep{2019MNRAS.486.2827D} all applied `a temperature floor' to the interstellar gas and/or artificially enhanced the gas pressure assuming
an effective equation of state, so that the (dense) gas cannot cool below $\sim 10^4$ K. The reason for this is twofold. First, simulating the cold phase is computationally expensive because of the small time-steps and the small Jeans lengths and masses that are readily reached in the dense, cold phase. Second, modelling the cold phase cannot be accomplished without accounting for physical processes that are important in this regime. These processes include \mbox{(self-)shielding} of gas from the extragalactic UV background and radiation from local sources, the formation and dissociation of molecules and cooling emission therefrom, and the formation and evolution of dust grains, including their interactions with the cold gas phase \citep[e.g.][]{doi:10.1146/annurev-astro-082812-141034}. It has also been argued that, unless an effective pressure floor is used, including cold gas in simulations that do not formally resolve the thermal Jeans mass in the cold ISM ($\lesssim 10^3~\mathrm{M_\odot}$) is problematic  \citep[e.g.][]{2008ApJ...680.1083R,2008MNRAS.383.1210S}, as it can lead to numerical artifacts such as artificial fragmentation \citep{1997MNRAS.288.1060B,1997ApJ...489L.179T}. However, \citet{2024MNRAS.528.2930P} recently showed that imposing an effective pressure floor is unnecessary, as in galaxy formation simulations with softened gravity the thermal Jeans mass criterion based on Newtonian gravity is inappropriate: in unresolved regimes, the Jeans mass scale is set by the softened Jeans mass, which substantially exceeds the Newtonian value.

Among the most recent simulations of cosmological volumes that allow the gas to enter the cold phase are \textsc{NewHorizon} \citep{2021A&A...651A.109D} and \textsc{Firebox} \citep{2023MNRAS.522.3831F}. The \textsc{Firebox} simulation used the \textsc{Fire2} galaxy formation model \citep{2018MNRAS.480..800H} and was run in a cosmological volume of ($22.1$ cMpc)$^{3}$ down to redshift $z = 0$ with gas and dark-matter (DM) particle masses of, respectively, $m_{\rm gas} = 6.3 \times 10^4~\mathrm{M_\odot}$ and $m_{\rm dm} = 3.3 \times 10^5~\mathrm{M_\odot}$. As with the \textsc{Fire2} simulations, \textsc{Firebox} was run with the \textsc{Gizmo} mesh-less finite-mass hydrodynamic and gravity solver \citep{2015MNRAS.450...53H}. The element ionization states were calculated based on tabulated equilibrium results from the simulations with the photoionization code \textsc{cloudy} \citep{1998PASP..110..761F}, including a shielding correction for cosmic UV background and local sources. The molecular-to-neutral gas fraction was computed on-the-fly by employing an analytical expression from \citet{2011ApJ...729...36K}. \textsc{Firebox} does not model the self-consistent evolution of dust grains but accounts for dust collisional heating/cooling and photo-electric heating using analytic expressions, assuming a constant dust-to-metal ratio (for further details, see \citealt{2018MNRAS.480..800H}). The \textsc{Firebox} model does not include a prescription for AGN feedback. As shown by \citet{2023MNRAS.522.3831F}, \textsc{Firebox} reproduces the mass -- metallicity relations for both the stellar and gas-phase components, as well as the star-forming main sequence and the relations between galaxy H~\textsc{i} and H$_2$ masses and stellar mass, although the $z=0$ \gsmf{} in \textsc{Firebox} is systematically higher than the observed \gsmf.
 
The domain of the \textsc{NewHorizon} simulation is a zoom-in region of $\sim$(16 cMpc)$^{3}$ taken from the larger, (142 cMpc)$^{3}$ volume of the \textsc{HorizonAGN} simulations \citep{2014MNRAS.444.1453D}. The \textsc{NewHorizon} simulation was run to redshift $z=0.25$ with a modified version of the \textsc{HorizonAGN} model, using the adaptive mesh refinement code \textsc{ramses} \citep{2002A&A...385..337T}. Inside the zoom-in region, the DM particle mass is equal to $1.2\times 10^6~\mathrm{M_\odot}$ and the cell size can reach $\approx 34$~pc in the densest environments. The cooling of metal-enriched gas in \textsc{NewHorizon} is based on tabulated equilibrium rates from \citet{1993ApJS...88..253S} at temperatures above $\approx 10^4$ K and \citet{1972ARA&A..10..375D} below $\approx 10^4$ K, which allows the gas to cool to $0.1$~K. Primordial species are assumed to be in ionization equilibrium under a homogeneous, redshift-dependent UV background, whose intensity is exponentially suppressed at densities $n_{\rm H} \gtrsim 0.01$~cm$^{-3}$ due to self-shielding. No dust evolution model is included. As shown by \citet{2021A&A...651A.109D}, \textsc{NewHorizon} agrees with the observed galaxy-averaged Kennicutt–Schmidt (KS) star formation law \citep[][]{1998ApJ...498..541K}, the black hole mass -- stellar mass relation, and the relations between stellar mass and the gas-phase and stellar metallicities, though, similarly to \textsc{Firebox}, galaxies in \textsc{NewHorizon} tend to be overmassive, resulting in a discrepancy between the predicted stellar-to-halo mass ratios and observational data.

The results from \textsc{NewHorizon} and \textsc{Firebox} demonstrate that simulations of galaxy formation that include a cold ISM are challenging but possible. In this companion paper to \citet{schaye2025colibreproject}, we present the calibration of the strengths of stellar and active galactic nucleus (AGN) feedback in the new galaxy formation model \colibre\footnote{\href{https://colibre-simulations.org/}{https://colibre-simulations.org}} \citep{schaye2025colibreproject}, which -- like \textsc{NewHorizon} and \textsc{Firebox} -- captures the multi-phase nature of the ISM. The \colibre{} model builds upon the \textsc{owls} \citep{2010MNRAS.402.1536S} and \textsc{eagle} \citep{2015MNRAS.446..521S} galaxy formation models with significant improvements on various fronts: in addition to the presence of the multi-phase ISM, \colibre{} includes non-equilibrium gas cooling, a live dust model coupled to the chemistry, and more sophisticated prescriptions for star formation and feedback from stellar evolution and AGN. Furthermore, in the initial conditions (ICs) of the \colibre{} simulations, we use four dark matter particles per gas particle to minimize spurious collisional heating of galaxies, which can negatively impact the sizes, kinematics, and morphologies of their stellar components \citep{2019MNRAS.488.3663L, 2021MNRAS.508.5114L,2023MNRAS.525.5614L,2023MNRAS.519.5942W}.

The need for calibration of galaxy formation simulations has been discussed extensively in the literature (see, e.g., \citealt{2010MNRAS.402.1536S,2013MNRAS.436.3031V,2015MNRAS.450.1937C,2017MNRAS.465.2936M}, and in particular $\S$2.1 of \citealt{2015MNRAS.446..521S}). Briefly, due to the finite resolution and large dynamic range of cosmological simulations, many astrophysical processes that occur on smaller scales -- such as stellar and AGN feedback -- are unresolved or only partially resolved. As a result, these processes must be implemented via \textit{subgrid} models, which typically involve free (also referred to as \textit{subgrid}) parameters that cannot be determined from first principles. The purpose of a subgrid model applied to an unresolved process is to reproduce the \textit{effective} impact of that process on the larger, resolved scales. The role of the free parameters is to control this effective impact while compensating for numerically induced effects, such as excessive radiative cooling of feedback-heated gas due to limited numerical resolution \citep[e.g.][]{2012MNRAS.426..140D}. In the case of supernova (SN) feedback, free parameters may include the (initial) mass loading and velocity of SN-driven winds \citep[e.g.][]{2003MNRAS.339..289S,2010MNRAS.402.1536S,2013MNRAS.436.3031V,2016MNRAS.462.3265D,2024MNRAS.527.1216S}, while for AGN feedback, they may include a boost to the accretion rate of supermassive black holes (SMBHs) or the energy released in a single AGN injection event \citep[e.g.][]{2009MNRAS.398...53B,2017MNRAS.465.2936M,2018MNRAS.479.5385H,2023MNRAS.526.6103K}. By comparing predictions from simulations with different subgrid parameters to observational data (such as the observed $z=0$ \gsmf), one can identify the parameter values that yield the best agreement with observations. This process is termed \textit{calibration}. A calibrated simulation loses its predictive power for the specific relations used in the calibration, but can still make genuine predictions at other redshifts and for quantities that were not used as calibration targets.

Generally, finding the values of subgrid parameters for which the simulation best reproduces a certain set of observational data is a cumbersome process. Given the number of `knobs to tune' in a galaxy formation model, the search for the best-fitting values may require running thousands of simulations for various values of the subgrid parameters. Such a blind search would be infeasible for \colibre{} because of the computational cost of running a new simulation at each step in the parameter space. Instead, a far more efficient approach is to use an emulator \citep{2010MNRAS.407.2017B,2023MNRAS.526.6103K}. 

In this work, we calibrate the SN and AGN feedback in the \colibre{} model at three resolutions: particle masses of $m_{\rm gas} \approx m_{\rm dm} \sim 10^7$ (m7), $10^6$ (m6), and $10^5~\mathrm{M_\odot}$ (m5). The calibration at m7 resolution is carried out by exploiting machine-learning techniques, following the approach taken by \citet{2023MNRAS.526.6103K} for the \textsc{flamingo} simulations \citep[][]{2023MNRAS.526.4978S}. Based on a modest number of simulations at m7 resolution in ($50~\mathrm{cMpc}$)$^{3}$ volumes, we train Gaussian process emulators that reconstruct the \gsmf{} and \ssm{} from the \colibre{} simulations as smooth functions of a small number of subgrid parameters. We then fit these emulators to the observed \gsmf{} and \ssm{} at $z=0$ in the stellar mass range $10^9 < M_*/\mathrm{M_\odot} < 10^{11.3}$ and obtain the best-fitting values for up to four subgrid parameters. Having calibrated the m7 \colibre{} model, we use it as a starting point to calibrate the m6 and m5 \colibre{} models. The calibration at m6 and m5 resolutions is performed through small, manual adjustments of the subgrid parameters of SN and AGN feedback relative to their best-fitting values at m7 resolution.

This work is structured as follows. In Section \ref{section: simulations} we describe the most relevant aspects of the \colibre{} galaxy formation model. In Section \ref{section: emulators} we present the details of the emulation. In Section \ref{Section: Calibration} we outline our strategy for calibrating the \colibre{} model at m7 resolution using emulators. In Section \ref{Section: Results} we present the results of the m7 calibration and its extensions to the higher resolutions, and we explore the effects of varying individual model parameters. In Section \ref{Section: Conclusions} we summarize our conclusions.
 
\section{Simulations}
\label{section: simulations}

The simulation methods are described in detail in \citet{schaye2025colibreproject}. Here we will provide a summary with an emphasis on the ingredients that are varied during the calibration of the \colibre{} model.

All simulations presented in this work were run with the astrophysical code \textsc{Swift}\footnote{\href{http://www.swiftsim.com}{www.swiftsim.com}} \citep{2024MNRAS.530.2378S}. \textsc{Swift} uses hybrid task-based parallelism to enable scalability to tens of thousands of cores. The equations of hydrodynamics are solved using the smoothed particle hydrodynamics (SPH) method with the density-energy scheme \textsc{Sphenix} \citep{2022MNRAS.511.2367B}, adopting its fiducial values for artificial viscosity and energy conduction. We use the quartic spline kernel with a resolution parameter $\eta = 1.2348$, which is the same value employed in the \textsc{eagle} simulations \citep[see][for details]{2015MNRAS.454.2277S}. For the quartic spline, $\eta = 1.2348$ yields an effective number of neighbours within the kernel support of $N_{\rm ngb} = 64.9$, which satisfies the requirement for accurate density reconstruction from \citet{2012MNRAS.425.1068D}, who showed that reconstruction errors remain low when $\eta$ is close to $1.2$, while much higher values can trigger the pairing instability. We also use $\eta = 1.2348$ for black hole (BH) particles\footnote{Although neither BH nor stellar particles experience hydrodynamic forces, they follow the SPH neighbour search algorithm to locate their gas neighbours, which is necessary for modelling AGN and stellar feedback processes.}, while for stellar particles $\eta$ is reduced to $1.1642$ to lower the computational cost. The gravity is solved with the use of the Fast Multiple Method \citep{1987JCoPh..73..325G} for short-range forces and a particle-mesh method solved in Fourier space for long-range forces. 

In this work, we use the simulation output at redshifts $z=0$ and $2$. To identify subhaloes in the simulation snapshots, we employ the publicly available subhalo finder HBT-HERONS \citep{2025arXiv250206932F}, which is an updated version of the Hierarchical Bound Tracing algorithm \citep[HBT+;][]{2018MNRAS.474..604H}. HBT-HERONS employs a history-based approach to identify subhaloes. The algorithm begins at the earliest simulation snapshot, using an iterative unbinding procedure to find self-bound subhaloes within spatial Friends-of-Friends (FoF) groups, and then processes each subsequent snapshot. Once a self-bound subhalo is identified, HBT-HERONS tracks its associated particles forward in time. Among these, the 10 most gravitationally bound tracer particles (dark matter or stars) are used to link subhaloes to a host FoF group, facilitating the identification of when subhaloes become satellites of a more massive host subhalo. The algorithm keeps track of the particles that were associated to satellite subhaloes before they became satellites. These tracked particles are used at later times to separate the satellite subhalo from the background of its larger host subhalo. At each simulation output time, all subhaloes are checked for self-boundness and phase-space overlap with neighbouring subhaloes, to determine whether they remain self-bound, merge with neighbouring subhaloes, or become disrupted.

We further process the HBT-HERONS output using the Spherical Overdensity and Aperture Processor \citep[SOAP;][]{2025JOSS...10.8252M} to compute a comprehensive set of subhalo properties. The properties that are used in this work include subhalo stellar and halo masses, star formation rates (SFRs), projected stellar half-mass radii, gas and stellar metallicities,  H~\textsc{i} and H$_2$ gas masses, and masses of the most massive BH particles in the subhaloes. Halo masses (for central subhaloes) are computed using the spherical overdensity definition from \citet[][their equation 6, which is a fitting formula to the results from \citealt{1996MNRAS.282..263E}]{1998ApJ...495...80B}, whereas all other galaxy properties are measured within 3D spherical apertures of radius 50 proper kpc (pkpc), considering only gravitationally bound particles associated with each subhalo.

\subsection{Initial conditions}
\label{subsection: ICs}

The ICs of our simulations are produced by the \textsc{monofonIC} code \citep{2020ascl.soft08024H,2021MNRAS.500..663M} using second-order Lagrangian perturbation theory. We follow the 2-field prescription from \citet{2021MNRAS.503..426H} to generate ICs for baryons and DM and take $z=63$ as the starting redshift of the simulations. We use the `3x2pt + all external constraints' cosmology from \citet{2022PhRvD.105b3520A}: $\Omega_{\rm m,0} = 0.306$, $\Omega_{\rm b, 0} = 0.0486$, $\sigma_8 = 0.807$, $h = 0.681$, $n_{s} = 0.967$. We assume a single massive neutrino species with a mass of $0.06$~eV.

The majority of the analysis in this work is based on simulations in a cosmological volume of (50 cMpc)$^{3}$, with a particle mass of $m_{\rm gas} = 1.47 \times 10^{7}~\mathrm{M_\odot}$ for baryons and $m_{\rm dm} = 1.94 \times 10^7~\mathrm{M_\odot}$ for DM. Unless otherwise stated, this volume and resolution (henceforth, m7 resolution) are assumed throughout. The associated Plummer-equivalent gravitational softening length $\varepsilon_{\rm soft}$, which we set to be the same for gas, stellar, BH, and DM particles, is equal to the minimum of $3.6$~ckpc and $1.4$~pkpc. The mass of a DM particle is comparable to that of a gas particle because in the ICs, there are four corresponding DM particles for each gas particle. In total, the ICs of a (50 cMpc)$^{3}$ volume simulation at m7 resolution include $376^3$ gas particles and $4 \times 376^3$ DM particles. As shown by \citet{2019MNRAS.488.3663L, 2021MNRAS.508.5114L, 2023MNRAS.525.5614L}, the finite number of particles in galaxy simulations makes galaxies prone to spurious energy transfer from dynamically hot DM to dynamically cold stars, which tend towards energy equipartition through gravitational interactions. Increasing the DM resolution (or, equivalently, the number of DM particles; e.g., by a factor of 4 relative to the number of gas particles in the ICs) reduces this spurious energy transfer. As a result, stellar particles experience less artificial heating from the DM, leading to more realistic structural and kinematic properties of the stellar components of galaxies.

The only exceptions where we consider volumes different from (50 cMpc)$^{3}$ and resolutions other than m7 are in $\S$\ref{subsection: three resolutions} and Appendix \ref{appendix: boxsize_effect}. In $\S$\ref{subsection: three resolutions}, we compare the fiducial, calibrated \colibre{} models at three resolutions -- m7 ($m_{\rm gas} = 1.47 \times 10^7~\mathrm{M_\odot}$, $m_{\rm dm} = 1.94 \times 10^7~\mathrm{M_\odot}$), m6 ($m_{\rm gas} = 1.8 \times 10^6~\mathrm{M_\odot}$, $m_{\rm dm} = 2.4 \times 10^6~\mathrm{M_\odot}$), and m5 ($m_{\rm gas} = 2.3 \times 10^5~\mathrm{M_\odot}$, $m_{\rm dm} = 3.0 \times 10^5~\mathrm{M_\odot}$) -- all in a (25 cMpc)$^3$ volume; while in Appendix \ref{appendix: boxsize_effect}, we investigate the effect of box size by comparing the fiducial \colibre{} model at m7 resolution in cosmological volumes of $25^3$, $50^3$, $100^3$, and $200^3$ cMpc$^3$. Simulations at m6 (m5) resolution contain 8 (64) times more gas and DM particles in the ICs than their m7 counterparts in the same cosmological volume. The gravitational softening length scales with resolution as $\varepsilon_{\rm soft} \propto m_{\rm gas}^{1/3}$. At a fixed resolution, increasing the cosmological volume by a factor of $2^3$ results in 8 times more gas and DM particles in the ICs. 

\subsection{The COLIBRE model}

In the following sections, we summarize the \colibre{} subgrid model at m7 resolution, which serves as the basis for building the emulators ($\S$\ref{section: emulators}) and performing the emulator-based calibration ($\S$\ref{Section: Calibration}). The \colibre{} models at m6 and m5 resolutions adopt slightly different subgrid parameter values for SN and AGN feedback compared to m7. The adjustments relative to the values reported in this section are detailed in $\S$\ref{subsection: three resolutions} and summarized in table 1 of \citet{schaye2025colibreproject}.

\subsubsection{Radiative cooling, chemistry, and dust}
\label{subsection: cooling_chemistry_and_dust}

The non-equilibrium abundances of hydrogen and helium species and associated free electrons, along with their radiative cooling and heating rates, are computed with the time-dependent thermochemistry solver \textsc{chimes} \citep{2014MNRAS.440.3349R,2014MNRAS.442.2780R}. Additionally, we explicitly track nine heavy elements that contribute most to the cooling rates: C, N, O, Ne, Mg, Si, S, Ca, and Fe. Their contributions to the cooling are provided by \textsc{hybrid-chimes} \citep{2025arXiv250615773P}. \textsc{hybrid-chimes} uses pre-computed cooling tables generated by \textsc{chimes} under the assumption of ionization equilibrium, but the rates are rescaled to account for the difference between the non-equilibrium and equilibrium free electron number densities. The tables also account for cooling due to free-free emission and molecular cooling (including from CO, H$_2$O, OH, HD), while the molecular cooling from H$_2$ is computed in non-equilibrium by \textsc{chimes}. Additionally, we include dust-associated heating and cooling processes using a live dust grain model coupled to \textsc{chimes}, and account for cosmic ray heating, as well as Compton cooling and heating from energy exchange between the gas and photons from the cosmic microwave background and other radiation fields included in \colibre{} (see below).

The cooling rates and the abundances of ions and molecules are evolved assuming the presence of a modified version of the uniform, redshift-dependent UV and X-ray background from \citet{2020MNRAS.493.1614F} (see appendix B of \citet{2020MNRAS.497.4857P} for details), a cosmic ray ionization background, and an interstellar radiation field (ISRF). The shape of the ISRF is constrained by that at the position of the Sun \citep{1987ASSL..134..731B}, combining the local interstellar radiation field \citep{Mathis1983} with the Galactic soft X-ray background \citep{Bregman1986}. For gas with temperatures below $10^4~\mathrm{K}$, the intensities of the cosmic ray background and ISRF scale as\footnote{The power of $1.4$ comes from the observed Kennicutt-Schmidt relation \citep{1998ApJ...498..541K}, $\Sigma_{\rm SFR}\propto \Sigma_{\rm gas}^{1.4}$, where $\Sigma_{\rm SFR}$ and $\Sigma_{\rm gas}$ are the galaxy-averaged star formation rate surface density and gas surface density, respectively.} $N_{\rm Jeans}^{1.4}$, where $N_{\rm Jeans}$ is the Jeans column density of gas with a 1D turbulent velocity dispersion of $6~\mathrm{km \, s^{-1}}$, and saturate at high column densities (see \citealt{2025arXiv250615773P} for details). Shielding by gas and dust is accounted for, assuming a shielding length equal to the Jeans length.

\colibre{} tracks the abundances of 12 individual elements: H, He, C, N, O, Ne, Mg, Si, Fe, Sr, Ba, and Eu\footnote{This set of elements is not identical to those used in the prescription for gas radiative cooling. In particular, the contributions of Sr, Ba and Eu to the cooling rates are neglected, while Ca and S -- the elements used in the radiative cooling calculations -- are not tracked in the \colibre{} chemistry. Instead, to reduce the memory footprint of the simulations, the abundances of Ca and S are assumed to have solar mass ratios relative to Si \citep{doi:10.1146/annurev.astro.46.060407.145222}. This is a reasonable approximation because Ca, S, and Si are all $\alpha$-elements that track each other relatively well \citep{2009MNRAS.399..574W}.}. All of the 12 elements are diffused among the neighbouring SPH gas particles following a velocity shear-based subgrid model for turbulent mixing described in Correa et al. (submitted). 

The \colibre{} simulations incorporate a subgrid model for the formation and evolution of interstellar dust grains, which is described in detail by \citet{2025arXiv250513056T}. Briefly, dust is treated as a scalar field, with gas particles tracking the fraction of their mass that resides in dust grains. The dust model distinguishes between three chemical species of dust grains: graphites and silicates, with the silicates further divided into Mg and Fe flavours. Dust grains are produced in the AGB phase of stellar evolution and in core collapse supernovae (CC SNe). Dust grains grow by accreting mass from the gas phase, while they are destroyed in SN feedback and lose mass in hot gas due to thermal sputtering. Additionally, the model includes two processes that alter grain sizes without changing their total mass: grain shattering and coagulation. We assume that all dust grains have spherical shapes and track two grain sizes: grains with radii of $0.01$ and $0.1$ $\mu \rm m$.

The dust abundances are coupled to the \textsc{chimes} solver, accounting for the distribution of dust mass between the two grain-size bins. The dust is used by \textsc{chimes} to calculate the formation rate of molecular hydrogen on dust grains and other reactions facilitated by dust, as well as to compute dust shielding and dust-associated heating and cooling processes, including dust radiative cooling and photoelectric heating. Additionally, the gas-phase metal abundances (and therefore the metal cooling and heating rates) account for depletion onto dust grains.

\subsubsection{Star formation, stellar evolution, and chemical enrichment}
\label{subsection: stellar evolution}

The \colibre{} prescription for star formation is detailed in \citet{2024MNRAS.532.3299N}. A gas element is labelled as `star-forming' if the gas is locally unstable against gravitational collapse. The instability condition is expressed by requiring that the (absolute) gravitational binding energy of a gas cloud represented by the gas element -- with mass equal to the mass of the gas element multiplied by the effective number of neighbours within the SPH kernel -- exceeds its kinetic energy from thermal and turbulent motions.

The finite resolution of our simulations prevents us from directly following gas collapse into stars. Instead, we convert star-forming gas particles into stellar particles stochastically. To compute the probability of a star-forming gas element becoming a stellar particle, we use the \citet{1959ApJ...129..243S} law with a star formation efficiency per free-fall time of $\varepsilon = 0.01$. 

A stellar particle physically represents a population of many stars that formed simultaneously from a single gas cloud with uniform chemical composition. We assume that all stellar particles are characterized by a \citet{Chabrier2003} stellar initial mass function (IMF) with minimum and maximum masses of $0.1$ and $100~\mathrm{M_\odot}$, respectively. Besides standard properties such as position, velocity, and metallicity, which are inherited from the parent gas particle, a stellar particle is characterized by its age and initial mass. 

Once formed, stellar particles enrich their surrounding gas with metals produced in six chemical enrichment channels: AGB stars, type-Ia SNe, CC SNe, neutron star mergers, common envelope jet SNe, and collapsars (see Correa et al., submitted, for further details).

The stellar feedback model includes three early stellar feedback processes from massive stars: stellar winds, direct radiation pressure, and H~\textsc{ii} regions. To determine the energies, momenta, and ionizing flux injected into the surrounding gas by these feedback processes, we use the Binary Population and Spectral Synthesis (\textsc{bpass}) tables \citep{BPASS2017,BPASS2018} version 2.2.1. These early feedback processes are not calibrated in this work; their numerical implementation and effects are presented in \citet{2025arXiv250925309B}.

\subsubsection{Core collapse supernova feedback}
\label{subsection: CC_SN_feedback}

The \colibre{} model for feedback from CC SNe is a modified version of the thermal-kinetic model of \cite{2023MNRAS.523.3709C}.

The amount of energy in CC SN feedback released by a stellar particle of initial mass $m_*$ over a time-step from $t$ to $t +\Delta t$ is calculated as

\begin{equation}
  \Delta E_{\mathrm{CCSN}} = 10^{51} \, \mathrm{erg} \, f_{\rm E} \, \, m_{\rm *}  \int_{m_{\rm d}(t+\Delta t)}^{m_{\rm d}(t)}  \, \Phi(m) \, \mathrm{d}m \, ,
    \label{eq: energy_per_dt}
\end{equation}
in which $\Phi(m)$ is the \citet{Chabrier2003} IMF and $m_{\rm d}(t)$ denotes the mass of the star(s) that explode as core-collapse SNe at age $t$. We use the metallicity-dependent stellar lifetime tables from \citet{Portinari1998} to compute $m_{\rm d}(t)$. The function $m_{\rm d}(t)$ is non-zero only for zero-age main sequence masses between $m_{\rm min,CCSN} = 8$ and $m_{\rm max,CCSN} = 100~\mathrm{M_\odot}$, which roughly correspond to stellar ages of $\approx 40$ and $3$ Myr, respectively. 

Unlike \cite{2023MNRAS.523.3709C}, we assume that the energy of a single SN in units of $10^{51}$ erg, $f_{\rm E}$, depends on the thermal pressure of the parent gas particle, $P_{\rm birth}$, measured in the time-step it turned into the stellar particle under consideration. The relation between $f_{\rm E}$ and $P_{\rm birth}$ has the following form

\begin{equation}
\label{eq: stellar_birth_pressure_vs_SN_energy}
    f_{\rm E}(P_{\rm birth}) = f_{\rm E, min} +  \frac{\displaystyle  f_{\rm E, max} - f_{\rm E, min}}{\displaystyle 1 +  \exp\left(-\frac{\log_{10} P_{\rm birth} - \log_{10} P_{\rm E, pivot}}{\sigma_{\rm P}}\right)} \, ,
\end{equation}
where $f_{\rm E, min}$ and $f_{\rm E, max}$ are, respectively, the minimum and maximum energies that can be injected by a single SN, in units of $10^{51}$ erg, $P_{\rm E,pivot}$ is a normalization constant, which we will call the pivot birth pressure, and the parameter $\sigma_{\rm P}$ defines the width of the transition from $f_{\rm E, min}$ to $f_{\rm E, max}$. The functional form of Eq. (\ref{eq: stellar_birth_pressure_vs_SN_energy}) implies that the SN feedback of stellar particles born in higher gas pressure environments will be more energetic. In our fiducial setting for m7 resolution, we take $f_{\rm E, min}=0.1$,  $f_{\rm E, max}=4$, and $\sigma_{\rm P}=0.3$, while the best value of $P_{\rm E,pivot}$ will be predicted by emulators (see below). In $\S$\ref{subsection: parameter variations}, we will show how variations in $f_{\rm E, min}$, $f_{\rm E, max}$, and $\sigma_{\rm P}$ affect the simulated galaxies, and discuss how the fiducial values of these three parameters were chosen.

Physically, the dependence of the SN energy on $P_{\rm birth}$ can be interpreted as non-universality of the stellar IMF, which is not unrealistic. In fact, a variable IMF has been suggested by multiple observational studies \citep[e.g.][]{2011MNRAS.415..545T,2012Natur.484..485C,2015MNRAS.447.1033M,2017ApJ...838...77L} and a pressure-dependent IMF has been employed in numerical simulations to successfully reproduce the observational trends \citep[e.g.][]{2018MNRAS.479.5448B,2019MNRAS.483..985B}. Values of $f_{\rm E}$ greater than one can be regarded as accounting for hypernovae \citep[e.g.][]{1999ApJ...516..788W}, and/or as compensation for some degree of numerical overcooling in high-density (and typically also high-pressure) gas \citep[e.g.][]{2006MNRAS.373.1074S,2012MNRAS.426..140D}.

The energy $\Delta E_{\mathrm{CCSN}}$ is injected stochastically into the gas within the SPH kernel of the stellar particle. As in \cite{2023MNRAS.523.3709C}, the parameter $f_{\rm kin}$ is used to split the energy $\Delta E_{\mathrm{CCSN}}$ between the two channels of energy injection: thermal and kinetic. A fraction $f_{\rm kin} \, \Delta E_{\mathrm{CCSN}}$ is injected kinetically, while the remainder, $(1-f_{\rm kin}) \Delta E_{\mathrm{CCSN}}$, is distributed within the gas in thermal form. The value of $f_{\rm kin}$ will be determined using emulators. 

As shown in \citet{2023MNRAS.523.3709C}, the thermal and kinetic channels of energy injection operate side by side with distinct roles: the former generates a hot phase of the ISM and launches strong galactic winds, while the latter drives turbulence in the ISM. This provides galaxies with two complementary means to regulate star formation: by maintaining turbulent support of the gas within the ISM and by ejecting gas from the galaxy.
 
\subsubsection{Thermal channel of energy injection}
\label{subsubsection: sn feedback thermal}

The thermal channel of CC SN feedback utilizes the stochastic model of \cite{2012MNRAS.426..140D}, which was employed in the \textsc{eagle} simulations \citep{2015MNRAS.446..521S}. In the \cite{2012MNRAS.426..140D} model, gas particles receive SN energy from nearby stellar particles with a certain probability, $p_{\rm SN, heat}$. The amount of the injected energy, $\Delta E_{\rm heat}$, is chosen such that following the injection, the gas particle's temperature is increased by a fixed, pre-defined amount, $\Delta T_{\rm SN}$. Mathematically, the relationship between $\Delta E_{\rm heat}$ and $\Delta T_{\rm SN}$ is expressed as

\begin{equation}
    \Delta E_{\rm heat}(m_{\rm gas},\Delta T_{\rm SN}) = \frac{k_{\rm B}\Delta T_{\rm SN}}{(\gamma-1)}\frac{m_{\rm gas}}{\mu m_{\rm p}} \, ,
    \label{eq: SN_heating_energy}
\end{equation}
where $m_{\rm p}$ is the proton mass, $k_{\rm B}$ is the Boltzmann constant, $m_{\rm gas}$ indicates the mass of the gas particle that is being heated, $\gamma=5/3$ is the ratio of specific heats for an ideal monatomic gas, and $\mu=0.59$ is the mean molecular weight of a fully ionized gas. The probability that a given stellar particle heats one of its gas neighbours in some time-step from $t$ to $t + \Delta t$, $p_{\rm SN, heat}$, is calculated as the ratio of the available SN energy in the time-step, $(1-f_{\rm kin}) \, \Delta E_{\mathrm{CCSN}}$, to the energy required to heat the gas mass contained within the stellar kernel, $\Delta E_{\rm heat}(m_{\rm ngb},\Delta T_{\rm SN})$, 

\begin{equation}
    p_{\rm SN,heat} = (1-f_{\rm kin}) \, \frac{\Delta E_{\mathrm{CCSN}}(t, \Delta t, m_*, f_{\rm E})}{\Delta E_{\rm heat}(m_{\rm ngb},\Delta T_{\rm SN})} \, ,
    \label{eq:prob_heat_paper4}
\end{equation}
where $m_{\rm ngb}$ is the sum of the masses of the gas neighbours found within the kernel of the stellar particle. Once $p_{\rm SN,heat}$ has been computed, we start drawing uniform random numbers, $r$, from the interval $0 \leq r < 1$. We draw the random numbers $N_{\rm ngb}$ times where $N_{\rm ngb}$ is the number of the stellar particle's gas neighbours. We then check how many times (out of $N_{\rm ngb}$) the random numbers happened to be smaller than $p_{\rm SN,heat}$. The number of such outcomes determines the number of energy injections the stellar particle will carry out in this time-step. To decide which gas particles within the stellar particle's kernel will receive these energy injections, we adopt the isotropic neighbour selection algorithm from \citet{2022MNRAS.514..249C} with the maximum number of rays set to 8. 

We note that in \textsc{eagle}, thermal energy injections were distributed among gas neighbours with an effectively mass-weighted neighbour selection scheme, as opposed to the isotropic method from \citet{2022MNRAS.514..249C} who showed that the former scheme is biased towards injecting SN energy into high-density gas, and for this reason leads to more radiative energy losses than the isotropic algorithm. A second significant change is the heating temperature increment $\Delta T_{\rm SN}$, which was constant in \textsc{eagle} but depends on the gas density in \colibre.

\subsubsection{A density-dependent heating temperature}
\label{subsubsection: dens_dep_dT}

Specifically, the CC SN feedback in the \textsc{eagle} simulations used a constant heating temperature of $\Delta T_{\rm SN}= 10^{7.5}$ K, with the detailed motivation provided by \citet{2012MNRAS.426..140D}. In short, values greater than $\sim 10^{7.5}$ K would lead to undersampling of SN feedback because the average number of energy injections distributed within the surrounding gas by a single stellar particle over its lifetime, $\langle N_{\rm heat, tot} \rangle$, would be less than $1$. The value of $\langle N_{\rm heat, tot} \rangle$ is computed as

\begin{align}
\langle N_{\rm heat, tot} \rangle  &= \frac{(1-f_{\rm kin}) \, E_{\rm CCSN, tot}(m_*, f_{\rm E}) \, }{\Delta E_{\rm heat}(\langle m_{\rm gas} \rangle,\Delta T_{\rm SN})} \nonumber \, , \\
&= 0.91 \, (1-f_{\rm kin}) \, f_{\rm E} \,  \, \left(\frac{m_{\rm *}}{\langle m_{\rm gas} \rangle}\right)\, \left(\frac{\Delta T_{\rm SN}}{\mathrm{10^{7.5} \, K}}\right)^{-1} \, ,
\label{eq: num_of_thermal_events}
\end{align}
where $E_{\rm CCSN, tot}(m_*, f_{\rm E})= 10^{51} \, \mathrm{erg} \, f_{\rm E} \, \, m_{\rm *}  \int_{m_{\rm min,CCSN}}^{m_{\rm max,CCSN}}  \, \Phi(m) \, \mathrm{d}m$ is the total CC SN energy released by the stellar particle over the course of its lifetime and $\langle m_{\rm gas} \rangle$ is the average mass of its neighbouring gas particles. Assuming that $m_* \approx \langle m_{\rm gas} \rangle$, $f_{\rm E} \sim 1$, and $f_{\rm kin} \ll 1$ and requiring that each star particle on average deposits at least one energy injection in its lifetime (i.e. $\langle N_{\rm heat, tot} \rangle \gtrsim 1$), gives a constraint on the heating temperature $\Delta T_{\rm SN} \lesssim 10^{7.5}~\mathrm{K}$. 

On the other hand, $\Delta T_{\rm SN}$ needs to be high enough to prevent the injected energy from being radiated away before doing work, which would lead to inefficient SN feedback, which is a consequence of the limited resolution \citep[e.g.][]{2006MNRAS.373.1074S,2012MNRAS.426..140D}. Assuming that at high temperatures radiative cooling is dominated by bremsstrahlung, \citet{2012MNRAS.426..140D} showed that the maximum density at which the feedback can remain efficient is

\begin{align}
    n_{\rm H, crit} &= 2 \, \text{cm$^{-3}$} \, \left(\frac{\Delta T_{\rm SN}}{10^{7.5} \, \mathrm{K}}\right)^{3/2} \left(\frac{f_{t}}{10}\right)^{-3/2} \left(\frac{\langle m_{\rm gas} \rangle}{1.5 \times 10^{7} \mathrm{M_\odot}}\right)^{-1/2} \times \nonumber \\ & \left(\frac{\langle N_{\rm ngb}\rangle}{65}\right)^{-1/2} \left( \frac{\mu}{0.6} \right)^{-9/4} \left(\frac{g(X_{\rm H})}{0.14}\right)^{3/2} \, ,
\label{eq: critical_density_SN_feedback}
\end{align}
where $f_t$ is the ratio of the radiative cooling time-scale of the heated gas element to the sound-crossing time-scale across the element and the function $g(X_{\rm H}) = X_{\rm H}^{2/3} (1+X_{\rm H})^{-1}(1+3X_{\rm H})^{-1}$ with $X_{\rm H}$ being the hydrogen mass fraction\footnote{From separate tests, we found that at m7 resolution, the requirement of $f_t=10$ proposed by \citet{2012MNRAS.426..140D} is sufficient but not necessary: somewhat lower values of $f_t$ are acceptable too, as long as $f_t\gtrsim 2$. For example, for $f_t=2$, the critical density for  $\Delta T_{\rm SN}=10^{7.5}~\mathrm{K}$ becomes $\approx 20$ cm$^{-3}$. For comparison, the median density in our simulations at which CC SNe take place is $\sim 1$ cm$^{-3}$.}.

In \colibre{} we exploit equation (\ref{eq: critical_density_SN_feedback}) to allow the heating temperature to vary within a certain range of values, $\Delta T_{\rm SN, min}  < \Delta T_{\rm SN}  < \Delta T_{\rm SN, max}$, monotonically increasing with the gas density. In our fiducial model at m7 resolution, we set $\Delta T_{\rm SN, min}$ and $\Delta T_{\rm SN, max}$ to $10^{6.5}$  and $10^{7.5}$ K, respectively. The use of values lower than $10^{7.5}~\mathrm{K}$ greatly increases the sampling of SN feedback events in low-mass galaxies where the number of stellar particles (and hence SN energy injections) may be small. Moreover, lower $\Delta T_{\rm SN}$ will make SN feedback less destructive in gas environments with relatively low densities, which potentially alleviates the problem of overly large SN-driven bubbles identified in the \textsc{eagle} simulations \citep{2016MNRAS.456.1115B}.

More precisely, we assume that the value of the heating temperature, $\Delta T_{\rm SN}$, depends on the average (physical) gas density at the location of the star particle, $\rho_{\rm SN}$, which is estimated in the time-step when the star particle does SN feedback. We compute $\rho_{\rm SN}$ as

\begin{equation}
    \rho_{\rm SN} = \sum_{i=1}^{N_{\rm ngb}} \, m_{\mathrm{gas},i} \,  W_{\rm}(|\boldsymbol{r}_* - \boldsymbol{r}_{\mathrm{gas},i}|, h_{\rm *}) \, ,
\label{eq: density_estimate_SPH}
\end{equation}
where the sum runs over all gas particles within the stellar kernel, $m_{\mathrm{gas},i}$ is the mass of gas particle $i$, $\boldsymbol{r}_*$ and $\boldsymbol{r}_{\mathrm{gas},i}$ are the coordinates of the stellar particle and gas particle $i$, respectively, and $W_{\rm}$ is the SPH kernel function with the stellar particle's smoothing length $h_{\rm *}$. After having computed $\rho_{\rm SN}$, we convert it to a hydrogen number density $n_{\rm H,SN}$ assuming primordial abundances, with the hydrogen mass fraction of $X_{\rm H}=0.756$, and calculate $\Delta T_{\rm SN}$ as 

\begin{equation}
\label{eq: heating_temperature_vs_gas_density}
   \Delta T_{\rm SN}(n_{\rm H,SN}) = \Delta T_{\rm SN, pivot} \, \left(\frac{n_{\rm H,SN}}{n_{\rm H, pivot}}\right)^{2/3} \, ,
\end{equation}
in which $\Delta T_{\rm SN, pivot}$ and $n_{\rm H, pivot}$ are free parameters and the slope of $2/3$ is motivated by the cooling argument from \citet{2012MNRAS.426..140D}, our equation (\ref{eq: critical_density_SN_feedback}). Because of the degeneracy between $n_{\rm SN, pivot}$ and $\Delta T_{\rm SN, pivot}$ ($\Delta T_{\rm SN, pivot} \propto n_{\rm H, pivot}^{2/3}$), we fix $\Delta T_{\rm SN, pivot}$, setting it to $10^{6.5}~\mathrm{K}$ at m7 resolution and only consider $n_{\rm H, pivot}$ in the following.

\subsubsection{Kinetic channel of energy injection}
\label{subsection: CC_SN_feedback_kin}

The remaining part of CC SN energy, which is not used up in the thermal channel, $f_{\rm kin} \, \Delta E_{\rm CCSN}$, is released in kinetic form, following a modified version of the stochastic kinetic model of \citet{2008MNRAS.387.1431D}. The full details of our algorithm for SN kinetic feedback are presented in \cite{2023MNRAS.523.3709C}. Briefly, stellar particles inject kinetic energy with a probability

\begin{equation}
    p_{\rm kick,pair} = f_{\rm kin} \frac{\Delta E_{\rm CCSN}(t, \Delta t, m_*, f_{\rm E}) }{2 \Delta E_{\rm kick}(m_{\rm ngb}, \Delta v_{\rm kick})} \nonumber \, ,
    \label{eq: prob_kick_paper4}
\end{equation}
where $\Delta E_{\rm kick}(m_{\rm ngb}, \Delta v_{\rm kick}) = m_{\rm ngb} \, \Delta v_{\rm kick}^2 / 2$ and $\Delta v_{\rm kick}$ is the desired kick velocity. \cite{2023MNRAS.523.3709C} showed that low values of $\Delta v_{\rm kick}$, such as $50$ km s$^{-1}$, help drive turbulence in the neutral ISM and improve the agreement with the observed spatially resolved relation between H~\textsc{i} velocity dispersion and SFR surface density in nearby galaxies. Based on these findings, in this work, we adopt $\Delta v_{\rm kick} = 50$ km s$^{-1}$. The effect of varying $\Delta v_{\rm kick}$ between $10$ and $10^3$ km s$^{-1}$ can be found in \cite{2023MNRAS.523.3709C}.

Similarly to the thermal SN feedback, once we know $p_{\rm kick,pair}$, we draw a random number $N_{\rm ngb}$ times from an interval of $0 \leq r < 1$. The number of kick events that the stellar particle will distribute in the time-step from $t$ to $t+\Delta t$ is equal to the number of times the condition $r < p_{\rm kick,pair}$ is found. In each kick event, the stellar particle kicks \textit{two} of its gas neighbours in opposite directions, which is necessary to conserve linear momentum. Additionally, the model ensures that angular momentum and energy in SN feedback are exactly conserved\footnote{The exact conservation of energy is realized by accounting for the relative motion between stellar particles and their gas neighbours. Owing to the relative velocity corrections, gas particles may experience velocity kicks that are greater or smaller than the desired kick velocity $\Delta v_{\rm kick}$.} and the injected energy is distributed statistically isotropically. For further details, we refer the reader to \cite{2023MNRAS.523.3709C}.

\subsubsection{Type-Ia supernovae}
\label{subsection: Ia_SN_feedback}

We implement type-Ia SN feedback as a purely thermal ($f_{\rm kin}=0$) isotropic stochastic feedback following the `isotropic' algorithm from \citet{2022MNRAS.514..249C}. We assume that the heating temperature in type-Ia SN feedback scales with the gas density in the same way as for CC SN feedback, following equation (\ref{eq: heating_temperature_vs_gas_density}), where the values of $\Delta T_{\rm SN, min}, \Delta T_{\rm SN, max}$, $\Delta T_{\rm SN, pivot}$, $n_{\rm H, pivot}$, and the maximum number of rays are set to those from CC SN feedback. 

To calculate the energy budget for type-Ia SN feedback executed by one stellar particle, we use a delay time distribution (DTD),
\begin{equation}
    \mathrm{DTD}(t) = \frac{\nu}{\tau} \exp\left(-\frac{t-t_{\rm delay}}{\tau}\right) \Theta(t-t_{\rm delay}) \, , 
    \label{eq:dtd_SNIa}
\end{equation}
in which $\nu = 1.54 \times 10^{-3} \, \rm M^{-1}_{\odot}$ is the total number of type-Ia SNe that will ever occur per unit initial stellar mass, $\tau = 2$ Gyr is the type-Ia SN time-scale, and $\Theta(x)$ is the Heaviside step function. Nobels et al. (in preparation) show that this form of DTD results in good agreement with the observed rates of type-Ia SNe. 

As was the case for CC SNe, the energy of type-Ia SNe released by one stellar particle corresponds to the combined energy from many individual type-Ia SNe that are not resolved in our simulations. We set the time $t_{\rm delay}$ to $40$ Myr, which marks the delay since the birth of the stellar particle before the first unresolved, individual type-Ia SN has gone off and contributed its energy to the stellar particle's total energy. The energy from all individual type-Ia SNe in a time-step $[t,t+\Delta t)$ is calculated by integrating equation (\ref{eq:dtd_SNIa}) from $t$ to $t+\Delta t$ and assuming an energy per individual type-Ia SN of $10^{51}$ erg. 

Energetically, type-Ia SN feedback is subdominant to that from CC SNe, and its presence has only a minor impact on the galaxy properties relevant for the calibration of the \colibre{} simulations  (see Nobels et al., in preparation, for more details). Unless stated otherwise, all discussions about SN feedback in the following text will refer entirely to CC SN feedback.

\subsubsection{Supermassive black holes}
\label{subsection: BH_and_AGN}

In galaxy simulations, supermassive black holes (SMBHs) are represented by collisionless BH particles, which can grow in mass by accreting surrounding gas and/or by merging with other BH particles \citep[e.g.][]{2005MNRAS.361..776S,2009MNRAS.398...53B}.

We employ an on-the-fly FoF group finder to seed BH particles in the simulation \citep[e.g.][]{2008ApJ...676...33D}. The FoF algorithm uses a linking length of $0.2$ times the mean dark matter inter-particle separation and is executed every $\Delta a = 0.00751 \, a$, starting at $a=0.05$, where $a$ is the cosmic scale factor. At m7 resolution, BHs are seeded in haloes whose FoF mass is greater than $5\times 10^{10}~\mathrm{M_\odot}$ and that do not already harbour a BH particle.

During seeding, we identify the densest gas particle in the FoF halo and convert it into a BH particle, which inherits the gas particle’s dynamical mass, velocity, and position. For all mass-dependent processes that are modelled in a subgrid fashion, such as gas accretion onto BHs and energy feedback, we use the BH subgrid mass, as opposed to the dynamical mass of the particle, in order to allow BH masses smaller than the particle mass \citep[e.g.][]{2005MNRAS.361..776S,2009MNRAS.398...53B}. The subgrid mass is initially equal to the seed mass, $m_{\rm BH, seed}$. The value of $m_{\rm BH, seed}$ will be calibrated using emulators.

The (instantaneous) mass accretion rate onto a BH particle is computed using a modified Bondi-Hoyle-Lyttleton formula \citep{Krumholz_et_al_2006},

\begin{equation}
\label{eq: SMBH_acc_rate}
    \dot{m}_\mathrm{accr} = 4\uppi\,G^2 \frac{m_\mathrm{BH}^2\,\rho_\mathrm{gas}}{c_\mathrm{sound}^3}\left[\frac{(1 + \mathscr{M}^2)^4}{1.1^2 + \mathscr{M}^2} + \frac{1}{(0.34\,f_\star)^2} \right]^{-1/2}\, ,
\end{equation}
where $\mathscr{M}^2 = (\sigma_\mathrm{turb}/c_\mathrm{sound})^2 + (v_\mathrm{gas}/ c_\mathrm{sound})^2$ is the Mach number squared, $f_{\star} = 1/[1 + (\omega\,r_\mathrm{Bondi}/ c_\mathrm{sound})^{0.9}]$ is the correction due to vorticity in the gas flow with $\omega$ being the vorticity, and $r_\mathrm{Bondi} = G m_\mathrm{BH}/c_\mathrm{sound}^2$ is the Bondi radius with $m_{\rm BH}$ being the subgrid mass of the BH particle. The magnitude of the gas bulk
velocity, $v_{\rm gas}\equiv|\boldsymbol{v}_\mathrm{gas}$|, the gas turbulent velocity dispersion $\sigma_\mathrm{turb}$, and the vorticity $\omega \equiv |\nabla \times \boldsymbol{v}_\mathrm{gas}|$ are all computed as mass-weighted averages over all gas neighbours within the kernel of the BH particle. We calculate the gas mass density, $\rho_{\rm gas}$, in the standard SPH way by applying equation (\ref{eq: density_estimate_SPH}) to the gas neighbours within the BH kernel. Finally, in \colibre{} the mass accretion rate, $\dot{m}_\mathrm{accr}$, is capped at 100 times the mass accretion rate at the Eddington luminosity.

Following \citet{2022MNRAS.516..167B}, two BH particles will merge if the distance between them, $\Delta r_{\rm BH}$, is less than three gravitational softening lengths, $\Delta r_{\rm BH} < 3 \varepsilon_{\rm soft}$; the less massive BH is within the kernel of the more massive BH; and if their relative velocity $\Delta v_{\rm BH}$ satisfies $\Delta v_{\rm BH} < \sqrt{2G(M + m)/\Delta r_{\rm BH}}$ where $M$ and $m$ are the dynamical masses of the larger and smaller merging BH particle, respectively. Once the merger criteria are simultaneously satisfied, the BHs are instantaneously merged. 

SMBHs are thought to be subject to significant dynamical friction, which causes them to lose their orbital energy and spiral in towards the centre of the host galaxy \citep[e.g.][]{1999ApJ...513..252O}. Because galaxy simulations of representative volumes lack the resolution to properly capture the effects of dynamical friction, SMBHs have to be `pushed' towards the centre of the galaxy with an ad hoc prescription \citep[e.g.][]{2008ApJ...676...33D,2022MNRAS.516..167B}. In \colibre{} we follow the method of \citet{2022MNRAS.516..167B} where at every time-step $\Delta t$, each BH particle searches for the gas particle within its SPH kernel that has the lowest gravitational potential. If this gas particle is also within three gravitational softening lengths of the BH and has a lower potential than at the BH's current position, the BH is immediately `re-positioned' to the location of that gas particle. The velocity of the BH remains unchanged during the re-positioning. As recommended by  \citet{2022MNRAS.516..167B}, when selecting the gas neighbour with the lowest gravitational potential, we exclude the contribution of the BH particle to the potential.

\subsubsection{Feedback from AGN}
\label{subsubsection: AGN_feedback}

The \colibre{} suite includes simulations with purely thermal AGN feedback as well as simulations with a hybrid AGN feedback mode that combines BH-spin dependent kinetic jets and thermal energy injections, with the largest \colibre{} volumes available only for the thermal models. This work focuses entirely on calibrating the \colibre{} simulations with purely thermal AGN feedback. The calibration of simulations with hybrid AGN feedback, which builds upon the results of this study, is described in \citet{2025arXiv250905179H}.

The purely thermal AGN feedback from SMBHs is implemented following \citet{2009MNRAS.398...53B} and is similar to that used in \textsc{eagle} \citep{2015MNRAS.446..521S}. Out of the total gas mass accreted by a BH particle over a given time-step from $t$ to $t+\Delta t$, $\dot{m}_\mathrm{accr} \Delta t$, the BH receives\footnote{If the updated subgrid mass of the BH particle, $m_{\rm BH}^{\rm new} = m_{\rm BH} + \Delta m_{\rm BH}$, is greater than its dynamical mass at the beginning of the time-step, $m_{\rm BH}^{\rm dyn}$, then the value of $m_{\rm BH}^{\rm dyn}$ is increased to $m_{\rm BH}^{\rm new}$. To ensure the conservation of mass, the mass deficit, $m_{\rm BH}^{\rm new} - m_{\rm BH}^{\rm dyn}$ is `nibbled' from the mass of the gas particles that reside within the BH kernel, following the method of \citet{2022MNRAS.516..167B}. Conversely, if $m_{\rm BH}^{\rm new}$ is less than $m_{\rm BH}^{\rm dyn}$, then we assume that the difference $m_{\rm BH}^{\rm dyn}-m_{\rm BH}^{\rm new} > 0$ represents a subgrid gas reservoir around the BH and all accreted mass comes therefrom. We then only reduce $m_{\rm BH}^{\rm dyn}$ by $\varepsilon_{\rm r} \Delta m_{\rm BH}$ to account for the energy that has been converted into radiation. No gas particle's mass is nibbled in this case.} a fraction
\begin{equation}
   \Delta m_{\rm BH} = (1-\varepsilon_{\rm r}) \dot{m}_\mathrm{accr} \Delta t\, ,
\end{equation}
where $\varepsilon_{\rm r}$ is the radiative efficiency. The remaining mass, $\varepsilon_{\rm r} \dot{m}_\mathrm{accr} \Delta t$, is assumed to have been converted into energy that escapes the BH as radiation, a fraction of which is coupled to the gas surrounding the BH. The energy received by the gas in the time-step $\Delta t$,  $\Delta E_{\rm AGN}$, is
\begin{equation}
\label{eq: AGN_energy}
    \Delta E_{\rm AGN} = \varepsilon_{\rm f} \varepsilon_{\rm r} \, \dot{m}_\mathrm{accr} \, c^2 \, \Delta t \, ,
\end{equation}
where $\varepsilon_{\rm f}$ is the coupling efficiency. For both $\varepsilon_{\rm r}$ and $\varepsilon_{\rm f}$, we adopt a value of $0.1$ at m7 resolution. The former is motivated by theoretical considerations \citep{1973A&A....24..337S}, while the latter was chosen to yield realistic $z=0$ SMBH masses in high-mass $z\approx 0$ galaxies.

As is the case with stellar feedback (see discussion in $\S$\ref{subsubsection: dens_dep_dT}), injecting the energy $\Delta E_{\rm AGN}$ into surrounding gas may result in numerical overcooling if $\Delta E_{\rm AGN}$ is insufficient to increase the temperature of the gas in which the energy is injected to values high enough for the cooling time to be long. Following \citet{2009MNRAS.398...53B}, we wait until a sufficiently large amount of energy has been accumulated by the accreting BH. Numerically, this is achieved by having each BH particle carry an energy reservoir, $E^{\rm reservoir}_{\rm AGN}$, which is empty upon BH seeding but whose energy is increased at every time-step by the value of $\Delta E_{\rm AGN}$ for that time-step. Once the energy in the reservoir exceeds a threshold energy $\Delta E_{\rm AGN,thr}$, we inject the energy $\Delta E_{\rm AGN,thr}$ into one of the gas particles within the SPH kernel of the BH particle and subtract an equivalent amount of energy from the reservoir. We define $\Delta E_{\rm AGN,thr}$ as the energy that results in a temperature increase of the heated gas neighbour by $\Delta T_{\rm AGN}$, $\Delta E_{\rm AGN,thr}\equiv \Delta E_{\rm heat}(\langle m_{\rm gas} \rangle,\Delta T_{\rm AGN})$, where $\langle m_{\rm gas} \rangle$ is the average gas particle mass in the BH's kernel and the expression for $\Delta E_{\rm heat}(\langle m_{\rm gas} \rangle,\Delta T_{\rm AGN})$ is given by equation (\ref{eq: SN_heating_energy}). If the BH particle accretes rapidly and/or its time-step is very long, the energy in the reservoir $E^{\rm reservoir}_{\rm AGN}$ may temporarily exceed $N_{\rm AGN} \, \Delta E_{\rm AGN,thr}$ where $N_{\rm AGN}$ is the maximum number of particles that can be heated (see below). In this case, the energy $\Delta E_{\rm AGN,thr}$ is injected into $N_{\rm AGN}$ gas neighbours, and $E^{\rm reservoir}_{\rm AGN}$ is reduced by $N_{\rm AGN} \, \Delta E_{\rm AGN,thr}$.

For all simulations used in the calibration of the \colibre{} model at m7 resolution, we employ $\Delta T_{\rm AGN} = 10^9~\mathrm{K}$. This value, which is the same as that used in the \textsc{eagle-recal} model \citep{2015MNRAS.446..521S}, ensures that AGN feedback is efficient and well-sampled in massive haloes. However, after completing the calibration simulations and identifying the best-fitting m7 model with $\Delta T_{\rm AGN} = 10^9~\mathrm{K}$, we found that a fixed value of $\Delta T_{\rm AGN} = 10^9~\mathrm{K}$ was not the optimal choice for higher resolutions. Instead, a BH mass-dependent $\Delta T_{\rm AGN}$ (equation~\ref{eq: variable_agn_temperature}) proved to be a better option. Therefore, while the calibration at m7 resolution (described in $\S$\ref{section: emulators} and $\S$\ref{Section: Calibration}) is based on $\Delta T_{\rm AGN} = 10^9~\mathrm{K}$, the fiducial \colibre{} m7 model adopts a variable $\Delta T_{\rm AGN}$, to be consistent with the fiducial models at higher resolutions. This choice is justified in more detail in $\S$\ref{subsection: three resolutions}, where we also present comparisons between the best-fitting m7 model with $\Delta T_{\rm AGN} = 10^9~\mathrm{K}$ and the fiducial \colibre{} model with variable $\Delta T_{\rm AGN}$ in Figs. \ref{fig: model_four_vs_final} and \ref{fig: model_four_vs_final_2}. To avoid confusion, the \colibre{} model with fixed $\Delta T_{\rm AGN} = 10^9~\mathrm{K}$ will consistently be referred to as such throughout the text. The final version, which employs a variable $\Delta T_{\rm AGN}$, will be referred to as the fiducial model, the \colibre{} model with variable $\Delta T_{\rm AGN}$, or simply the \colibre{} model.

To select the gas neighbours that will receive the energy $\Delta E_{\rm AGN,thr}$, we employ the `Minimum Distance' algorithm from \citet{2022MNRAS.514..249C}. If a BH particle needs to distribute $N_{\rm AGN}\geq 1$ energy injections among its neighbours, then the $N_{\rm AGN}$ closest neighbours each receive one energy injection. As in \citet{2022MNRAS.516..167B}, the maximum number of gas neighbours a BH particle can heat in a single time-step, $N_{\rm BH, max}$, is set to $50$\footnote{In rare events where $N_{\rm AGN}$ is greater than $N_{\rm BH, max}$, we increase $\Delta T_{\rm AGN}$ by $N_{\rm AGN}/N_{\rm BH, max}$ and heat the $N_{\rm BH, max}$ closest neighbours using the updated value of $\Delta T_{\rm AGN}$. Additionally, if the number of gas particles within the kernel of the BH, $N_{\rm ngb}$, is smaller than $\mathrm{min}(N_{\rm AGN},N_{\rm BH, max})$, then $\Delta T_{\rm AGN}$ is raised by $\mathrm{min}(N_{\rm AGN},N_{\rm BH, max})/N_{\rm ngb}$ and all $N_{\rm ngb}$ particles receive the energy corresponding to the updated $\Delta T_{\rm AGN}$.}.

\section{Emulators}
\label{section: emulators}

We use Gaussian process emulators to determine the optimal values of the subgrid parameters for SN and AGN feedback in the \colibre{} model at m7 resolution and to demonstrate that simplified prescriptions for SN feedback with reduced numbers of free parameters cannot provide an equally good fit to the target observational data. We construct Gaussian process emulators using the python package \textsc{swift-emulator} \citep{Kugel2022} and follow the method\footnote{However, unlike \citet{2023MNRAS.526.6103K}, we do not consider the bias parameters for stellar mass and cosmic variance, which \citet{2023MNRAS.526.6103K} found to have a negligible effect on the calibration of the \textsc{flamingo} simulations.} from \citet{2023MNRAS.526.6103K}, who employed Gaussian process emulators to calibrate the large (up to $2.8^3~\mathrm{cGpc}^3$) cosmological simulations \textsc{flamingo} \citep{2023MNRAS.526.4978S}, but which have lower resolution and use a simpler galaxy formation model than \colibre. 

We set up $\approx 200$ simulations that sample the part of the \colibre{} parameter space of interest here at unique points, utilizing the Latin hypercube sampling technique (see $\S$\ref{subsection: training set}). These simulations are used to train emulators in order to `interpolate' to other points in the parameter space. That is, the emulators will provide a continuous reconstruction of the parameter space, without requiring us to run any additional simulations. With the trained emulators, the search for the best-fitting values of subgrid parameters will be simplified to the minimization of the error between the emulator predictions and the target observational data.

We apply this emulation method exclusively to the m7 resolution, where we run all simulations for the emulator training set in a ($50~\mathrm{cMpc}$)$^3$ volume. Emulators are not used for calibrating the m6 and m5 \colibre{} models, as running simulations at m6 and m5 resolutions in the same cosmological volume would be prohibitively computationally expensive. Reducing the cosmological volume to compensate for the increased computational cost is not a viable solution either, as the absence of relatively massive galaxies ($M_{\rm halo} \gtrsim 10^{13}~\mathrm{M_\odot}$) in smaller volumes would prevent the emulators from properly calibrating the strength of AGN feedback. Instead, at m6 and m5 resolutions, we chose to calibrate the model manually, using the subgrid parameter values of the calibrated m7 model as an initial guess. The ability to start from the best-fitting parameter values determined at m7 resolution -- combined with the relatively good convergence of the \colibre{} model with resolution and insights into the model's response to subgrid parameter variations gained through the emulators -- makes manual calibration at m6 and m5 resolutions feasible.

We note that an alternative approach to calibrating higher-resolution models is to perform the emulator-based calibration at m7, m6, and m5 resolutions \textit{simultaneously}, treating the gas particle mass as an additional emulator parameter. In this setup, if the model exhibits reasonable convergence with resolution, the higher-resolution simulations can be run in progressively smaller volumes to reduce the computational cost of generating training data. The emulator then learns the properties of massive haloes from the lower-resolution simulations and, by using the (lower-mass) haloes present at all three resolutions, can potentially still infer how the properties of those massive haloes -- absent from the higher-resolution training sets -- vary with resolution. Compared to the manual approach adopted in this work, this alternative method has the advantage of incorporating resolution effects directly into the emulator’s dependence on a single parameter, $m_{\rm gas}$, making it straightforward to study resolution effects and enabling the simultaneous determination of best-fitting parameter values across all resolutions. Its main drawback (and the reason we did not adopt it) is that it requires significantly more prior knowledge of the model, as each iteration of narrowing the parameter space in search of the best-fitting model becomes substantially more computationally expensive due to the inclusion of higher-resolution simulations, even when these are run in smaller cosmological volumes.

\subsection{Setup}
\label{subsection: emulator_setup}

Consider a smooth mapping $y = f(x,\boldsymbol{\theta})$, where an output $y$ depends on an input scalar variable $x$ and a parameter vector $\boldsymbol{\theta}$\footnote{As we will show later in $\S$\ref{subsection: selection_theta}, these parameters will be combinations of various subgrid parameters of the \colibre{} SN and AGN feedback.}. Assuming that we know the true relation $y = f(x,\boldsymbol{\theta})$ only at a \textit{finite} number of points $N$, denoted $\{x_n,\boldsymbol{\theta}_n,y_n=f(x_n,\boldsymbol{\theta}_n)\}_{n=1}^{N}$, our goal is to use this limited information to approximate $y = f(x,\boldsymbol{\theta})$ throughout the joint input space of $x$ and $\boldsymbol{\theta}$. When $f$ represents a complex, computationally expensive model (such as the \colibre{} galaxy formation model, where the data are generated by running numerical simulations), this approximation process is termed \textit{emulation}. 

We will write a hat above `$f$' to distinguish an emulator, $y = \hat{f}(x,\boldsymbol{\theta})$, from the true relation, $y = f(x,\boldsymbol{\theta})$. The size of vector $\boldsymbol{\theta}$ is equal to the number of parameters on which the emulator depends. In the following, we will write $N_{\rm param}$ as a short-hand notation for the length of $\boldsymbol{\theta}$.

A Gaussian process with zero mean is fully specified by its covariance function \citep[e.g.][]{books/lib/RasmussenW06}. As in \citet{2023MNRAS.526.6103K}, we construct the covariance function using the squared exponential kernel,
\begin{equation}
\label{eq: covariance_function}
    k(\boldsymbol{X},\boldsymbol{X}^\prime) = \exp\left[ -\frac{(\boldsymbol{X} -\boldsymbol{X}^\prime)^{\rm T} \mathbf{C}^{-1} (\boldsymbol{X} -\boldsymbol{X}^\prime)}{2}\right] \, ,
\end{equation}
where the vectors $\boldsymbol{X} = (x, \boldsymbol{\theta})$ and $\boldsymbol{X}^\prime = (x^\prime, \boldsymbol{\theta}^\prime)$ correspond to two different points in the $N_{\rm param}+1$-dimensional parameter space, and $\mathbf{C}$ is a diagonal matrix that sets the length scale for each dimension of the parameter space. We do not opt for more sophisticated kernels because the relations we emulate vary smoothly with $\boldsymbol{X}$ across the entire parameter space. The entries of the matrix $\mathbf{C}$ are optimised during training of the Gaussian process emulators by maximizing the log marginal likelihood of the Gaussian process \citep[see, e.g.,][]{books/lib/RasmussenW06}.

\subsection{The emulated relations}
\label{subsection: simulated_relations}

We emulate two relations that are used to calibrate the \colibre{} SN and AGN feedback:
\begin{itemize}
    \item \textit{Galaxy stellar mass function} (\gsmf) at $z=0$. Here the input variable $x$ is the galaxy stellar mass, $M_*$, and the output variable $y$ is the number of galaxies per unit volume, $\mathrm{d}n$, per logarithmic bin of stellar mass, $\mathrm{d}\log_{10} M_*$. Because the stellar mass $M_*$ can span many orders of magnitude, we perform the emulation in log space, adopting $x=\log_{10} M_*$ as opposed to $x = M_*$. Likewise, for the output $y$, we take $y\equiv f(x)=\log_{10}(\mathrm{d}n/\mathrm{d}\log_{10} M_*)$, as opposed to $\mathrm{d}n/\mathrm{d}\log_{10} M_*$.

    \item \textit{Size -- stellar mass relation} (\ssm) at $z=0$. Here the input is again $x=\log_{10} M_*$, while the output is $\log_{10}$ of the median projected stellar half-mass radius of the simulated galaxies whose stellar mass is $M_*$.
\end{itemize}

For each subhalo, the stellar mass $M_*$ is computed as the sum of the masses of gravitationally bound stellar particles within a 3D spherical aperture of 50 pkpc, centred on the position of the most bound particle (of any type). By conducting mock observations of galaxies from the \textsc{eagle} simulations, \citet{2022MNRAS.511.2544D} found that this choice of aperture yields results similar to the masses inferred from fitting S\'ersic profiles, which is a method frequently used in observations. For both emulated relations, we take the $x$ values from the simulations and arrange them in bins of equal size of $\Delta \log_{10} (M_* / \mathrm{M_\odot}) = 0.2$. For the \ssm, we then compute the median projected galaxy half-mass radius in each bin, while for the \gsmf, we count the number of objects in each mass bin and divide it by the simulated volume and by the logarithmic width of the bin. Projected stellar half-mass radii, like $M_*$, are calculated in 50 pkpc 3D apertures, considering only stellar particles that are gravitationally bound to the subhalo.

Before binning the simulated data, we shift all stellar masses by $\Delta M_*$ where $\Delta M_*$ is drawn from a lognormal distribution with zero mean and a standard deviation of $0.1$~dex. This adjustment accounts for the \citet{1913MNRAS..73..359E} bias, which affects observational data and thus must also be applied in the simulation. The choice of $0.1$~dex reflects a typical uncertainty in stellar mass measurements \citep[e.g.][]{2020MNRAS.495..905R}. Throughout this work, we apply this correction not only to the \gsmf{} and \ssm, but also to all other simulation-predicted relations where the independent variable is galaxy stellar mass.

The uncertainties in the $y$ values of the simulated data are accounted for as follows. For the \ssm, the error $\Delta y$ is defined as half the difference between the 84$^{\rm th}$ and 16$^{\rm th}$ percentiles of the distribution of $y$ values within each stellar mass bin. For the \gsmf, the error $\Delta y$ is given by the Poisson uncertainty. We use the square of $\Delta y$ to define the diagonal entries of a noise covariance matrix (which is otherwise zero). This noise matrix is added to the Gaussian process covariance matrix, constructed using the kernel function (equation~\ref{eq: covariance_function}) and the training data, to form the total covariance for the Gaussian process. Including this noise term prevents overfitting to the training data, yielding smoother (i.e., less oscillatory) emulated relations. To optimise the smoothness, we introduce a hyperparameter that rescales the noise matrix. The value of this hyperparameter is set during training of the Gaussian process emulators, alongside the optimisation of the diagonal elements of the matrix $\mathbf{C}$ from equation~(\ref{eq: covariance_function}).

Additionally, we emulate \textit{the stellar-to-halo mass relation} (\shmr) for central subhaloes. This relation is not used to calibrate the \colibre{} model due to its weak observational constraints, but serves as a diagnostic tool. Our halo mass, $M_{\rm halo}$, follows the spherical overdensity definition of \citet{1998ApJ...495...80B}. The emulator input consists of $x = \log_{10} M_{\rm halo}$ and $y = \log_{10} (M_*/M_{\rm halo})$. The $y$ values represent the median stellar-to-halo mass ratios computed in halo mass bins of 0.2 dex width, with uncertainties given by half the difference between the 84$^{\rm th}$ and 16$^{\rm th}$ percentiles of the distribution of individual subhalo values in each mass bin. 

\subsection{Target observational data}
\label{subsection: target_observational_data}

We now describe the observational data to which the \gsmf{} and \ssm{} emulators from $\S$\ref{subsection: simulated_relations} will be fit. The fitting will be performed over the galaxy stellar mass range $10^9 < M_* / \mathrm{M_\odot} < 10^{11.3}$. The lower bound is set by resolution constraints: at m7 resolution, $M_* = 10^9~\mathrm{M_\odot}$ corresponds to only $\sim 100$ stellar particles. The upper bound is set by the relatively small number of galaxies with $M_* > 10^{11.3}~\mathrm{M_\odot}$ in the (50 cMpc)$^3$ volume simulations that are used to construct the emulator training data.

\subsubsection{Galaxy stellar mass function at $z=0$}

The \gsmf{} provides one of the most stringent constraints on the evolution of stellar mass in the Universe: it determines not only the total stellar mass formed in the Universe but also the relative abundance of low- and high-mass galaxies. We constrain the simulated $z=0$ \gsmf{} by matching it to the $z=0$ \gsmf{} from \citet{2022MNRAS.513..439D} (their table 6, column ‘all’, including the $0.0807$ dex correction accounting for the re-normalization to the Sloan Digital Sky Survey (SDSS) and evolution to precisely $z=0$). The \citet{2022MNRAS.513..439D} \gsmf{} was derived from the Galaxy And Mass Assembly, Data Release 4 (GAMA DR4) survey, which provides multi-wavelength observations for over $200,000$ galaxies with spectroscopically confirmed redshifts. The \citet{2022MNRAS.513..439D} \gsmf{} spans $\sim 5$~dex in stellar mass, extending down to $M_*\sim 10^7~\mathrm{M_\odot}$, and is presented precisely at $z=0$, as the authors account for redshift evolution between the median redshift of their sample ($z\approx 0.079$) and $z=0$. The stellar masses were computed assuming a \citet{Chabrier2003} IMF.

\subsubsection{Galaxy size -- stellar mass relation at $z=0$}

Reproducing the observed \gsmf{} does not guarantee that other properties of the simulated galaxies, besides stellar mass, will be realistic. For example, while the \gsmf{} provides constraints on the total stellar masses of galaxies, it says nothing about how that stellar mass is distributed within the galaxies. Indeed, \citet{2015MNRAS.450.1937C} showed that depending on the subgrid model adopted in the numerical simulation, simulated galaxies may be described by the same \gsmf, but differ drastically in their stellar half-mass radii. 

We take the galaxy stellar mass -- size relation from \citet{2022MNRAS.509.3751H} (their table C1, column `Half-mass radius') as a secondary constraint on our simulations. The \citet{2022MNRAS.509.3751H} data come from the eXtended GALEX Arecibo SDSS Survey \citep[xGASS;][]{2018MNRAS.476..875C}, which contains $\sim 1200$ galaxies selected from the SDSS DR7 catalogue. The sample has a flat stellar mass distribution spanning $M_* \approx 10^9$ to $10^{11.5}~\mathrm{M_\odot}$ and redshifts in the range $0.01 < z < 0.05$. \citet{2022MNRAS.509.3751H} provide the median galaxy half-mass radii in stellar-mass bins of $0.2$~dex width. The half-mass sizes were estimated by applying the \citet{2009MNRAS.400.1181Z} mass-to-light conversion to the stellar light profiles, which themselves are S\'ersic fits preformed by \citet{2019MNRAS.490.4060C}. As shown by \citet{2022MNRAS.509.3751H}, the half-mass sizes are smaller by $\approx 0.1$~dex than the corresponding half-light sizes in the r-band, which is in agreement with \citet{2022MNRAS.511.2544D}, who found a similar difference between mass- and luminosity-weighted galaxy sizes in the \textsc{eagle} simulations, using the SDSS pipeline for the analysis.

\subsection{The search for the best-fitting parameter values}
\label{subsection: MCMC}

Each emulated relation $y = \hat{f}(x,\boldsymbol{\theta})$ changes smoothly with $x$ and $\boldsymbol{\theta}$. Once we have constructed $y = \hat{f}(x,\boldsymbol{\theta})$ by training the emulator on the simulation data  $\{x_n,\boldsymbol{\theta}_n,y_n\}_{n=1}^{N}$, our goal is to find the values of the parameters $\boldsymbol{\theta}$ that result in the best agreement between the emulator predictions and observational data. To quantify how well Gaussian process emulators can fit the observational data, we use Bayesian analysis.

\subsubsection{Prior}

We start by setting up a prior on the emulator parameters $\boldsymbol{\theta}$, $\mathcal{P}_{\rm prior}(\boldsymbol{\theta})$. We assume that each parameter $\theta_i$ has a uniform prior within some range from $\theta_{i,\rm min}$ to $\theta_{i,\rm max}$, and is otherwise zero,

\begin{equation}
    \mathcal{P}_{i,\mathrm{prior}}(\theta_i) = \begin{cases} 1, & \mbox{if } \theta_{i,\rm min} \leq \theta_i \leq \theta_{i,\rm max} \\ 0, & \mbox{otherwise} \, . \end{cases}
\label{eq: prior}
\end{equation}
The total prior for $\boldsymbol{\theta}$ is then the product $\mathcal{P}_{\rm prior}(\boldsymbol{\theta}) = \prod_i^{N_{\rm param}}\mathcal{P}_{i,\mathrm{prior}}$. 

We opt for such a prior because of our limited knowledge about the parameters $\boldsymbol{\theta}$. The vector $\boldsymbol{\theta}$ contains the subgrid parameters of the SN and AGN feedback model. Probing $N$ random realizations of $\boldsymbol{\theta}$ requires running $N$ independent simulations, each of which may take a long time to complete. Therefore, given $N$ simulations that we can afford to run, our training data contain $N$ unique values for each subgrid parameter $\theta_i$. The values of $\theta_i$ are distributed within a certain interval, whose lower and upper bounds define, respectively, $\theta_{i,\rm min}$ and $\theta_{i,\rm max}$ in equation (\ref{eq: prior}). We set the prior to zero outside the domain sampled by the simulations because the errors of a Gaussian process emulator become large when it is used for extrapolation.

\subsubsection{Likelihood}
\label{subsection: likelihood}

We compute the total log-likelihood function, $\ln \mathcal{L}(\boldsymbol{\theta})$, as the sum of individual log-likelihood functions for the emulated \gsmf{} and \ssm,

\begin{equation}
    \ln \mathcal{L}(\boldsymbol{\theta}) = \ln \mathcal{L}_{\rm GSMF} (\boldsymbol{\theta})  + \ln \mathcal{L}_{\rm SSMR} (\boldsymbol{\theta}) \, ,
\label{eq: likelihood}
\end{equation}
which means that the \gsmf{} and \ssm{} contribute equally to the total likelihood.

The likelihood of each emulated relation is computed assuming that the statistical errors in the emulator prediction and the observational data are Gaussian distributed and independent,

\begin{equation}
\label{eq: likelihood_function}
    \ln \mathcal{L}_{R}(\boldsymbol{\theta}) = -\frac{\langle N_{\rm obs}\rangle}{N_{R,\mathrm{obs}}}\frac{1}{2}\sum_{n=1}^{N_{R,\mathrm{obs}}}\left[ \frac{\hat{f}_{R}(x_{R,n},\boldsymbol{\theta}) - y_{R,n}}{\sqrt{\sigma^2_{R,n} + \varepsilon_{R,\mathrm{emu}}^2}}\right]^2 \, ,
\end{equation}
where the subscript $R$ is a placeholder for \gsmf{} or \ssm. Next, $x_{R,n}$, $y_{R,n}$, and $\sigma_{R,n}$ are, respectively, the $x$ values, $y$ values, and errors on the $y$ values of the observational data used to constrain the emulated relation $R$ (see $\S$\ref{subsection: target_observational_data}), and $N_{R,\mathrm{obs}}$ is the number of observational data points over which the sum is computed. $\hat{f}_{\rm R}(x_{R,n},\boldsymbol{\theta})$ is the prediction of the emulator of the relation $R$ evaluated at $x_{R,n}$ and for the parameter vector $\boldsymbol{\theta}$. The \ssm{} and \gsmf{} likelihood functions are normalized by $N_{R,\mathrm{obs}}/\langle N_{\rm obs}\rangle$ where $\langle N_{\rm obs}\rangle = (N_{\rm GSMF,obs} + N_{\rm SSMR,obs})/2$ is the average number of data points contained in the observational data for the \gsmf{} and \ssm{} emulators. This normalization ensures that differences in the number of data points between the \gsmf{} and \ssm{} datasets do not affect their relative contributions to the total likelihood. Note that since the emulators are constructed using simulation data in log-log space (see $\S$\ref{subsection: simulated_relations}), the observational data used in equation (\ref{eq: likelihood_function}) -- including both the 
$x$ and $y$ values, as well as the errors on the $y$ values -- are also logarithmic.

Finally, $\varepsilon_{R,\mathrm{emu}}$ is the uncertainty in the emulator predictions for the relation $R$.  To estimate $\varepsilon_{R,\mathrm{emu}}$, we train the emulators on all but one simulation from the training data of a given model (see Table \ref{table: hypercubes} and $\S$\ref{subsection:models}), and ask the emulator to predict the \gsmf{} and \ssm{} for the simulation that was left out. We repeat this procedure for each simulation in the training data of each model and record the differences between the emulator predictions and the simulation data. For both \gsmf{} and \ssm, this gives us $N_{\rm runs}$ vectors with emulator errors where entries of each vector correspond to different stellar mass bins and the number $N_{\rm runs}$ is the total number of simulations in the training data (see $\S$\ref{subsection: training set}). We concatenate all $N_{\rm runs}$ vectors into a single list and compute $\varepsilon_{R,\mathrm{emu}}$ as the standard deviation of the entries in this list. The value of $\varepsilon_{R,\mathrm{emu}}$ slightly changes depending on the relation and the model for which the emulator is constructed but averages to $\approx 0.07$~dex\footnote{Because the emulator error, $\varepsilon_{R,\mathrm{emu}}$, varies slightly between different emulated relations $R$, it affects the relative contribution of each relation to the total likelihood. To estimate the impact of this, we tested alternative choices for $\varepsilon_{R,\mathrm{emu}}$ in equation (\ref{eq: likelihood_function}), such as using an average error over the two emulated relations — \gsmf{} and \ssm{} — or an error further averaged over all models used in the emulation. We did not find any advantage of using these alternatives, nor any significant impact on the best-fitting parameter values of the final model.}.

\subsubsection{Posterior}

The log posterior is the sum of the log likelihood and the log prior,
\begin{equation}
     \ln \mathcal{P}_{\rm posterior}(\boldsymbol{\theta}) = \ln \mathcal{L}(\boldsymbol{\theta}) +  \ln \mathcal{P}_{\rm prior}(\boldsymbol{\theta}) \, ,
\end{equation}
from which we obtain the values of the parameters of the best-fitting model, $\boldsymbol{\theta}_{\rm best}$, as
\begin{equation}
\label{eq: best-fit-param-defintion}
    \ln \mathcal{P}_{\rm posterior}(\boldsymbol{\theta}_{\rm best}) = \max\left(\ln \mathcal{P}_{\rm posterior}(\boldsymbol{\theta})\right) \, .
\end{equation}

To find the maximum of the posterior distribution, $\ln \mathcal{P}_{\rm posterior}(\boldsymbol{\theta})$, we use the Markov chain Monte Carlo (MCMC) python package \textsc{emcee} \citep{2013PASP..125..306F}. We run MCMC for $5,000$ steps using 30 independent walkers, which is more than sufficient for convergence, as indicated by the MCMC trace plot (not shown here). The walkers are initialized at random positions within the region of parameter space where the prior is non-zero. In the analysis of the posterior distribution, we remove the first 200 steps for each walker to avoid the `burn-in' phase. To generate proposal steps for the random walk through the parameter space, we employ the `stretch move' algorithm \citep{2010CAMCS...5...65G} with a stretch scale parameter of $2$. Lastly, we note that it is not necessary to normalize the posterior to find the best-fitting parameter values, as can be seen from equation (\ref{eq: best-fit-param-defintion}).

\section{Calibration with emulators}
\label{Section: Calibration}

This section describes the calibration strategy of the \colibre{} model with fixed $\Delta T_{\rm AGN}=10^9~\mathrm{K}$ at m7 resolution, which makes use of the method of emulators detailed above. The objective of the calibration is to maximize the agreement between the simulation and target observational data, which is achieved by adjusting the subgrid parameters of the model, or the model itself.

\subsection{Models with simplified supernova feedback}
\label{subsection:models}

As the starting point of the calibration, we take a model of galaxy formation with a significantly simplified version of SN feedback, compared to the fiducial \colibre{} prescription presented in $\S$\ref{subsection: CC_SN_feedback}. We will call this model the \texttt{Basic} model. The other aspects of the galaxy formation physics in the \texttt{Basic} model will be the same as described in Section \ref{section: simulations}.

We do not commence with calibrating the fiducial  \colibre{} SN feedback because it is not obvious \textit{a priori} whether using a more complex model will lead to a better fit of the simulation to the observational data. Only failing to match the target observational data with the simplified model will indicate that a more sophisticated model is necessary.

Besides the \texttt{Basic} model, we consider two other simplified prescriptions for SN feedback: \texttt{ThermalKinetic} and \VardT{}, each of which is detailed below. We emphasize that the only difference between the \colibre{} model with fixed $\Delta T_{\rm AGN}$ and its simplified versions -- \texttt{Basic}, \texttt{ThermalKinetic}, 
 and \VardT{} -- is the treatment of SN feedback, while all other parts of the galaxy formation physics remain identical, including the heating temperature $\Delta T_{\rm AGN}=10^9~\mathrm{K}$ in AGN feedback. Unlike SN feedback, we do not consider simplified prescriptions for AGN feedback because BH particles heating gas neighbours with a constant temperature and using fixed radiative and coupling efficiencies (see $\S$\ref{subsubsection: AGN_feedback}) already constitutes a basic AGN algorithm.

\subsubsection{The Basic model}

Relative to the fiducial \colibre{} prescription for SN feedback from Section \ref{section: simulations}, we make the following simplifications in the \texttt{Basic} model: 

\begin{enumerate}
\item The energy of a single SN, in units of $10^{51}$ erg, $f_{\rm E}$, is constant rather than dependent on the stellar birth pressure, $P_{\rm birth}$ (see equation~\ref{eq: stellar_birth_pressure_vs_SN_energy}). The value of the constant $f_{\rm E}$ will be determined using emulators.

\item All energy released by SNe is injected thermally; that is, the fraction of SN energy injected in kinetic form, $f_{\rm kin}$, is set to $0$, meaning the kinetic channel of SN feedback is not used.
\item The heating temperature in the thermal (CC and type-Ia) SN feedback, $\Delta T_{\rm SN}$, is set to a constant value of $10^{7.5}~\mathrm{K}$ -- the value used in the \textsc{eagle} simulations -- rather than being density-dependent (equation~\ref{eq: heating_temperature_vs_gas_density}).
\end{enumerate}

\subsubsection{The thermal-kinetic model}

In addition to the \texttt{Basic} model, we consider a modification in which the prescription for SN feedback includes both kinetic and thermal channels of energy injection (i.e. $f_{\rm kin}$ is no longer necessarily $0$). We refer to this model as \texttt{ThermalKinetic}. As is the case in the \colibre{} model with fixed $\Delta T_{\rm AGN}$, the kinetic channel of the \texttt{ThermalKinetic} model uses the desired kick velocity parameter of $\Delta v_{\rm kick} = 50$ km s$^{-1}$. Otherwise, \texttt{ThermalKinetic} is the same as \texttt{Basic}, including the constant energy in SN feedback and the constant heating temperature of $\Delta T_{\rm SN} = 10^{7.5}~\mathrm{K}$ in the thermal channel of energy injection. 

Compared to \texttt{Basic}, the \texttt{ThermalKinetic} model introduces one additional free parameter: the fraction of SN energy injected in kinetic form, $f_{\rm kin}$, which will be determined using emulators. We note that for $f_{\rm E}=2$ and $f_{\rm kin}=0.1$, the CC SN feedback in the \texttt{ThermalKinetic} model becomes identical to that in the fiducial model used in the simulations of isolated disc galaxies by \cite{2023MNRAS.523.3709C}.

\subsubsection{The thermal-kinetic model with a variable heating temperature}

Our final simplified model is \VardT. As the name suggests, compared to \texttt{ThermalKinetic}, \VardT{} adopts the density-dependent heating temperature for thermal SN feedback (for both CC and type-Ia SNe) detailed in $\S$\ref{subsubsection: dens_dep_dT}, while \texttt{Basic} and \texttt{ThermalKinetic} use a constant value of $\Delta T_{\rm SN} = 10^{7.5}~\mathrm{K}$.

Of the simplified models, \VardT{} is the closest to the \colibre{} model with fixed $\Delta T_{\rm AGN}=10^{9}~\mathrm{K}$. The only difference is that the \colibre{} fiducial prescription for SN feedback adopts an $f_{\rm E}$ that depends on the stellar birth gas pressure, following equation (\ref{eq: stellar_birth_pressure_vs_SN_energy}), whereas \VardT{} uses a constant $f_{\rm E}$.

\subsection{Selection of subgrid parameters for emulator-based calibration}
\label{subsection: selection_theta}

We next describe the selection of the subgrid parameters that will be calibrated using emulators. These parameters, which we denote by the vector $\boldsymbol{\theta}$, enter the emulators defined in $\S$\ref{subsection: simulated_relations} and are optimized with the methods of Bayesian statistics ($\S$\ref{subsection: MCMC}), such that the simulation provides the best match to the observational data ($\S$\ref{subsection: target_observational_data}).

We will use the emulators to optimize only the subgrid parameters that govern the strengths of SN and AGN feedback. Parameters related to other aspects of galaxy formation physics (such as star formation, chemical enrichment, or radiative cooling) will not be considered, either because they have little to no impact on the galaxy properties relevant to our calibration (i.e., \gsmf{} and \ssm), or because their values are well constrained by fundamental physics or inferred from independent observations.

\subsubsection{AGN feedback parameters}
\label{subsubsection: AGN_feedback_params}

The notable parameters of the \colibre{} model with fixed $\Delta T_{\rm AGN}$ related to AGN feedback are (i) the AGN heating temperature, $\Delta T_{\rm AGN}$; (ii) the seed mass of BH particles, $m_{\rm BH, seed}$; and (iii) the minimum FoF mass of a halo in which BH particles can be seeded, $M_{\rm FoF, seed}$. 

\begin{table*}
\caption{Latin hypercubes used to train the emulators. Column (1): the name of the model for which the Latin hypercube is created (see $\S$\ref{subsection:models}); column (2): the hypercube level in the hierarchy (level 2 is a subregion of level 1 with finer sampling; the \texttt{Basic} model only has level 1); column (3): the number of simulations included in the Latin hypercube at a given level; column (4): the energy per single CC SN in units of $10^{51}$ erg; column (5): the fraction of SN energy that is injected in kinetic form; column (6): the pivot density in the relation between the SN heating temperature and the gas density (equation~\ref{eq: heating_temperature_vs_gas_density}); column (7) the pivot birth pressure in the relation between the energy in CC SN feedback and the stellar birth gas pressure (equation~\ref{eq: stellar_birth_pressure_vs_SN_energy}); column (8) the BH seed mass. The two numbers in each cell of columns $4-8$ specify the interval over which each parameter is varied in the Latin hypercube. For a given model, a cell left blank indicates that the model does not include the corresponding parameter.}
	\centering
	\begin{tabular}{llrrrrrr} 
   \hline
   Model name & Hypercube level & $N^{\rm L1,2}_{\rm runs}$ & \multicolumn{5}{c}{Model parameters that are varied in the Latin hypercube}   \\ 
   & & & $f_{\rm E}$  & $f_{\rm kin}$ & $n_{\rm H, pivot}$   & $\log_{\rm 10} P_{\rm E,pivot}/k_{\rm B}$ & $\log_{\rm 10} m_{\rm BH, seed}$ \\
   & & & &  &  [cm$^{-3}$] & [K cm$^{-3}$]  & [$\rm M_\odot$] \\
	    \hline
    \texttt{Basic} & 1 & $24$ & [$0.1, 5$]  & -- &   -- & -- & [3, 6] \\
      & -- & -- & -- & -- &  -- & -- & -- \\
    \texttt{ThermalKinetic} & 1 & $32$ & [$0.3$, $2.3$] & [$0$, $1$] &  -- &  -- & [$3.5$, $6$] \\
      & $2$ & $40$ & [$0.3$, $2.3$] & [$0$, $0.5$] &  -- &  -- & [$4.2$, $5.5$] \\
     \VardT{} &  $1$ & $40$ & [$0.3, 2.3$] & [$0$, $0.5$] & [$0.05$, $2.5$] &  --  & [$4.2$, $5.5$] \\
      & $2$ & $8$ & [$1.25$, $1.55$] & [$0$,  $0.14$] & [$0.4$, $0.65$] &  --  & [$4.4$, $4.6$] \\
     \colibrefixedagntemp{} & $1$ & $40$  & -- & [$0$, $0.5$]  & [$0.05, 2$] & [$3.3$, $4.5$] & [$4.2$, $5.5$]\\  
     & $2$ & $8$  & -- & [$0.07$,  $0.14$]  & [$0.4$, $0.65$] &[$3.8$,   $4.0$]  & [$4.7$, $5$]\\   \hline
\end{tabular}
\label{table: hypercubes}
\end{table*}

\begin{itemize}
    \item The AGN heating temperature, $\Delta T_{\rm AGN}$, determines the thermal energy received by a gas particle in a single AGN energy injection event. In other words, $\Delta T_{\rm AGN}$ is a measure of the `burstiness' of AGN feedback. In principle, higher (lower) values of $\Delta T_{\rm AGN}$ tend to yield stronger (weaker) AGN feedback. However, owing to the ability of SMBHs to self-regulate \citep{2009MNRAS.398...53B,2010MNRAS.405L...1B}, we expect the exact value of $\Delta T_{\rm AGN}$ to have only a minor impact on the $z=0$ \gsmf{} and \ssm{} \citep[e.g.][]{2017MNRAS.465.2936M}, provided that the temperature of the heated gas remains sufficiently high for the injected thermal energy not to be rapidly radiated away. We will therefore not consider $\Delta T_{\rm AGN}$ as one of the subgrid parameters for calibration and instead keep it fixed at $10^9~\mathrm{K}$. In $\S$\ref{subsection: parameter variations} we will confirm that (modest) variations in $\Delta T_{\rm AGN}$ indeed have only a minor effect on the $z=0$ \gsmf{} and \ssm, and that these small differences can be compensated for by adjusting other model parameters.

    \item The BH seed mass, $m_{\rm BH, seed}$, determines how quickly BHs can grow over time (see equation \ref{eq: SMBH_acc_rate}). Higher $m_{\rm BH, seed}$ will cause faster BH growth, leading to more energetic AGN feedback in lower-mass galaxies and at higher redshifts \citep[e.g.][]{2009MNRAS.398...53B}. Because both the \gsmf{} and \ssm{} depend sensitively on the strength of AGN feedback, we will include $m_{\rm BH, seed}$ in the set of subgrid parameters for optimization with emulators, $\boldsymbol{\theta}$.

    \item The minimum halo FoF mass in which BHs are seeded, $M_{\rm FoF, seed}$, has a prominent effect on the calibrated relations too, because, similarly to $m_{\rm BH, seed}$, $M_{\rm FoF, seed}$ determines how early BHs can start growing in mass \citep[e.g.][]{2009MNRAS.398...53B}. However, for the same reasons, $M_{\rm FoF, seed}$ is strongly degenerate with $m_{\rm BH, seed}$. For example, increasing $M_{\rm FoF, seed}$ will delay the growth of BHs, but a similar effect can be achieved by decreasing $m_{\rm BH, seed}$. Owing to this degeneracy, which will be shown in $\S$\ref{subsection: parameter variations}, we will not include $M_{\rm FoF, seed}$ in our set of parameters for optimization. As already explained in $\S$\ref{subsection: BH_and_AGN}, at m7 resolution we set $M_{\rm FoF, seed}=5\times 10^{10}~\mathrm{M_\odot}$.
\end{itemize}

\subsubsection{Supernova feedback parameters}
\label{subsubsection: SN_feedback_params}

Because we consider four different prescriptions for SN feedback, we will, for clarity, describe the SN feedback parameters for each model separately.

\begin{itemize}
    \item In the \texttt{Basic} model, the only free parameter is the energy per single SN in units of $10^{51}$ erg, $f_{\rm E}$. Therefore, the full parameter vector $\boldsymbol{\theta}$ for the \texttt{Basic} model, including the AGN feedback parameters from $\S$\ref{subsubsection: AGN_feedback_params}, is given by $\boldsymbol{\theta} = (m_{\rm BH, seed}, f_{\rm E})$.

    \item The \texttt{ThermalKinetic} model contains an additional free parameter: the fraction of SN energy injected in kinetic form, $f_{\rm kin}$. This makes the total number of parameters entering the parameter vector $\boldsymbol{\theta}$ equal to three: $\boldsymbol{\theta} = (m_{\rm BH, seed}, f_{\rm E}, f_{\rm kin})$.

    \item The \VardT{} model employs a density-dependent heating temperature for SN feedback (equation~\ref{eq: heating_temperature_vs_gas_density}), which is described by four parameters: the pivot gas density $n_{\rm H, pivot}$, the heating temperature at the pivot density $\Delta T_{\rm SN, pivot}$, and the minimum and maximum heating temperatures, $\Delta T_{\rm SN,min}$ and $\Delta T_{\rm SN,max}$. As documented in $\S$\ref{subsubsection: dens_dep_dT}, at m7 resolution we set $\Delta T_{\rm SN,pivot}$, $\Delta T_{\rm SN,min}$ and $\Delta T_{\rm SN,max}$ to $10^{6.5}~\mathrm{K}$, $10^{6.5}~\mathrm{K}$, and $10^{7.5}~\mathrm{K}$, respectively, which leaves us with only one free parameter: $n_{\rm H, pivot}$. Therefore, the final form of the parameter vector $\boldsymbol{\theta}$ for the \VardT{} model is $\boldsymbol{\theta} = (m_{\rm BH, seed}, f_{\rm E}, f_{\rm kin}, n_{\rm H, pivot})$.

    \item Finally, the \colibre{} model with fixed $\Delta T_{\rm AGN}=10^9~\mathrm{K}$ (henceforth, the \colibrefixedagntemp{} model) uses a stellar birth pressure dependent energy for CC SN feedback (equation~\ref{eq: stellar_birth_pressure_vs_SN_energy}). This comes with another set of four parameters, $f_{\rm E, min}$, $f_{\rm E, max}$, $\sigma_{\rm P}$, and $P_{\rm E, pivot}$, which together replace the parameter $f_{\rm E}$ that specifies the constant energy in CC SN feedback in the three simplified models. As explained in $\S$\ref{subsection: CC_SN_feedback}, the values of three out of the four extra parameters are fixed: $f_{\rm E, min}=0.1$, $f_{\rm E, max}=4$, and $\sigma_{\rm P}=0.3$. Thus, in the emulation, the dependence of $f_{\rm E}$ on the stellar birth pressure will be described with a single free parameter, $P_{\rm E,pivot}$, resulting in the parameter vector for the \colibrefixedagntemp{} model, $\boldsymbol{\theta} = (m_{\rm BH, seed}, P_{\rm E,pivot}, f_{\rm kin}, n_{\rm H, pivot})$.
    
    We note that due to the functional degeneracies between $P_{\rm E,pivot}$, $f_{\rm E, min}$, and $f_{\rm E, max}$ (see equation~\ref{eq: stellar_birth_pressure_vs_SN_energy}), a very low (high) $P_{\rm E,pivot}$ preferred by the emulator will suggest that our chosen value of $f_{\rm E, max}$ ($f_{\rm E, min}$) may be too low (high). In $\S$\ref{subsection: parameter variations} we will show that $P_{\rm E,pivot}$, $f_{\rm E, min}$, and $f_{\rm E, max}$ are indeed degenerate with one another (and also with $\sigma_{\rm P}$).
\end{itemize}

\subsection{Training data for emulators}
\label{subsection: training set}

\begin{table*}
\caption{The best-fitting values of the parameters identified by the emulator by matching the model to the $z=0$ observational data: the galaxy stellar mass function (\gsmf) from \citet{2022MNRAS.513..439D} and size -- stellar mass relation (\ssm) from \citet{2022MNRAS.509.3751H}. Column (1): the name of the model; column (2): the observational data to which the model was fit; column (3): the $\chi^2_\nu$ value of the best-fitting model. The remaining columns indicate the best-fitting values of the model parameters and the corresponding $1\sigma$ errors. The model parameters are arranged in the same way as in Table \ref{table: hypercubes}. For a given model, an empty cell means that the corresponding parameter does not exist in the model.}
	\centering
	\begin{tabular}{lrrrrrrr} 
    \hline
     Model name & Emulator was fit to & $\chi^2_\nu$ &\multicolumn{5}{c}{Best-fitting values of model parameters}   \\ 
     & & & $f_{\rm E}$  & $f_{\rm kin}$ & $n_{\rm H, pivot}$    & $\log_{\rm 10} P_{\rm E,pivot}/ k_{\rm B}$ & $\log_{\rm 10} m_{\rm BH, seed}$ \\
    & & &  &  &  [cm$^{-3}$] & [K cm$^{-3}$]  & [$\rm M_\odot$] \\
	    \hline
     \texttt{Basic} & \gsmf{} and \ssm{} & $9.3$ & $1.0^{+0.05}_{-0.1}$  & -- &   -- & -- & $4.0 \pm 0.1$ \\
    \texttt{ThermalKinetic} & \gsmf{} and \ssm{} & 4.7 & $1.0 \pm 0.1$  & $0.3 \pm 0.05$ &   -- & -- & $4.7^{+0.1}_{-0.05}$ \\
    \VardT & \gsmf{} and \ssm{}  & $2.9$ & $1.3 \pm 0.1$  & $0.1 \pm 0.05$ &   $0.5^{+0.1}_{-0.2}$ & -- & $4.6^{+0.05}_{-0.1}$ \\
    \colibrefixedagntemp{} & \gsmf{} and \ssm{}  & $0.8$ & --  & $0.1 \pm 0.05$ &  $0.6^{+0.1}_{-0.2}$ & $3.9  \pm 0.1$ & $4.8^{+0.05}_{-0.1}$ \\
    \\
     \texttt{ThermalKinetic} & \gsmf{} & $0.3$ & $1.3 \pm 0.1$  & $0.6 \pm 0.1$ &   -- & -- & $4.8 \pm 0.2$ \\
     \texttt{ThermalKinetic} & \ssm{} & $0.9$ & $1.0^{+0.1}_{-0.05}$ & $0^{+0.05}_{-0}$ &   -- & -- & $4.3 \pm 0.1$\\
        \hline
\end{tabular}
\label{table: best-fitting models}
\end{table*}

For the \colibrefixedagntemp{} model described in Section~\ref{section: simulations}, as well as its three simplified counterparts introduced in Section~\ref{subsection:models}, we construct emulators of both the \gsmf{} and the \ssm{} as defined in Section~\ref{subsection: simulated_relations}, resulting in a total of eight independent emulators. Additionally, for diagnostic purposes, we construct emulators of the \shmr{} for the \texttt{Basic} and \texttt{ThermalKinetic} models, yielding two more independent emulators. Each emulator must be trained before it can be used for parameter inference or diagnostics.
 
To build the training datasets, we run a set of simulations. For a given model, each simulation represents a unique combination of values of the subgrid parameters $\boldsymbol{\theta}$ (i.e. it is a unique sampling point in the parameter space). To evenly sample the parameter space given our uniform priors, we make use of the \textit{Latin hypercube} sampling technique \citep{ef76b040-2f28-37ba-b0c4-02ed99573416}. The main advantage of Latin hypercube sampling over random sampling is that it requires significantly fewer sampling points to achieve the desired accuracy in emulator predictions. Given a target of $N$ sampling points in an $N_{\rm param}$-dimensional parameter space, this is achieved by first creating a grid that divides the space into $N^{N_{\rm param}}$ equal-volume cells. Sampling is then performed by placing $N$ points into $N$ cells such that, when projected onto any single parameter dimension, there is exactly one point in each of the $N$ equally spaced intervals, with each point positioned randomly within its assigned cell. As a result, Latin hypercube sampling covers the parameter space more evenly, allowing the number of simulations needed to train the emulators for each model to remain relatively modest.

To further enhance emulator accuracy, we construct the Latin hypercubes hierarchically at two levels: a coarse level 1 with broad parameter variations and level 2, a refined subregion within level 1 with finer sampling. The boundaries of level 2 are determined by first training the emulator using only the simulations from level 1 and identifying the subregion of the parameter space that most likely contains the best-fitting model. At level 1, we run $N^{\rm L1}_{\rm runs}=24$ simulations for the \texttt{Basic} model, $32$ for the \texttt{ThermalKinetic} model, and $40$ for \VardT{} and \colibrefixedagntemp. The number of simulations in the Latin hypercube increases with the size of the parameter vector $\boldsymbol{\theta}$, whose dimensions for the four models are, respectively, 2, 3, 4, and 4. At level 2, we use $N^{\rm L2}_{\rm runs} = 40$ simulations for the \texttt{ThermalKinetic} model, $8$ for \VardT{}, and $8$ for \colibrefixedagntemp. We do not use level 2 for the \texttt{Basic} model because its simplicity allows the emulator to accurately determine the best-fitting parameter values using only level 1. Level 2 of the \texttt{ThermalKinetic} model contains significantly more simulations than those of \VardT{} and \colibrefixedagntemp, as we fit \texttt{ThermalKinetic} not only to the observed \gsmf{} and \ssm{} together but also separately to each observable. This results in \textit{three} best-fitting models located in different regions of the parameter space (see $\S$\ref{subsubsection: vary_fitting_constraints}). Furthermore, since the \texttt{Basic} model is equivalent to the \texttt{ThermalKinetic} model with $f_{\rm kin} = 0$, we incorporate 24 simulations from the \texttt{Basic} model's hypercube into that of the \texttt{ThermalKinetic} model. This improves the accuracy of the emulator predictions for \texttt{ThermalKinetic} near the hypercube boundary where $f_{\rm kin} = 0$. 

In total, the Latin hypercubes for \texttt{Basic}, \texttt{Thermal}\-\texttt{Kinetic}, \VardT{}, and \colibrefixedagntemp{} contain $N^{\rm total}_{\rm runs} = 24$, $96$, $48$, and $48$ simulations, respectively. Table \ref{table: hypercubes} summarizes the properties of these Latin hypercubes, including the parameter ranges $\boldsymbol{\theta}$ explored at each level. The level 1 ranges were preselected to ensure that the peaks of the posterior distributions of $\boldsymbol{\theta}$ fall within the hypercube domain\footnote{These ranges were determined by running simulations in small cosmological volumes (25$^3$ cMpc$^3$) with much broader parameter variations.}. In the training set, no distinction is made between the simulations from level 1 and level 2; the emulator is trained using simulations from both levels simultaneously.

All simulations in the Latin hypercubes were run to $z=0$ in (50 cMpc)$^3$ volumes at m7 resolution, with initial numbers of gas and DM particles of $376^3$ and $4\times 376^3$, respectively (see $\S$\ref{subsection: ICs} for more details on the ICs). Each simulation was run on 128 cores and took on average $\approx 5$ days\footnote{From additional tests, we found that decreasing the number of DM particles in the ICs from $4\times 376^3$ to $376^3$ (i.e., matching the initial number of gas particles) reduces the total wall-clock time to reach $z=0$ by about 30 per cent.} to reach $z=0$. For the entire set of simulations, this translates to $\approx 3\times 10^{6}$ core hours. The values of the subgrid parameters that are not part of $\boldsymbol{\theta}$ do not change between different simulations from the same Latin hypercube. In Appendix \ref{appendix: boxsize_effect}, we demonstrate that a (50 cMpc)$^3$ volume is sufficient to produce a \gsmf{} and \ssm{} within the stellar mass range considered in the calibration, $10^9 < M_* / \mathrm{M_\odot} < 10^{11.3}$, that are consistent with those obtained from simulations in larger volumes.

\begin{figure*}
    \centering
    \includegraphics[width=0.9\textwidth]{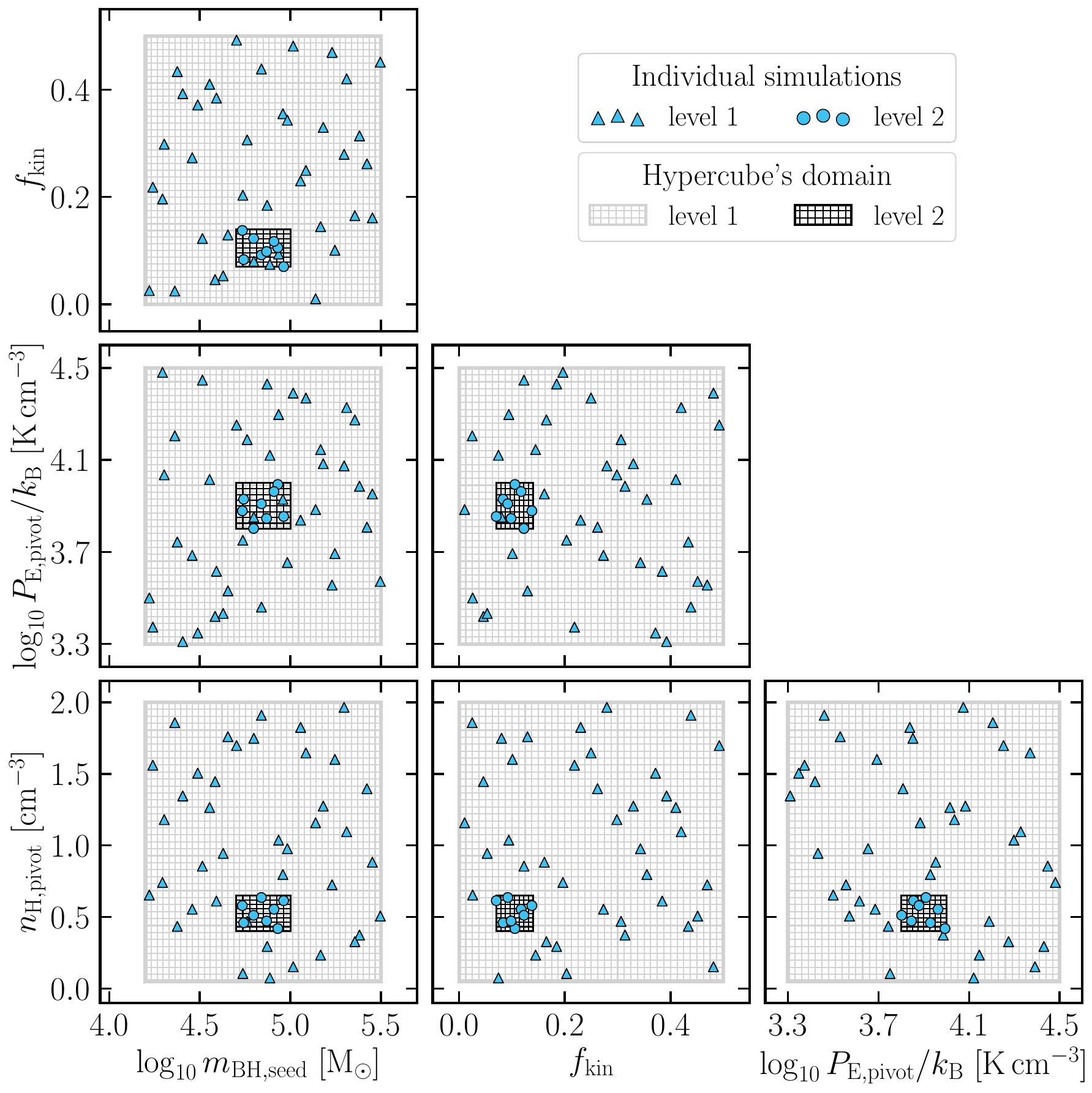}
    \caption{The Latin hypercube for the \colibrefixedagntemp{} model. The axes of the panels correspond to different parameters of the model: $m_{\rm BH, seed}$, $f_{\rm kin}$, $n_{\rm H,pivot}$, and $P_{\rm E,pivot}$ (see $\S$\ref{subsection: selection_theta} for details). The grey (black) hatched rectangle marks level 1 (level 2) of the hypercube's domain, while \colibrefixedagndtcolor{} triangles (circles) indicate the sampling for level 1 (level 2), consisting of 40 (8) individual simulations. Together, the simulations from levels 1 and 2 form the training dataset for the \colibrefixedagntemp{} model, which is used to train the emulators for the $z=0$ \gsmf{} and \ssm.}
\label{fig:basic_and_thermalkinetic_models_hypercubes}
\end{figure*}

As an illustration, Fig.~\ref{fig:basic_and_thermalkinetic_models_hypercubes} shows the Latin hypercube for the \colibrefixedagntemp{} model. The axes correspond to different hypercube parameters: $m_{\rm BH, seed}$, $f_{\rm kin}$, $n_{\rm H,pivot}$, and $P_{\rm E,pivot}$. Grey and black hatched rectangles denote levels 1 and 2 of the hypercube’s domain, while \colibrefixedagndtcolor{} triangles and circles represent 40 and 8 individual simulations, respectively, sampling the parameter space at these levels. These 48 simulations are used to train the emulators to predict the $z=0$ \gsmf{} and \ssm{} as functions of the hypercube's parameters. Once the \gsmf{} and \ssm{} emulators have been trained, we can fit the \colibrefixedagntemp{} model to the observed \gsmf{} and \ssm, as described in $\S$\ref{subsection: MCMC}, to find the best-fitting values of the model parameters. The Latin hypercubes for the other three models -- \texttt{Basic}, \texttt{ThermalKinetic}, and \VardT{} -- look qualitatively similar, except that the parameter spaces of the first two models have fewer dimensions. 

\subsection{Best-fitting parameter values and simulations with best-fitting models}

The (rounded) best-fitting values of the parameters of the \colibrefixedagntemp{} model, as well as its three simplified analogues, are presented in Table \ref{table: best-fitting models}. For each model, the table lists only the values of those parameters that were optimized by the emulators by fitting the model to the observational data. Additionally, the table provides (rounded) $1\sigma$ errors on the parameter values. Finally, we show the values of the reduced $\chi^2$ (i.e. $\chi^2_\nu$), which quantify the goodness of fit to the observational data. We compute $\chi^2$ as the sum of the squared differences between the predictions of the best-fitting model and the observational data to which the model was calibrated, normalized by the combined uncertainties from the observations and the emulator (see $\S$\ref{subsection: likelihood}). However, since the emulator uncertainty, $\varepsilon_{R,\mathrm{emu}}$, varies slightly between models and emulated relations, we adopt a fixed value of $\varepsilon_{R,\mathrm{emu}} = 0.07$ -- the average across the four models and the two emulated relations (\gsmf{} and \ssm) -- for computing $\chi^2$. This normalization places all best-fitting models on equal footing, enabling a direct comparison of their $\chi^2_\nu$ values.

We round the best-fitting values of the model parameters to one digit to the right of the decimal point, as we found no significant improvement in the accuracy of the fits when using more precise values. The direction of rounding -- up or down -- is determined not only by the proximity of the best-fitting value to its rounded counterpart, but also to ensure that the model with the rounded values of the subgrid parameters remains within the 68 per cent credibility interval of the posterior (i.e. within $1\sigma$ from the peak if the posterior has a Gaussian form)\footnote{There are, however, two exceptions to this rule: (i) the best-fitting value of $f_{\rm E}$ in the \texttt{Basic} model is $\approx 0.94$, but it is rounded to $1$ instead of $0.9$, as we found no improvement in using the more precise value; (ii) the best-fitting value of $n_{\rm H, pivot}$ in the \colibrefixedagntemp{} model is $\approx 0.53$ cm$^{-3}$, but it is rounded to $0.6$ cm$^{-3}$ instead of $0.5$ cm$^{-3}$ because at the time we finalized the model parameters, we were using a slightly different (older) version of the emulator training data, which preferred $0.6$ over $0.5$ cm$^{-3}$. In practice, this difference has a negligible impact, as the \colibrefixedagntemp{} 
models with both $n_{\rm H, pivot} = 0.6$ and $0.5~\mathrm{cm}^{-3}$ lie within the 68 per cent credibility interval of the posterior and yield similar $\chi^2_\nu$ values.}.

The \texttt{ThermalKinetic} model appears three times in Table \ref{table: best-fitting models}, as we fit it to the observed \gsmf{} and \ssm{} separately and jointly (see $\S$\ref{subsubsection: vary_fitting_constraints}). For each model and each set of best-fitting parameter values (different rows in Table \ref{table: best-fitting models}), we run a separate numerical simulation in a (50 cMpc)$^3$ volume, resulting in a total of six new simulations. In these simulations, we use the rounded best-fitting parameter values as reported in Table \ref{table: best-fitting models}, except that, for convenience, the rounded value of $P_{\rm E,pivot}/k_{\rm B}$ is specified in scientific notation as $8 \times 10^3$ K cm$^{-3}$ ($\approx 10^{3.903}$ K cm$^{-3}$), rather than as $10^{3.9}$ K cm$^{-3}$.

\begin{figure*}
    \centering
    \includegraphics[width=0.99\textwidth]{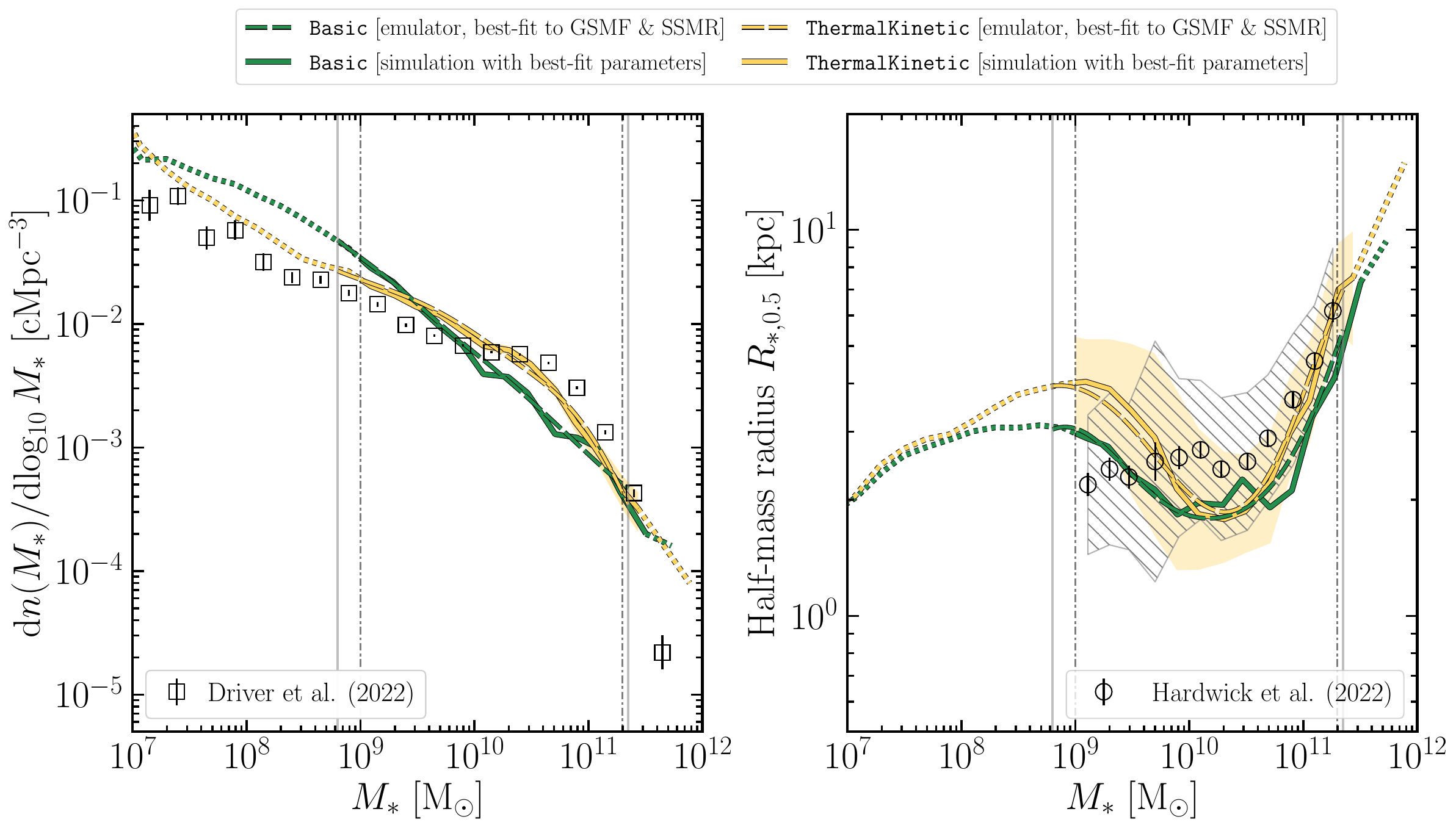}
    \caption{The galaxy stellar mass function (\gsmf; \textit{left}) and the median size -- stellar mass relation (\ssm; \textit{right}) at $z=0$, for the \texttt{Basic} (\modelbasiccolor) and \texttt{ThermalKinetic} (\modelthermalkineticcolor) models fit to the observed \gsmf{} and \ssm. The dashed and solid curves indicate, respectively, the best-fitting predictions of the emulator and the corresponding simulations with the best-fitting parameters from Table \ref{table: best-fitting models}. The shaded \modelthermalkineticcolor{} regions in the left and right panels indicate the Poisson uncertainty for the \gsmf{} and the $16^{\rm th}$ to $84^{\rm th}$ percentile scatter for the \ssm{} in the simulation using the \texttt{ThermalKinetic} model, respectively. We convert the \modelbasiccolor{} and \modelthermalkineticcolor{} solid curves into dotted curves where galaxies are poorly resolved ($M_* < 10^9~\mathrm{M_\odot}$) and where the number of galaxies is strongly limited by the finite simulated volume (the number of objects per bin is less than $5$). The vertical solid (dash-dotted) lines show the mass range within which the emulators were trained on the simulations (fit to observational data). The target observational data from \citet{2022MNRAS.513..439D} and  \citet{2022MNRAS.509.3751H} are shown as black squares and circles, respectively. Additionally, the grey hatched region in the right panel indicates the galaxy population-wide scatter in the \ssm{} from \citet{2022MNRAS.509.3751H}. Although the \texttt{ThermalKinetic} model produces a combined fit to the observed \gsmf{} and \ssm{} that is better than the fit with the \texttt{Basic} model, neither model is particularly satisfactory.}
\label{fig:basic_and_thermal_kinetic_model_calibration}
\end{figure*}

\section{Results}
\label{Section: Results}

We begin this section by evaluating the performance of the two simplest best-fitting models to the $z=0$ observed \gsmf{} and \ssm, \texttt{Basic} and \texttt{ThermalKinetic}, focusing on their predictions for the \gsmf, \ssm, and \shmr, as well as assessing the accuracy of the emulators relative to the simulations ($\S$\ref{subsubsection: calibration_simplest}). We then examine the $z=0$ \gsmf{} and \ssm{} in the more advanced best-fitting models, \VardT{} and \colibrefixedagntemp{} ($\S$\ref{subsection: more_advanced_models}). Next, we compare the simulations using the four best-fitting models to observations of various galaxy properties not considered during the emulation-based calibration ($\S$\ref{subsection: other_galaxy_properties}).

After showing that the best-fitting \colibrefixedagntemp{} model outperforms its three counterparts with more simplified SN feedback, we apply it to the higher \colibre{} resolutions, m6 and m5 ($\S$\ref{subsection: three resolutions}). We discuss that, at m5 resolution, \colibrefixedagntemp{} requires an undesirably low $m_{\rm BH, seed}$. We show that to allow for a higher $m_{\rm BH, seed}$ while maintaining a good fit to the observed \gsmf{} and \ssm, the fixed $\Delta T_{\rm AGN} = 10^9~\mathrm{K}$ in \colibrefixedagntemp{} needs to be replaced with a variable $\Delta T_{\rm AGN}$. We then compare the best-fitting \colibrefixedagntemp{} model with its variable $\Delta T_{\rm AGN}$ modification, showing that both achieve similar level of agreement with the observed \gsmf, \ssm, and other galaxy properties. Consequently, we establish the latter as the fiducial \colibre{} model and demonstrate that the \colibre{} simulations successfully reproduce the observed \gsmf{} and \ssm{} not only at m7 resolution but also at m6 and m5. Finally, in $\S$\ref{subsection: parameter variations}, we explore how variations in individual subgrid parameters, including those not optimized by the emulators, impact the calibrated galaxy properties in the \colibre{} fiducial model at m7 resolution.

\subsection{Calibration diagnostics for \texorpdfstring{$\protect\boldsymbol{\protect\mathtt{Basic}}$ and $\protect\boldsymbol{\protect\mathtt{ThermalKinetic}}$}{Basic and ThermalKinetic} models
}
\label{subsubsection: calibration_simplest}

\subsubsection{\gsmf{} and \ssm{} with the best-fitting parameters}

Fig.~\ref{fig:basic_and_thermal_kinetic_model_calibration} shows the $z=0$ \gsmf{} and \ssm{} for the \texttt{Basic} (\modelbasiccolor) and \texttt{ThermalKinetic} (\modelthermalkineticcolor) models. The dashed curves are the best-fitting predictions of the emulators that were trained on the Latin hypercubes and fit to the observed \gsmf{} from \citet{2022MNRAS.513..439D} and the observed \ssm{} from \citet{2022MNRAS.509.3751H}. The solid curves are the \gsmf{} and \ssm{} from the simulations that use the best-fitting parameter values found by the emulators (see Table \ref{table: best-fitting models}). The shaded \modelthermalkineticcolor{} region designates the scatter in the simulation with the \texttt{ThermalKinetic} model: the Poisson uncertainty for the \gsmf{} and the $16^{\rm th}$ to $84^{\rm th}$ percentile scatter for the \ssm. We change the style of the solid curves to dotted in the stellar mass range where galaxies become poorly resolved ($M_* < 10^9~\mathrm{M_\odot}$) and where the number of galaxies per bin drops below 5 (roughly corresponding to $M_* \gtrsim 10^{11.5}~\mathrm{M_\odot}$). The vertical solid lines indicate the edges of the stellar mass interval used in the training of the emulators: $ 10^{8.8} < M_*/\mathrm{M_\odot} < 10^{11.35}$. The vertical dash-dotted lines show the mass range within which the trained emulators were fit to the observational data: $ 10^{9} < M_*/\mathrm{M_\odot} < 10^{11.3}$. The observed \gsmf{} from \citet{2022MNRAS.513..439D} is shown in the left panel as black squares, and the observed \ssm{} from \citet{2022MNRAS.509.3751H} is shown in the right panel as black circles. The error bars in the observed \gsmf{} indicate the Poisson uncertainty, while in the observed \ssm, they show the $1\sigma$ error on the median. The grey hatched region in the right panel additionally shows the galaxy population-wide scatter in the \ssm{} from \citet{2022MNRAS.509.3751H}. 

By comparing the solid curves to the dashed curves of the same colour, we find that the differences between the \gsmf{} and \ssm{} predicted by the emulators and resulting from the simulations are negligibly small. Specifically, there are no systematic differences between the simulations and emulators, and the emulator errors in different stellar-mass bins of the \ssm{} and \gsmf{} range between $0$ and $\approx 0.1$~dex, which is comparable to the intrinsic scatter in simulations such as ours due to their stochastic nature \citep[e.g.][]{2023MNRAS.tmp.2803B}.

By comparing the solid curves to the black squares in the left panel and the black circles in the right panel, we find that the \texttt{ThermalKinetic} model is closer to the observational data than the \texttt{Basic} model is. In particular, the \texttt{Basic} model severely underpredicts the number of galaxies with stellar masses $M_* \gtrsim 10^{10}~\mathrm{M_\odot}$ and overpredicts it at $M_*\lesssim 10^{9.5}~\mathrm{M_\odot}$. In fact, the shape of the \texttt{Basic} model's \gsmf{} resembles a power-law, which disagrees with the shape of the observed \gsmf, which is known to be described by a single- or double-component \citet{1976ApJ...203..297S} function, featuring an exponential down-turn at high stellar mass. Although the \texttt{ThermalKinetic} model matches the observed \gsmf{} better than \texttt{Basic}, the discrepancy between its \gsmf{} and the observed data is still significant. Moreover, both models perform poorly in matching the observed galaxy sizes: the \ssm{} in \texttt{ThermalKinetic} features a prominent dip around $M_*\sim 10^{10.5}~\mathrm{M_\odot}$, while in the \texttt{Basic} model, the sizes of galaxies with stellar mass $M_*\gtrsim 10^{9.5}~\mathrm{M_\odot}$ are consistently lower than the observed relation by $\approx 0.1-0.2$~dex. 

Overall, the combined fit to the observed \gsmf{} and \ssm{} is better in the \texttt{ThermalKinetic} model than in \texttt{Basic}, but still not satisfactory, motivating a more sophisticated model.

\subsubsection{Posterior distributions of the model parameters}

\begin{figure}
    \centering
    \includegraphics[width=0.49\textwidth]{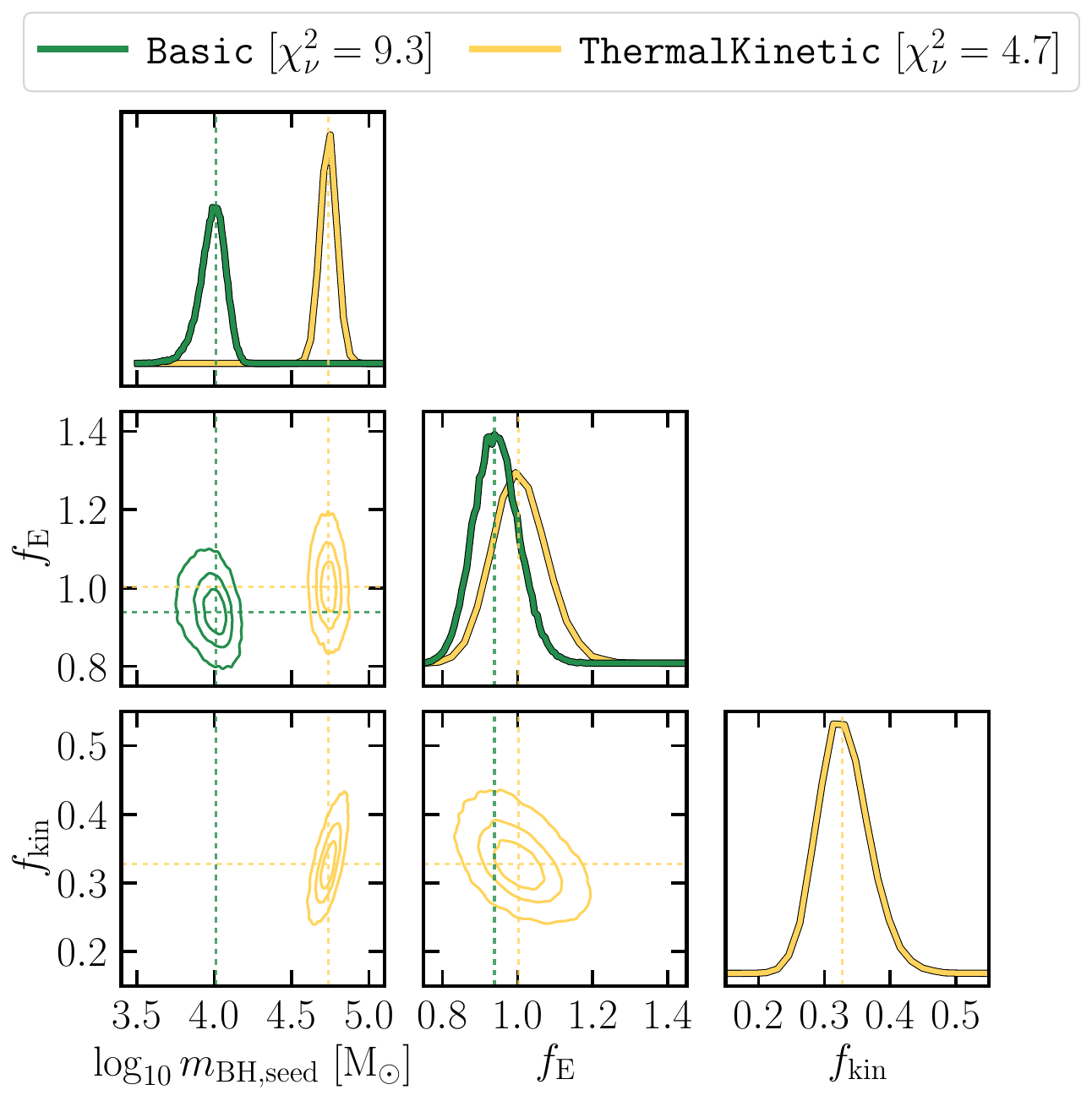}
    \caption{Posterior distribution of the parameters for the \texttt{Basic} (\modelbasiccolor) and \texttt{ThermalKinetic} (\modelthermalkineticcolor) models fit to the observed $z=0$ \gsmf{} and \ssm. The $\chi^2_\nu$ values of the fits are shown in the legend. The three contours of each colour indicate $34$, $68$, and $95$ per cent credibility levels. The vertical and horizontal dotted lines indicate the best-fitting values of the model parameters, corresponding to the maximum of the posterior.}
    \label{fig:basic_and_thermal_kinetic_models_posterior}
\end{figure}

Fig.~\ref{fig:basic_and_thermal_kinetic_models_posterior} shows the posterior distributions of the parameters of the \texttt{Basic} (\modelbasiccolor) and \texttt{ThermalKinetic} (\modelthermalkineticcolor) models resulting from fitting the emulators to the observed \gsmf{} and \ssm, as explained in $\S$\ref{subsection: MCMC}. The three contours of the same colour signify the 34, 68, and 95 per cent credibility levels of the posterior. Additionally, we show one-dimensional projections of the posterior distribution for each subgrid parameter. Because the \texttt{Basic} model does not include the kinetic channel of SN feedback, this model is not displayed in the bottom row where the values of the kinetic feedback-related parameter, $f_{\rm kin}$, are plotted. 

First, we observe that the regions of parameter space explored by the Latin hypercubes encompass the peaks of the posterior distribution in the two models, covering it by more than $\pm 2\sigma$ (the 95 per cent credibility levels). This is reassuring in that it implies our results are not driven by the boundaries of the chosen prior. Second, both models prefer a value of the dimensionless SN energy parameter $f_{\rm E}$ of order unity, implying that a single CC SN releases $\sim 10^{51}$ erg of energy, which is consistent with standard theoretical expectations. Third, the best-fitting \texttt{ThermalKinetic} model has a BH seed mass of $m_{\rm BH, seed} \approx 10^{4.7}~\mathrm{M_\odot}$, whereas for the \texttt{Basic} model, $m_{\rm BH, seed}$ is nearly an order of magnitude lower, $m_{\rm BH, seed} \approx 10^{4.0}~\mathrm{M_\odot}$. This is likely because the prescription for SN feedback in the \texttt{Basic} model is too simplistic, such that the model's only way to improve agreement with the observed \gsmf{} at the massive end -- without worsening the match at the low-mass end even more -- is to reduce the strength of AGN feedback. The AGN feedback is suppressed by lowering $m_{\rm BH, seed}$, as there is no other AGN feedback-related parameter available for tuning in the \texttt{Basic} model. Fourth, the posterior of the \texttt{ThermalKinetic} model peaks at the kinetic energy fraction in SN feedback of $f_{\rm kin}\approx 0.3$, indicating the importance of SN kinetic feedback in bringing this model closer to the observational data.

Finally, in the legend next to the names of the models, we show the $\chi^2_\nu$ value of their fits to the observational data, which is $9.3$ for the \texttt{Basic} model and $4.7$ for \texttt{ThermalKinetic}. These values are in line with our conclusions from Fig.~\ref{fig:basic_and_thermal_kinetic_model_calibration}: that the \texttt{ThermalKinetic} model outperforms the \texttt{Basic} model, but neither model is a good fit to the data. 

\subsubsection{The effect of changing the model parameters}

\begin{figure*}
    \centering
    \includegraphics[width=0.99\textwidth]{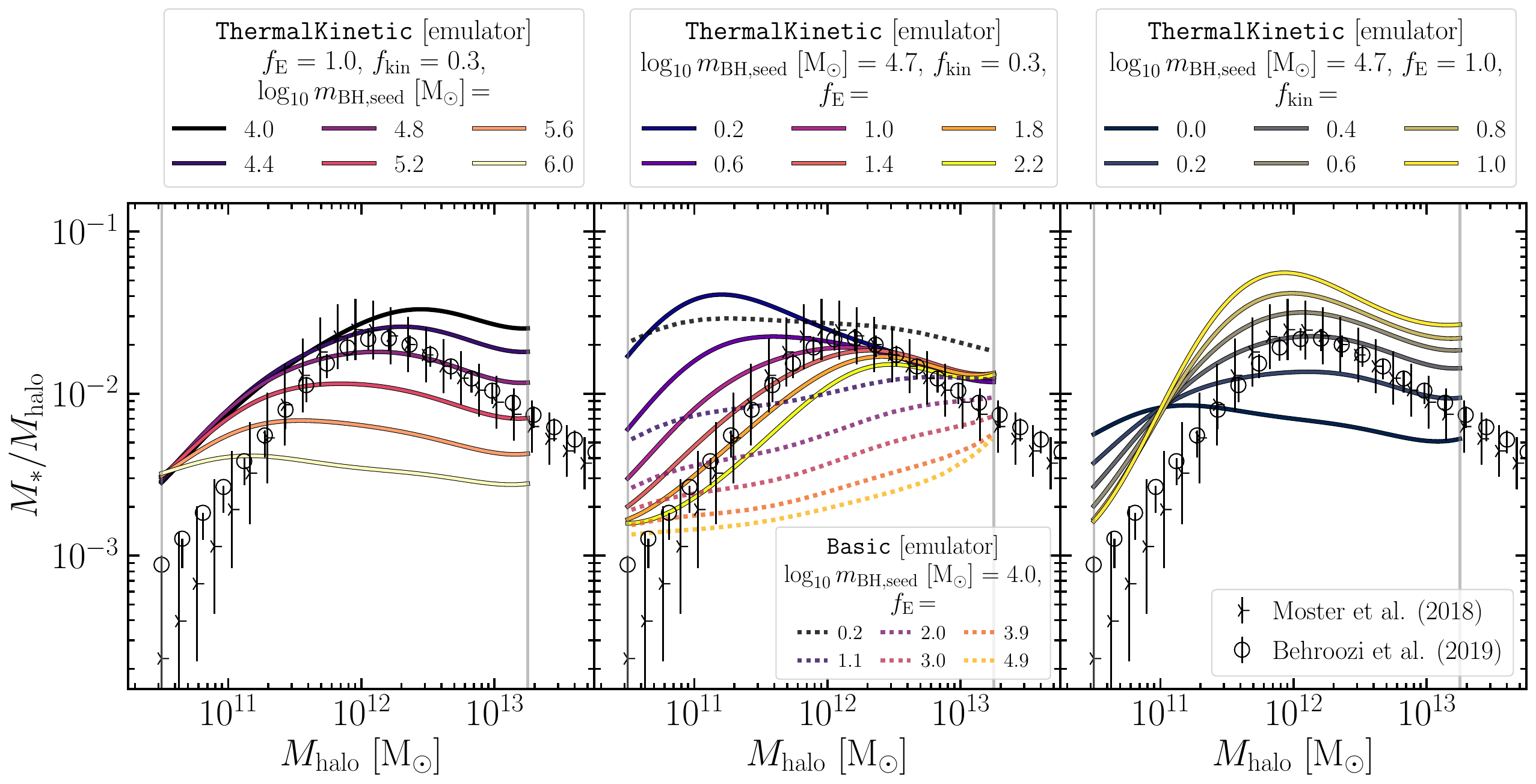}
    \caption{The stellar to halo mass relation (\shmr) at $z=0$ predicted by the trained emulators. The results are shown for the \texttt{ThermalKinetic} model fit to the observed \gsmf{} and \ssm. The individual panels show how the emulated \shmr{} varies with the BH seed mass ($m_{\rm BH, seed}$; \textit{left}), the energy in SN feedback in units of $10^{51}$ erg ($f_{\rm E}$; \textit{middle}), and the fraction of SN energy injected in kinetic form ($f_{\rm kin}$; \textit{right}). Different colours correspond to different values of each parameter. Only one parameter is varied at a time, while the other parameters are fixed to their best-fitting values as indicated in the legends. The vertical solid lines designate the mass range within which the emulators were trained on the simulations. For reference, each panel shows the data from the semi-empirical models of \citet{2018MNRAS.477.1822M} and \citet{2019MNRAS.488.3143B}, displayed as black points. Additionally, the middle panel shows the \shmr{} in the \texttt{Basic} model, also for different values of $f_{\rm E}$ (thin dotted curves). Regardless of the value of $f_{\rm E}$, the \shmr{} in the \texttt{Basic} model is always too flat compared to the data. This problem is resolved in the \texttt{ThermalKinetic} model for high enough $f_{\rm kin}$.}
    \label{fig:smhm_emulator_thermalkinetic}
\end{figure*}

Fig.~\ref{fig:smhm_emulator_thermalkinetic} shows the $z=0$ stellar-to-halo mass relation for central subhaloes in the \texttt{ThermalKinetic} model. As discussed in $\S$\ref{subsection: simulated_relations}, we do not fit the emulator to the \shmr{} because it is not directly observed. Here we only use the \shmr{} to predict how varying the model parameters affects the galaxy stellar mass at fixed halo mass. Displaying the \shmr{} as opposed to the \gsmf{} makes it easier to visually distinguish the impact of different subgrid parameters because, compared to the \gsmf, the \shmr{} varies over a smaller dynamical range and includes a characteristic change in the sign of the slope of the relation.

In each panel of Fig.~\ref{fig:smhm_emulator_thermalkinetic}, differently coloured solid curves correspond to different \shmrs{} predicted by the emulator in which two of the three model parameters are fixed to their best-fitting values, and the remaining parameter is varied. The BH seed mass is varied in the left panel, the SN energy in the middle panel, and the fraction of SN energy injected in kinetic form in the right panel. In addition, in the middle panel, we display the \shmr{} predicted by the emulator of the \texttt{Basic} model and how it changes with $f_{\rm E}$ (thin dotted curves in different colours). The only other parameter of the \texttt{Basic} model, $m_{\rm BH, seed}$ is equal to its best-fitting value, $10^{4.0}~\mathrm{M_\odot}$. The \shmr{} from the semi-empirical models of \citet{2018MNRAS.477.1822M} and \citet{2019MNRAS.488.3143B} are shown for reference only (black points).

First, by examining the left panel, we find that, as expected, $m_{\rm BH, seed}$ predominantly affects the stellar mass of massive haloes ($M_{\rm halo} \gtrsim 10^{11.5}~\mathrm{M_\odot}$). In lower mass haloes ($M_{\rm halo} \lesssim 10^{11.5}~\mathrm{M_\odot}$), BHs grow less efficiently, resulting in a lack of AGN feedback and, consequently, a much weaker dependence of the \shmr{} on $m_{\rm BH, seed}$, unless $m_{\rm BH, seed}$ is very high ($m_{\rm BH, seed}\gtrsim 10^{5.5}~\mathrm{M_\odot}$). Furthermore, we observe that the value of the BH seed mass determines the halo mass at which the \shmr{} reaches its peak. The curve with $m_{\rm BH, seed} = 10^{4.8}~\mathrm{M_\odot}$, which is the value nearest to $m_{\rm BH, seed} \approx 10^{4.7}~\mathrm{M_\odot}$ in the best-fitting \texttt{ThermalKinetic} model to the \gsmf{} and \ssm, yields an \shmr{} that is closest in shape and normalization to the \shmr{} inferred from the data by the semi-empirical models of \citet{2018MNRAS.477.1822M} and \citet{2019MNRAS.488.3143B}. This is expected, since constraints on the \gsmf{} and \shmr{} are correlated: fitting the model to either relation should improve the agreement with the other.

We next move to the middle panel of Fig.~\ref{fig:smhm_emulator_thermalkinetic}, which shows the effect of varying $f_{\rm E}$. Unlike the left panel, here we display the results for both the \texttt{ThermalKinetic} and \texttt{Basic} models\footnote{Note that the ranges over which $f_{\rm E}$ is varied are different for the two models.}. In essence, increasing (decreasing) $f_{\rm E}$ moves the bulk of the \shmr{} down (up) in both models, as the SN feedback becomes stronger (weaker), leading to less (more) stellar mass formed by $z=0$. Crucially, in the \texttt{Basic} model, the shape of the \shmr{} exhibits minimal dependence on $f_{\rm E}$, which renders it impossible for this model to match the \shmr{} of the semi-empirical models solely by adjusting $f_{\rm E}$. Setting $f_{\rm E}$ to $2$ or $3$ to achieve realistic stellar-to-halo mass ratios at $M_{\rm halo} \sim 10^{11}~\mathrm{M_\odot}$ (pink through violet dotted curves) while simultaneously lowering 
$m_{\rm BH, seed}$ from its best-fitting value of $\approx 10^{4.0}~\mathrm{M_\odot}$ to better match the peak of the \shmr{} of the semi-empirical models, cannot help either because doing so will either result in excessively high stellar masses for the most massive haloes ($M_{\rm halo} \gtrsim 10^{13}~\mathrm{M_\odot}$) or undershoot the peak of the \shmr.

The agreement with the semi-empirical models' \shmrs{} is strongly improved in the \texttt{ThermalKinetic} model, which exploits the kinetic channel of SN feedback with low-energy kicks, corresponding to the kick velocity of 50 km s$^{-1}$. The right panel of the figure shows that increasing $f_{\rm kin}$ reduces the galaxy stellar mass at low $M_{\rm halo}$ and increases it at high $M_{\rm halo}$, thereby steepening the slope of the \shmr. This helps the \texttt{ThermalKinetic} model obtain a better fit to the observed \gsmf, as we have seen in Fig.~\ref{fig:basic_and_thermal_kinetic_model_calibration}, and correspondingly, to the \shmr, as is seen in the current figure. Such a behaviour of the \shmr{} with $f_{\rm kin}$ can be expected: higher $f_{\rm kin}$ implies that more SN energy is injected kinetically through numerous $50$ km s$^{-1}$ kicks and that less energy is distributed thermally via large, rare energy injections corresponding to a gas temperature increase of $\Delta T_{\rm SN} = 10^{7.5}~\mathrm{K}$. The kinetic channel is especially efficient in low-mass galaxies, in which the escape velocity is comparable to or lower than the kick velocity used by the kinetic channel. At the same time, the kinetic channel is too weak to push the gas out of more massive objects because of their deeper gravitational potential wells. Conversely, the thermal channel can drive vigorous outflows in galaxies as massive as the Milky Way \citep{2023MNRAS.523.3709C} but is hindered by poor sampling in low-mass objects (see $\S$\ref{subsubsection: dens_dep_dT}).

\subsubsection{Fitting to the observed \gsmf{} and \ssm{} separately and simultaneously}
\label{subsubsection: vary_fitting_constraints}

The best-fitting models that we have discussed so far fit simultaneously the observed \gsmf{} and \ssm. We now investigate the effect of fitting the models separately to either the \gsmf{} or the \ssm. Mathematically, this means setting the log likelihood function in equation (\ref{eq: likelihood}),  $\ln \mathcal{L}(\boldsymbol{\theta})$, to either $\ln \mathcal{L}_{\rm GSMF} (\boldsymbol{\theta})$ or $\ln \mathcal{L}_{\rm SSMR} (\boldsymbol{\theta})$, instead of the sum of the two.

\begin{figure*}
\includegraphics[width=0.99\textwidth]{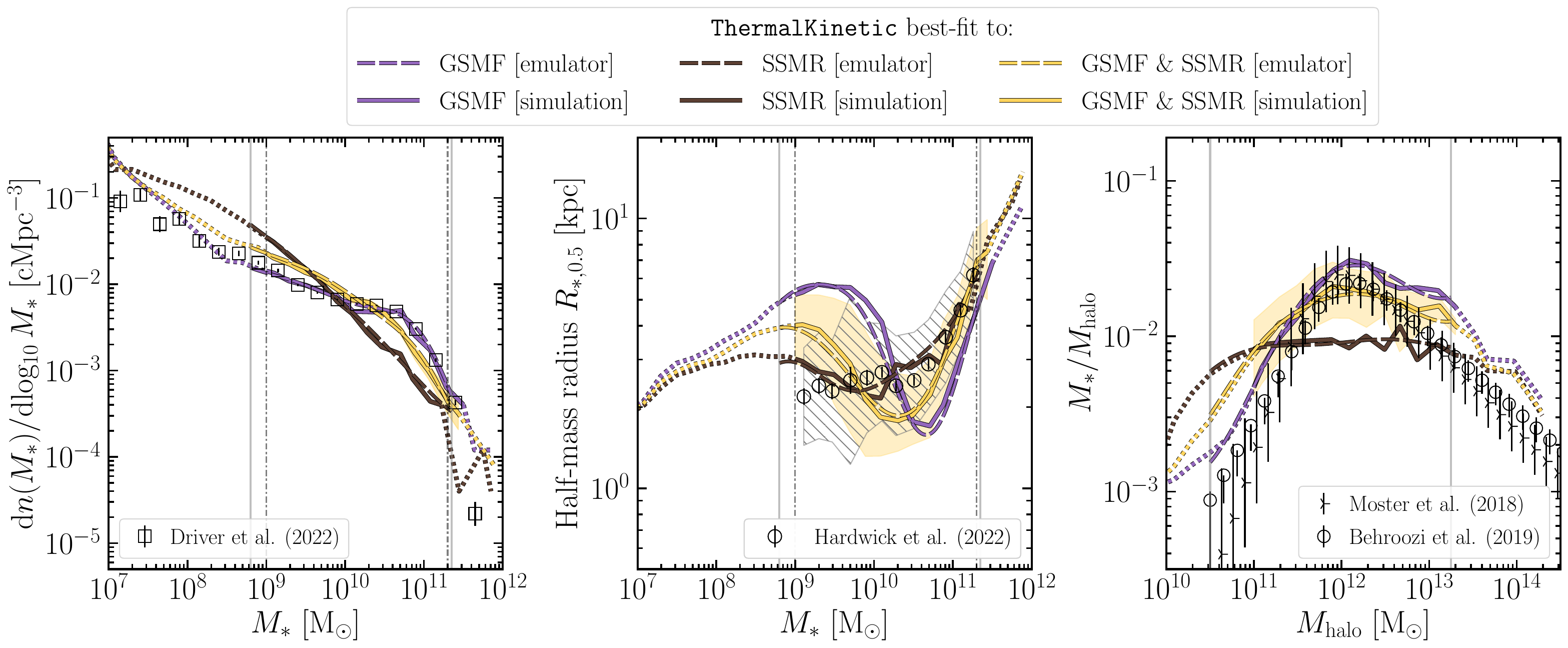}
\caption{Predictions of the best-fitting \texttt{ThermalKinetic} model fit to the observed galaxy stellar mass function (\gsmf, purple), galaxy size -- stellar mass relation (\ssm, brown), or to both the \gsmf{} and \ssm{} (\modelthermalkineticcolor). We show the $z=0$ \gsmf{} (\textit{left}), the $z=0$ \ssm{} (\textit{middle}), and the $z=0$ stellar to halo mass relation (\shmr; \textit{right}). The emulator predictions are shown as dashed curves, and the results from simulations using the best-fitting parameters are shown as solid curves. The solid curves become dotted at stellar (or halo) masses where galaxies are poorly resolved or where the number of galaxies is small due to the finite simulation volume. The vertical solid and dash-dotted lines carry the same meaning as in Fig.~\ref{fig:basic_and_thermal_kinetic_model_calibration}. There are no vertical dash-dotted lines in the right panel because we do not fit the model to the \shmr. Fitting only to the observed \gsmf{} (\ssm) results in a good match to the observed \gsmf{} (\ssm) but a poor match to the \ssm{} (\gsmf). Fitting to both observed relations at the same time produces only a reasonable match to the two constraints.}
\label{fig: thermal_kinetic_model_with_var_constraints}
\end{figure*}

Fig.~\ref{fig: thermal_kinetic_model_with_var_constraints} compares the \texttt{Thermalkinetic} model with three different sets of the best-fitting parameters, obtained from fitting the emulator to three different sets of observational data: the \gsmf{} (purple), the \ssm{} (brown), or both the \gsmf{} and \ssm{} (\modelthermalkineticcolor). The dashed curves indicate the best-fitting predictions of the emulators and the solid curves correspond to simulations with the best-fitting parameters. The shaded \modelthermalkineticcolor{} region shows the $1\sigma$ scatter in the simulation whose model was simultaneously fit to the \gsmf{} and \ssm.

The left panel shows the $z=0$ \gsmf, the middle panel shows the $z=0$ \ssm, and the right panel shows the $z=0$ \shmr. As in Fig.~\ref{fig:basic_and_thermal_kinetic_model_calibration}, the solid lines become dotted in the mass range where galaxies are poorly resolved ($M_* < 10^9~\mathrm{M_\odot}$) and where the number of galaxies per bin drops below $5$. In the right panel, which plots the halo mass instead of stellar mass, the mass below which the solid lines turn to dotted is $M_{\rm halo} = 10^{11}~\mathrm{M_\odot}$. In each panel, the vertical solid lines indicate the mass range within which the emulators were trained on the Latin hypercubes, while the vertical dash-dotted lines, if present, specify the mass range where the trained emulators were fit to the observational data. As in the previous figures, the comparison data from \citet{2018MNRAS.477.1822M}, \citet{2019MNRAS.488.3143B}, \cite{2022MNRAS.509.3751H}, and \citet{2022MNRAS.513..439D} are displayed as black points.

Examining the left and middle panels of Fig.~\ref{fig: thermal_kinetic_model_with_var_constraints}, we see that the \texttt{ThermalKinetic} model with the best-fitting parameters matches the observed \gsmf{} (\ssm) well if it is fit to the \gsmf{} (\ssm) alone. This, however, comes at the expense of a poor fit to the other observed relation, which was left out of the fitting. In contrast, fitting the model to both the \gsmf{} and \ssm{} at the same time forces the emulator to find a compromise solution. In this case, the best-fitting model provides a mediocre match to the observed \gsmf{} while also being less far off from (but still not close to) the observed \ssm. We conclude that although the \texttt{ThermalKinetic} model can match either of the two observed relations well, the model is too limited to be able to reproduce both relations at the same time. To succeed in doing so, a more complex model is required.

The right panel of Fig.~\ref{fig: thermal_kinetic_model_with_var_constraints}, which shows the \shmr, confirms what we have seen in the left panel, but the differences between the different cases appear more striking. The galaxies in the \texttt{ThermalKinetic} model fit to the \ssm{} follow a nearly flat \shmr, which is clearly wrong. The two other cases, in which the observed \gsmf{} was used as a constraint to the model, have \shmrs{} whose shape resembles that in \citet{2018MNRAS.477.1822M} and \citet{2019MNRAS.488.3143B}. Interestingly, the model that is constrained only by the \gsmf{} produces an \shmr{} whose peak is $\approx 0.1$~dex higher than in the two semi-empirical models. 

Similar to Fig.~\ref{fig:basic_and_thermal_kinetic_model_calibration}, for all cases of observational constraints and for all emulated relations shown in Fig.~\ref{fig: thermal_kinetic_model_with_var_constraints}, the simulations closely follow the predictions of the emulators, thereby validating our emulator-based approach.

\begin{figure}
    \centering
    \includegraphics[width=0.49\textwidth]{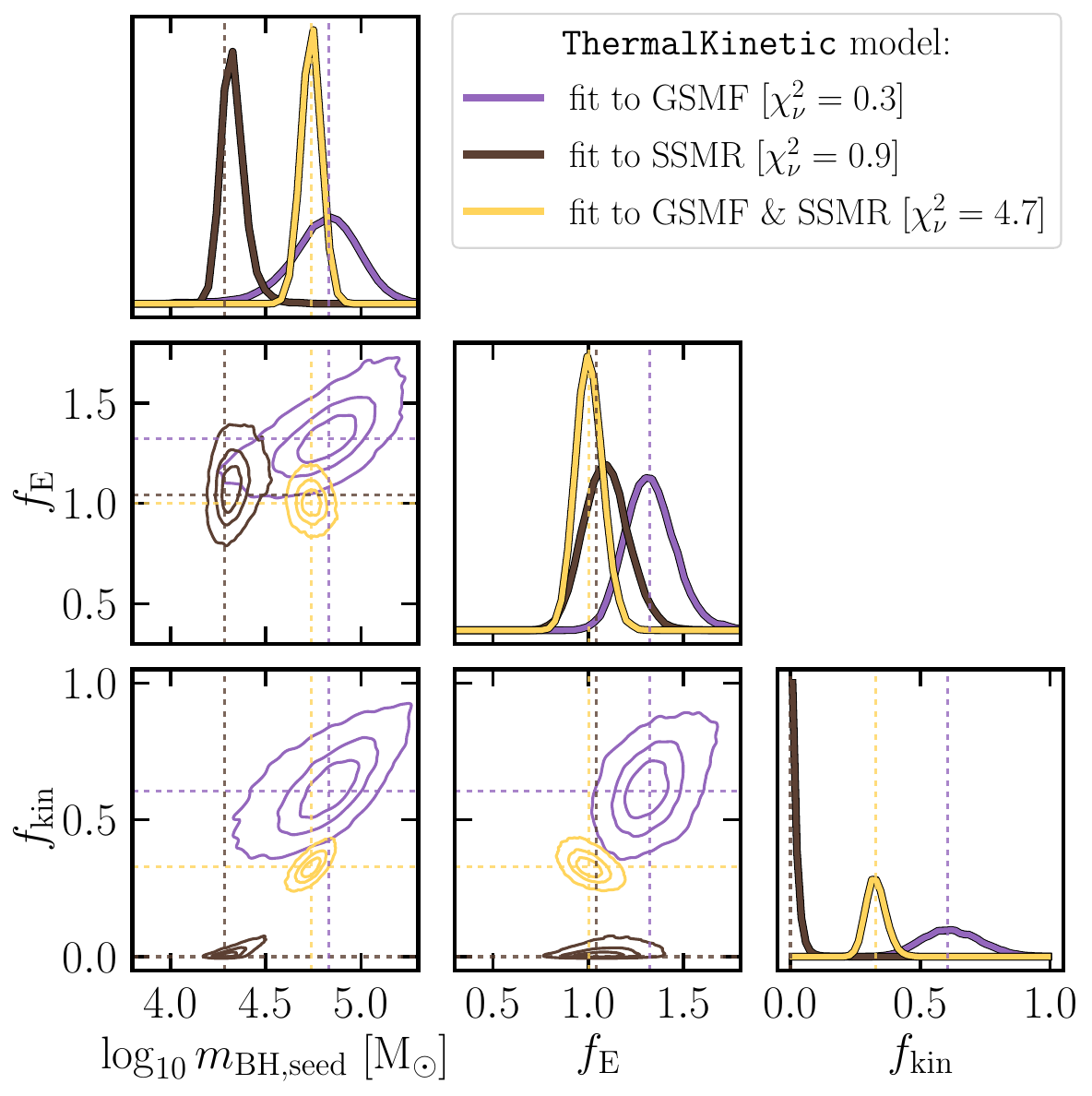}
    \caption{Posterior distributions of the parameters of the \texttt{ThermalKinetic} model resulting from fitting the emulator to the observed \gsmf{} (purple), the observed \ssm{} (brown), or both the \gsmf{} and \ssm{} (\modelthermalkineticcolor), with the $\chi^2_\nu$ value of each fit indicated in the legend. The three contours of each colour correspond to $34$, $68$, and $95$ credibility levels. The vertical and horizontal lines indicate the values of the best-fitting parameters for each case. The best-fitting parameter values of the model fit to the \gsmf{} and the model fit to the \ssm{} belong to very different regions of the parameter space. The model fit to both the \gsmf{} and \ssm{} (\modelthermalkineticcolor) is located in between the models fit to the \gsmf{} and \ssm{} separately.}
    \label{fig:thermalkinetic_model_posterior}
\end{figure}

Fig.~\ref{fig:thermalkinetic_model_posterior} shows the posterior distributions of the subgrid parameters for the \texttt{ThermalKinetic} model after fitting the emulator to the observed \gsmf{} (purple), to the observed \ssm{} (brown), or to both (\modelthermalkineticcolor). The three contours of each colour indicate the $34$, $68$, and $95$ per cent credibility regions of the posterior distributions. In the legend, we quote the values of $\chi^2_\nu$ for each fit. Fitting to either the \gsmf{} or \ssm{} yields a $\chi^2_\nu$ of order unity or less, indicating that the model is a good fit to the data or slightly overfits them, given the emulator uncertainties. In contrast, fitting simultaneously to the \gsmf{} and \ssm{} yields a $\chi^2_\nu$ of $4.7$, indicating the model lacks the necessary complexity to match both observed relations simultaneously. 

Examining the peaks of the posterior distributions reveals that, in each case, the models calibrated to only the \gsmf{} or only the \ssm{} occupy different regions of parameter space. Specifically, the model best-fitting the \ssm{} (brown) prefers a BH seed mass of $m_{\rm BH, seed} \approx 10^{4.3}~\mathrm{M_\odot}$, an SN energy per event of $f_{\rm E} \approx 1.0$, and no kinetic SN feedback ($f_{\rm kin} \approx 0$). In contrast, the model best-fitting the \gsmf{} (purple) yields $m_{\rm BH, seed} \approx 10^{4.8}~\mathrm{M_\odot}$, $f_{\rm E} \approx 1.3$, and $f_{\rm kin} \approx 0.6$. The best-fitting parameter $m_{\rm BH, seed}$ ($f_{\rm E}$) of the model calibrated to both the \gsmf{} and \ssm{} is close to that of the model calibrated only to the \gsmf{} (\ssm), while the value of $f_{\rm kin}$ lies in between the two, at $f_{\rm kin} \approx 0.3$.

\subsection{Calibration diagnostics for \texorpdfstring{$\protect\boldsymbol{\protect\mathtt{ThermalKinetic\_var\Delta T_{\protect\mathtt{SN}}}}$ and 
$\protect\boldsymbol{\protect\mathtt{ThermalKinetic\_var\Delta T_{\protect\mathtt{SN}}varf_{\protect\mathtt{E}}}}$}{ThermalKinetic\_varDelta T\_SN and ThermalKinetic\_varDelta T\_SN varf\_E} models
}
\label{subsection: more_advanced_models}

Having learned that neither the \texttt{Basic} model nor the \texttt{ThermalKinetic} model can simultaneously fit the observed \gsmf{} and \ssm, we turn our attention to the more complex models: \VardT{} and \colibrefixedagntemp.

\subsubsection{Galaxy stellar mass function and galaxy sizes}

\begin{figure*}
    \centering
    \includegraphics[width=0.99\textwidth]{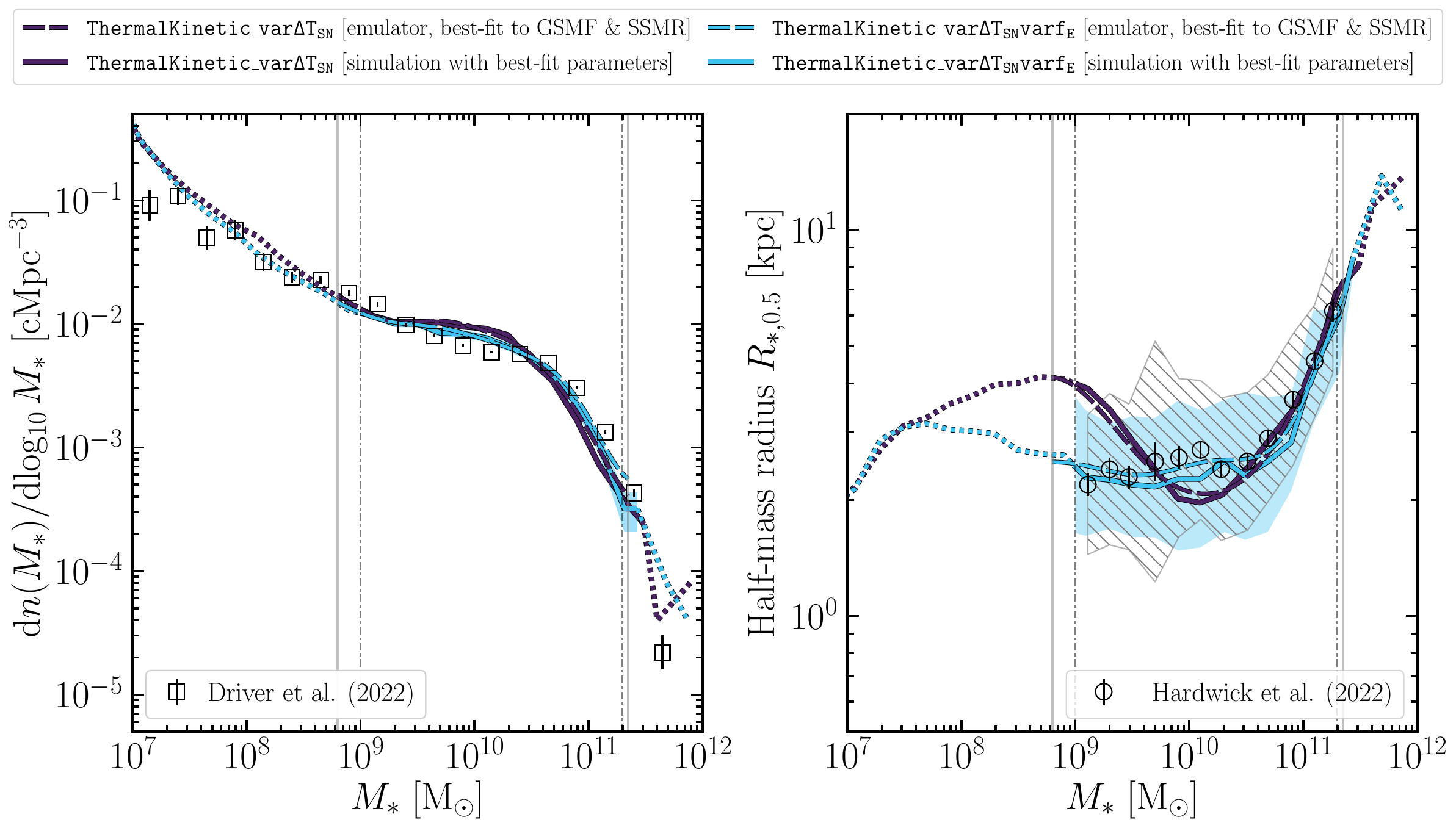}
    \caption{As Fig.~\ref{fig:basic_and_thermal_kinetic_model_calibration} but showing the $z=0$ galaxy stellar mass function (\gsmf; \textit{left}) and median size -- stellar mass relation (\ssm; \textit{right}) for the \VardT{} (\modelvardtcolor) and \colibrefixedagntemp{} (\colibrefixedagndtcolor) models. While the \VardT{} model shows only reasonably good agreement with the observed \gsmf{} and disagrees in shape with the observed \ssm{} at $M_* \lesssim 10^{10.5}~\mathrm{M_\odot}$, the \colibrefixedagntemp{} model successfully reproduces both observational constraints across the entire fitting range ($10^9 < M_*/\mathrm{M_\odot} < 10^{11.3}$).}
\label{fig: models_three_and_four}
\end{figure*}

Fig.~\ref{fig: models_three_and_four} shows the \gsmf{} and \ssm{} at $z=0$ for the best-fitting \VardT{} (\modelvardtcolor) and \colibrefixedagntemp{} (\colibrefixedagndtcolor) models. The different symbols and line styles have the same meaning as in Fig.~\ref{fig:basic_and_thermal_kinetic_model_calibration}.

It is evident that both the \VardT{} and \colibrefixedagntemp{} models outperform the \texttt{Basic} and \texttt{ThermalKinetic} models, whose \gsmf{} and \ssm{} were shown in Fig.~\ref{fig:basic_and_thermal_kinetic_model_calibration}. The \gsmf{} in the \VardT{} model agrees with the observed \gsmf{} for $M_* \lesssim 10^{9.5}~\mathrm{M_\odot}$, overshoots it by up to $\approx 0.15$~dex in the range $10^{9.5} \lesssim M_*/\mathrm{M_\odot} \lesssim 10^{10.5}$, and undershoots it by up to $\approx 0.25$~dex at higher $M_*$. The best performance is achieved for the \colibrefixedagntemp{} model whose \gsmf{} closely follows the observed \gsmf{} across more than 4 dex in $M_*$, with deviations within the fitting range ($10^{9} < M_*/\mathrm{M_\odot} < 10^{11.3}$) remaining within $\approx 0.1$~dex.

The agreement between the \ssm{} in the \VardT{} model and the observed data is less satisfactory than for the \gsmf. Similar to the \texttt{ThermalKinetic} model shown in Fig.~\ref{fig:basic_and_thermal_kinetic_model_calibration}, the \VardT{} model overshoots the observed sizes of galaxies with $M_* \lesssim 10^{9.5}~\mathrm{M_\odot}$ and exhibits a dip in the \ssm{} at $M_* \approx 2 \times 10^{10}~\mathrm{M_\odot}$ -- albeit less pronounced than \texttt{ThermalKinetic} -- which is not present in the observed relation. In contrast, this dip is absent in the \ssm{} of the \colibrefixedagntemp{} model, which closely follows the observed \ssm{} for all stellar masses in the fitting range, $10^9< M_* / \mathrm{M_\odot} < 10^{11.3}$, reproducing both the median galaxy half-mass size and its scatter at fixed $M_*$. For stellar masses $M_*<10^9~\mathrm{M_\odot}$, where the models are not fit to observational data, the median half-mass radius does not drop below $2$ kpc in either model. This `floor' likely arises due to the relatively low numerical resolution of the simulations: at m7 resolution, a galaxy with a stellar mass $10^9~\mathrm{M_\odot}$ is sampled with only $\sim 100$ stellar particles. 

To sum up, among the four considered models -- \texttt{Basic}, \texttt{ThermalKinetic}, \VardT, and \colibrefixedagntemp{} -- \colibrefixedagntemp{} is the only model that simultaneously reproduces both the observed \gsmf{} and \ssm. Furthermore, as in previous figures that showed results from emulators and simulations for the same model parameters, Fig.~\ref{fig: models_three_and_four} confirms that the emulator errors are negligibly small.

\begin{figure*}
    \centering
    \includegraphics[width=0.999\textwidth]{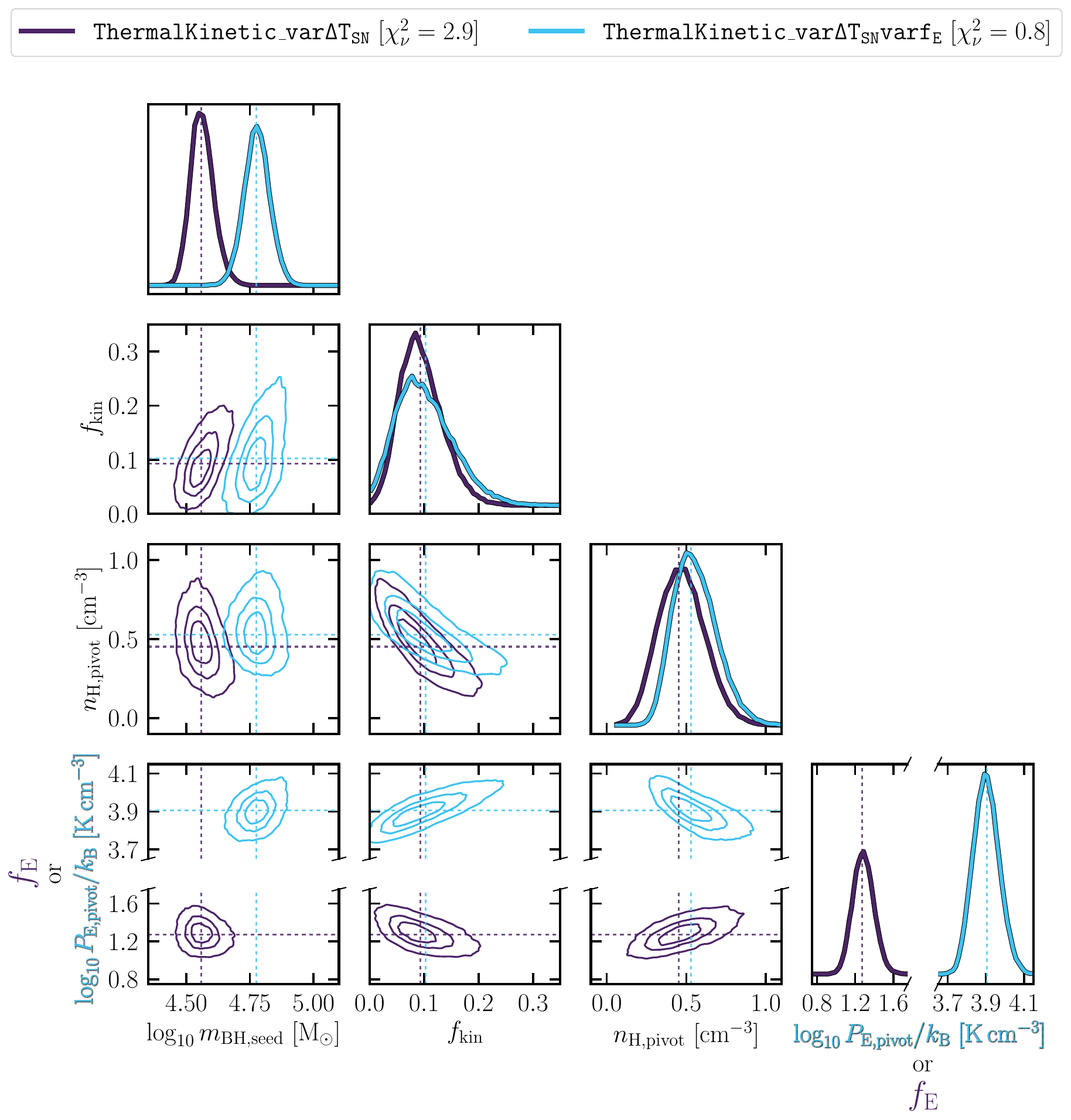}
    \caption{Posterior distributions of the parameters of the \VardT{} model (\modelvardtcolor) and the \colibrefixedagntemp{} model (\colibrefixedagndtcolor), obtained by fitting the emulator to the observed $z=0$ \gsmf{} and \ssm. The contours of the same colour indicate the $34$, $68$, and $95$ per cent credibility regions of the posterior distributions. Vertical and horizontal dotted lines mark the best-fitting parameter values for each model, corresponding to the maxima of their respective posterior distributions. While the \colibrefixedagntemp{} (\VardT) model favours a BH seed mass of $\approx 10^{4.8}~\mathrm{M_\odot}$ ($10^{4.6}~\mathrm{M_\odot}$), the fraction of SN energy injected in kinetic form, $f_{\rm kin}$, and the pivot density in SN thermal feedback, $n_{\rm SN, pivot}$, are similar in both models: $f_{\rm kin} \approx 0.1$ and $n_{\rm H,pivot} \approx 0.5$ cm$^{-3}$. In the bottom row, we show the parameters that are unique to each model: $f_{\rm E}$ for \VardT{}, and $P_{\rm E,pivot} / k_{\rm B}$ for \colibrefixedagntemp{}; their best-fitting values are approximately $1.3$ and $10^{3.9}~\mathrm{\rm K \, cm^{-3}}$, respectively.}
    \label{fig:reference_model_posterior}
\end{figure*}

\subsubsection{Posterior distributions of the model parameters}

Fig.~\ref{fig:reference_model_posterior} displays the posterior distribution of the parameters in \VardT{} (\modelvardtcolor) and \colibrefixedagntemp{} (\colibrefixedagndtcolor) after each model was fit to both the observed \gsmf{} and \ssm. Because one of the four parameters from \VardT{} does not exist in \colibrefixedagntemp, and vice versa, we plot the unique parameters of the two models in the bottom row of the figure at the same time: $f_{\rm E}$ for \VardT{} and $P_{\rm E,pivot}$ (in log) for \colibrefixedagntemp. We show the same range of values for both parameters but attach two different labels to the panel axes, which for clarity are shown in the colours of the corresponding models (\modelvardtcolor{} and \colibrefixedagndtcolor).

The \VardT{} model prefers a BH seed mass of $\approx 10^{4.6}~\mathrm{M_\odot}$, while the \colibrefixedagntemp{} model favours a slightly higher value of $m_{\rm BH,seed} \approx 10^{4.8}~\mathrm{M_\odot}$. The fraction of SN energy injected in kinetic form is close to 10 per cent in both models, which is lower than the $f_{\rm kin} \approx 0.3$ in the \texttt{ThermalKinetic} model calibrated to the same observational data. This is likely because, unlike in \texttt{ThermalKinetic}, the heating temperature $\Delta T_{\rm SN}$ in \VardT{} and \colibrefixedagntemp{} can vary between $10^{6.5}$ and $10^{7.5}~\mathrm{K}$. Thermal SN feedback with low $\Delta T_{\rm SN}$ can reproduce some of the effects of kinetic feedback with low $\Delta v_{\rm kick}$, reducing the need for a high $f_{\rm kin}$. However, the thermal feedback cannot replace kinetic feedback completely because for low $\Delta T_{\rm SN}$ and/or in high-density gas, radiative energy losses will inevitably become high, rendering the thermal feedback inefficient.

In both models, the best-fitting value of the pivot density in the SN thermal feedback with a variable heating temperature is $n_{\rm SN, pivot} \approx 0.5$ cm$^{-3}$, with \colibrefixedagntemp{} (\VardT) favouring slightly higher (lower) values. At $\Delta T_{\rm SN}=\Delta T_{\rm SN, pivot}=10^{6.5}~\mathrm{K}$, $n_{\rm SN, pivot}=0.5~\mathrm{cm}^{-3}$ corresponds to the \cite{2012MNRAS.426..140D} critical density for $f_t\approx 2.5$ (see equation~\ref{eq: critical_density_SN_feedback}). The best-fitting value of $f_{\rm E}$ in the \VardT{} model is $\approx 1.3$, which is slightly higher than in the \texttt{ThermalKinetic} model ($f_{\rm E}\approx 1.0$). The increase in $f_{\rm E}$ likely originates from the fact that the average $\Delta T_{\rm SN}$ in \VardT{} is lower than $10^{7.5}$ K, which results in weaker SN thermal feedback compared to \texttt{ThermalKinetic}, as more energy is radiated away due to the enhanced radiative cooling rates of gas heated to lower $\Delta T_{\rm SN}$. To compensate for the weaker SN feedback, the energy per SN in units of $10^{51}$ erg is increased from $\approx 1.0$ to $1.3$.

The best-fitting value of the pivot birth pressure in the \colibrefixedagntemp{} model is $P_{\rm E,pivot}/k_{\rm B}\approx 10^{3.9}~\mathrm{K \, cm}^{-3}$, which is close to the median stellar birth pressure in the simulation with this best-fitting model: $\approx 10^{3.75}$ $\rm K \, cm^{-3}$. Substituting this median value into equation (\ref{eq: stellar_birth_pressure_vs_SN_energy}), along with the other model parameters that were not considered in the calibration ($f_{\rm E, min} = 0.1$, $f_{\rm E, max} = 4$, and $\sigma_{\rm P} = 0.3$), we obtain an SN energy at the median stellar birth pressure of $f_{\rm E}(P_{\rm birth}/k_{\rm B} = 10^{3.75}~\mathrm{K \, cm}^{-3}) \approx 1.6$. The value of $f_{\rm E}$ averaged over all stellar particles formed in the simulation is slightly higher\footnote{We note that, as described in $\S$\ref{subsection: CC_SN_feedback}, \colibre{} computes the energy in CC SN feedback per stellar particle by integrating the \citet{Chabrier2003} IMF from $m_{\rm min,CCSN} = 8~\mathrm{M_\odot}$ to $m_{\rm max,CCSN} = 100~\mathrm{M_\odot}$, whereas \textsc{eagle} used a lower integration limit of $m_{\rm min,CCSN} = 6~\mathrm{M_\odot}$. As a result, $f_{\rm E} = 1.8$ in \colibre{} corresponds to $f_{\rm E} \approx 1.2$ in \textsc{eagle}.}: $\langle f_{\rm E} \rangle \approx 1.8$. Both of these values of $f_{\rm E}$ are comparable to the best-fitting (constant) value in \VardT, $f_{\rm E} = 1.3$, indicating that both models favour an (average) SN feedback energy slightly higher than the theoretical expectation, $f_{\rm E} = 1$, corresponding to $10^{51}$ erg. In Appendix \ref{appendix: redshift_evolution_of_sn_energy}, we provide further details on $f_{\rm E}$, including the redshift evolution of the median values of $f_{\rm E}$ and $P_{\rm birth}$ as measured in the simulation using the best-fitting \colibrefixedagntemp{} model.

\begin{figure*}
    \centering
\includegraphics[width=0.999\textwidth]{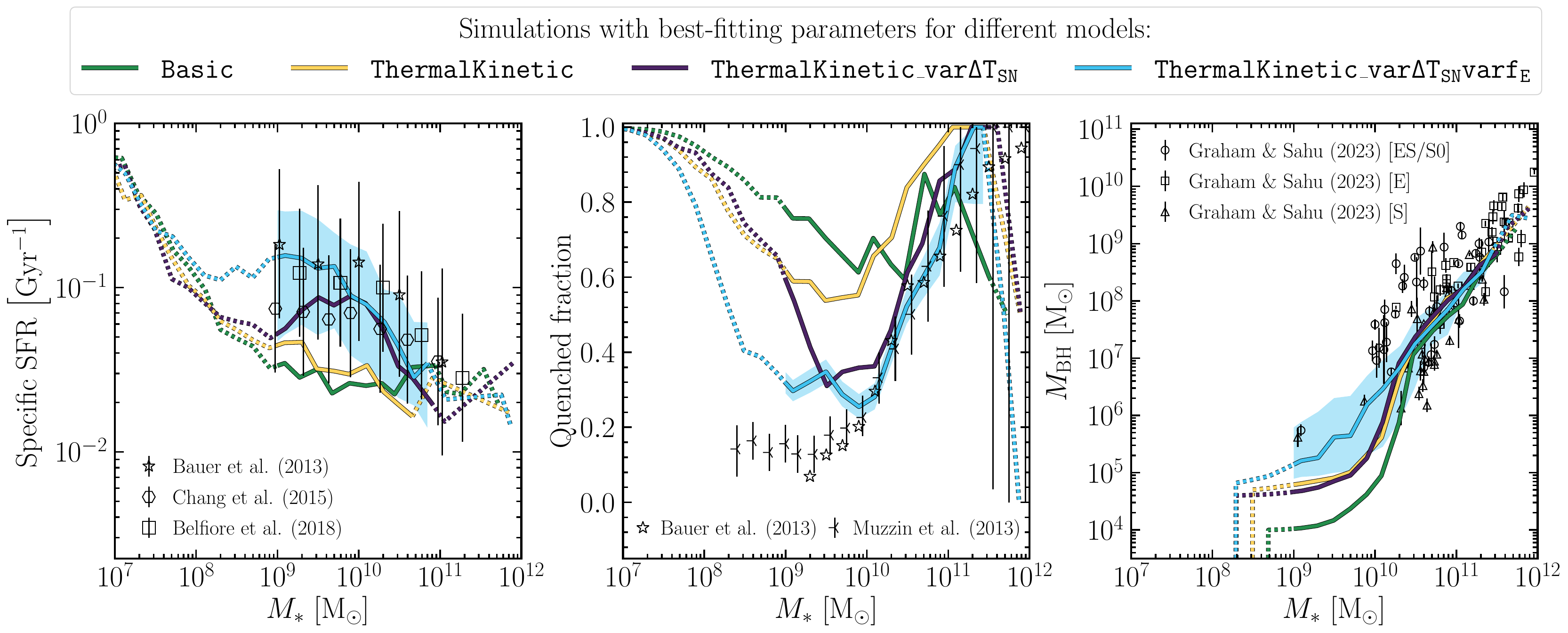}
    \caption{The median specific star formation rate (sSFR) of active galaxies ($\mathrm{sSFR} > 10^{-2}$~Gyr$^{-1}$) versus stellar mass (\textit{left}), the fraction of quenched galaxies versus stellar mass (\textit{middle}), and the median mass of supermassive black holes versus stellar mass (\textit{right}), all shown at $z=0$. Differently coloured solid curves show the results from the simulations with the best-fitting parameter values for the \texttt{Basic} (\modelbasiccolor), \texttt{ThermalKinetic} (\modelthermalkineticcolor), \VardT{} (\modelvardtcolor), and \colibrefixedagntemp{} (\colibrefixedagndtcolor) models. All models were fit to the observed $z=0$ \gsmf{} and \ssm. The solid curves turn into dotted curves at stellar mass below $10^{9}~\mathrm{M_\odot}$ indicating that those galaxies are poorly resolved, and when the number of objects per bin is less than 5 indicating the limit due to the simulated volume. The shaded \colibrefixedagndtcolor{} region shows the $1\sigma$ scatter for the sSFR -- $M_*$ relation and the \bhmsm{}, and the $1\sigma$ confidence interval for the quenched fraction -- $M_*$ relation, all for the \colibrefixedagntemp{} model. The median BH mass drops to zero at $M_* \lesssim 10^{8.5}~\mathrm{M_\odot}$ in all models because the corresponding haloes are not massive enough to be seeded with a BH particle. A compilation of observational data is shown as black symbols. Both the \VardT{} and \colibrefixedagntemp{} models show a good agreement with the comparison data for all three relations, with the \colibrefixedagntemp{} model exhibiting marginally better sSFR and quenched fractions at $M_* \lesssim 10^{10}~\mathrm{M_\odot}$.}
    \label{fig: four_models_ssfr}
\end{figure*}

The legend of Fig.~\ref{fig:reference_model_posterior} lists the $\chi^2_\nu$ values for the fits to the observational data: $2.9$ for the \VardT{} model and $0.8$ for \colibrefixedagntemp. This confirms that both models outperform the \texttt{Basic} and \texttt{ThermalKinetic} models, with \colibrefixedagntemp{} providing the best match to the \gsmf{} and \ssm. We will use these results in $\S$\ref{subsection: three resolutions} to define the fiducial \colibre{} model at m7 and higher resolutions.

Lastly, we note that based on isolated galaxy simulations at much higher resolution ($m_{\rm gas} = 10^5~\mathrm{M_\odot}$), \citet{2023MNRAS.523.3709C} found that $f_{\rm kin}\approx 0.1$ (together with the kick velocity of $\Delta v_{\rm kick} = 50~\mathrm{km \, s}^{-1}$) allows reproducing the relation between spatially resolved H~\textsc{i} velocity dispersion and the galaxy SFR surface density, as well as the observed spatially-resolved KS star-formation law \citep{2007ApJ...671..333K}. The latter was confirmed by \citet{2024MNRAS.532.3299N}, who showed that the observed KS relation is reproduced for the range of mass resolutions from $m_{\rm gas} = 1.25 \times 10^4$ to $5.12\times 10^7~\mathrm{M_\odot}$. These findings are reassuring given that the \colibrefixedagntemp{} best-fitting model prefers $f_{\rm kin} \approx 0.1$. 

\subsection{Predictions for galaxy properties that were not included in emulator-based calibration}
\label{subsection: other_galaxy_properties}

In this section, we explore simulation predictions for galaxy properties that have not been previously discussed and therefore were not included in the emulator-based calibration of the model parameters. The following figures present the results for the four best-fitting models, \texttt{Basic}, \texttt{ThermalKinetic}, \VardT, and \colibrefixedagntemp, which have all been fit to both the observed \gsmf{} and \ssm. We show only the results from the simulations, as the emulators were constructed exclusively for the $z=0$ \gsmf{}, \ssm{}, and \shmr.

\subsubsection{Star formation rates and quenched fraction}

Fig.~\ref{fig: four_models_ssfr} displays the specific star formation rates (sSFR) of active galaxies, the fraction of quenched galaxies, and the BH masses. All relations are shown at $z=0$ and plotted versus galaxy stellar mass. All quantities are measured within 3D spherical apertures of radius 50~pkpc, and galaxy SFRs are computed using the instantaneous SFRs of gas particles. In a given stellar mass bin, we show the median sSFR and the median (subgrid) mass of the BHs. If a subhalo contains multiple BH particles, then the mass of the most massive BH is used to compute the median. We define a galaxy as ‘active’ if its instantaneous sSFR exceeds $10^{-2}~\mathrm{Gyr^{-1}}$ \citep[e.g.][]{2012MNRAS.424..232W}; otherwise, it is considered quenched. The solid curves give the results from the simulations with the four best-fitting models: \texttt{Basic} (\modelbasiccolor), \texttt{ThermalKinetic} (\modelthermalkineticcolor), \VardT{} (\modelvardtcolor) and \colibrefixedagntemp{} (\colibrefixedagndtcolor). The shaded \colibrefixedagndtcolor{} region represents the uncertainty in the \colibrefixedagntemp{} model, computed as the 16$^{\rm th}$ to 84$^{\rm th}$ percentiles for the relations between sSFR and $M_*$ and between BH mass and $M_*$, and using the Clopper–Pearson interval at the 68 per cent confidence level for the quenched fraction -- $M_*$ relation.

For comparison, we use the sSFR -- stellar mass relations for star-forming galaxies from \citet{2013MNRAS.434..209B}, \citet{2015ApJS..219....8C}, and \citet{2018MNRAS.477.3014B}. The relation from \citet{2013MNRAS.434..209B} is based on $\sim 10^5$ galaxies from GAMA DR1 \citep{2011MNRAS.413..971D} within the redshift range $0.05 < z < 0.32$, where galaxies were classified as star-forming based on the flux and equivalent width of the H~$\upalpha$ line. \citet{2015ApJS..219....8C} used $\sim 10^6$ SDSS galaxies with $z < 0.2$, combined with four-band WISE photometry \citep{2010AJ....140.1868W}, and identified star-forming galaxies using colour–colour diagram cuts. Finally, the relation from \citet{2018MNRAS.477.3014B} is based on $\sim 10^4$ galaxies at $0.01 < z < 0.15$ from the MaNGA survey \citep{Bundy2015}, where star-forming galaxies were classified using the Baldwin–Phillips–Terlevich diagram \citep{Baldwin1981}. For all three datasets, the $y$ values represent the median sSFR, and the error bars indicate the $1\sigma$ scatter. For quenched fractions, we use $z\approx 0.1$ data from GAMA DR1 presented by \citet{2013MNRAS.434..209B}, which were re-calculated by \citet{2019MNRAS.488.3143B} using our threshold for quiescent galaxies of $\mathrm{sSFR} < 10^{-2}~\mathrm{Gyr}^{-1}$, as well as data from \citet{2013ApJ...777...18M} at $0.2 < z < 0.5$, based on the COSMOS/UltraVISTA galaxy catalogue, where quiescent galaxies were defined using UVJ colour selection. For the BH mass -- stellar mass relation (\bhmsm), we adopt the measurements\footnote{Following the erratum \citet{2024MNRAS.530.3429G}, we applied a $-0.15$~dex correction to the stellar masses from \citet{2023MNRAS.518.2177G}.} from \citet{2023MNRAS.518.2177G}, whose sample is subdivided by morphological type: E, ES/S0, and S (shown with different black symbols as indicated in the legend). Where necessary, we correct for differences in the assumed stellar IMF by converting all data to the \citet{Chabrier2003} IMF.

Overall, the agreement between the simulations and comparison data improves with increasing model complexity. First, the \texttt{Basic} and \texttt{ThermalKinetic} models predict a $z=0$ sSFR -- stellar mass relation with an unrealistically flat shape, offset by more than $0.4$~dex toward lower values at low and intermediate stellar masses ($M_* \lesssim 10^{10.5}~\mathrm{M_\odot}$) compared to observed trends. The fraction of quenched galaxies at similar stellar masses is significantly overestimated. This discrepancy in both models arises from the use of a constant heating temperature for SN thermal feedback, $\Delta T_{\rm SN} = 10^{7.5}$ K, which results in excessively large energy injections from (clustered) SNe. By $z=0$, this overly powerful SN feedback in low- and intermediate-mass galaxies has likely disrupted, heated, and/or ejected most of the cold gas that would otherwise contribute to star formation, leading to lower sSFRs and higher quenched fractions at fixed $M_*$.

Switching to a density-dependent heating temperature in the \VardT{} and \colibrefixedagntemp{} models significantly improves agreement with the data. Both models reproduce the observed quenched fraction for galaxies with stellar masses $M_* \gtrsim 10^{9.5}~\mathrm{M_\odot}$ and the sSFRs of active galaxies for $M_* \gtrsim 10^{9}~\mathrm{M_\odot}$, with \colibrefixedagntemp{} exhibiting slightly better agreement for $M_* \sim 10^9~\mathrm{M_\odot}$. At lower stellar masses ($M_* \lesssim 10^{9}~\mathrm{M_\odot}$, corresponding to $\lesssim 100$ stellar particles), all four models show an increasing quenched fraction with decreasing stellar mass, which is likely driven by resolution effects. At m7 resolution, the ISM of such low-mass galaxies is sampled by only a few star-forming gas particles, if any, which makes them appear more quenched on average \citep[e.g.][]{2015MNRAS.446..521S}, resulting in lower median sSFRs and higher quenched fraction at $M_* \lesssim 10^{9}~\mathrm{M_\odot}$. In contrast, in the same mass range, the observed quenched fraction may be biased low due to selection effects, which are known to become progressively more important at lower $M_*$ \citep[e.g.][]{2025MNRAS.538..153K}.

\begin{figure}
        \centering
    \includegraphics[width=0.49\textwidth]{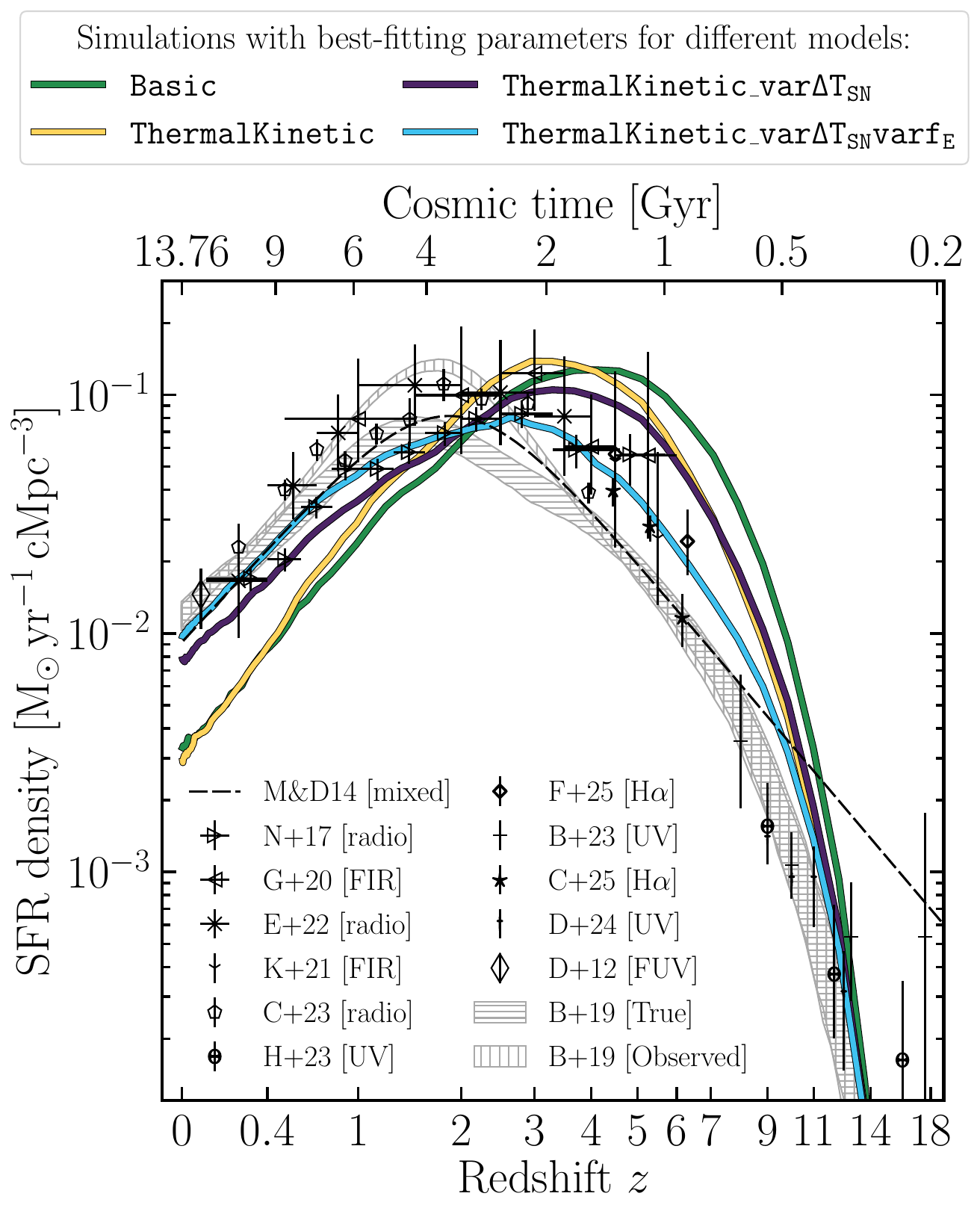}
    \caption{Cosmic star formation rate density (SFRD) versus redshift from the simulations with the best-fitting models to the $z=0$ \gsmf{} and \ssm:  \texttt{Basic}, \texttt{ThermalKinetic}, \VardT{}, and \colibrefixedagntemp{} (differently coloured solid curves). For comparison, the black points show a compilation of observational data from \citet{2012MNRAS.427.3244D,2017A&A...602A...5N,2020A&A...643A...8G,2021A&A...649A.152K,2022ApJ...927..204E,2023MNRAS.523.6082C,2023MNRAS.523.1009B,2023ApJS..265....5H,2024MNRAS.533.3222D,2025A&A...694A.178C,fu2025}, the grey hatched regions indicate the intrinsic (labelled as `true') and observed SRFD from the \textsc{UniverseMachine} \citep{2019MNRAS.488.3143B}, and the black dashed curve shows the best-fitting observed SFRD from \citet{2014ARA&A..52..415M}. The use of a variable heating temperature in the SN feedback of the \VardT{} and \colibrefixedagntemp{} models greatly improves the agreement with the comparison data at $z<1$. The inclusion of a stellar birth pressure-dependent SN energy in the \colibrefixedagntemp{} model results in a lower SFRD at $z>2$, thereby further improving the agreement with the comparison data.}
    \label{fig: four_models_SFH}
\end{figure}

Focusing on the right panel, we find that in all four models, galaxies with stellar masses $M_* \gtrsim 10^{10.5}~\mathrm{M_\odot}$ host SMBHs that have grown to masses $\gtrsim 10^7~\mathrm{M_\odot}$. The masses of these BHs follow a tight relation with the host galaxy's stellar mass, with a slope and normalization that closely match the observed scaling from \citet{2023MNRAS.518.2177G}. The agreement with the observational data is expected, as the AGN feedback coupling efficiency, $\varepsilon_{\rm f} = 0.1$ (see equation~\ref{eq: AGN_energy}), which determines the normalization of the \bhmsm{}, was chosen to produce realistic $z=0$ SMBH masses in the high-mass galaxies for which BH masses can be measured observationally. The value of $\varepsilon_{\rm f}$ was set \textit{independently} of the emulator-based calibration of SN and AGN feedback to the observed \gsmf{} and \ssm{}. Due to the self-regulating nature of SMBHs, variations in $\varepsilon_{\rm f}$ primarily influence SMBH masses, while having little to no effect on other galaxy properties such as the \gsmf{} and \ssm{} \citep[e.g.][]{2009MNRAS.398...53B}.

We do not observe any large differences between the models, which is expected since all models use the same numerical prescription for BH growth and AGN feedback. The only AGN-related parameter that differs between the models is the best-fitting value of the BH seed mass, $m_{\rm BH, seed}$. Among the four models, the \texttt{Basic} model uses the lowest seed mass, while the \colibrefixedagntemp{} model adopts the highest. A higher (lower) $m_{\rm BH, seed}$ results in somewhat faster (slower) BH growth in the latter (former) model, lasting until the BH enters the self-regulating regime where AGN feedback overtakes stellar feedback. As a consequence, at $M_* \lesssim 10^{10.5}~\mathrm{M_\odot}$, the median BH mass in the \colibrefixedagntemp{} model is higher than in the \texttt{Basic} model, whereas at $M_* \gtrsim 10^{10.5}~\mathrm{M_\odot}$ the two relations converge. The BH masses in the other two models, \texttt{ThermalKinetic} and \VardT{}, fall between those in \texttt{Basic} and \colibrefixedagntemp.

\subsubsection{Cosmic star formation rate density}
\label{subsection: cosmic_star_formation_rate_history}

\begin{figure*}
        \centering
    \includegraphics[width=0.99\textwidth]{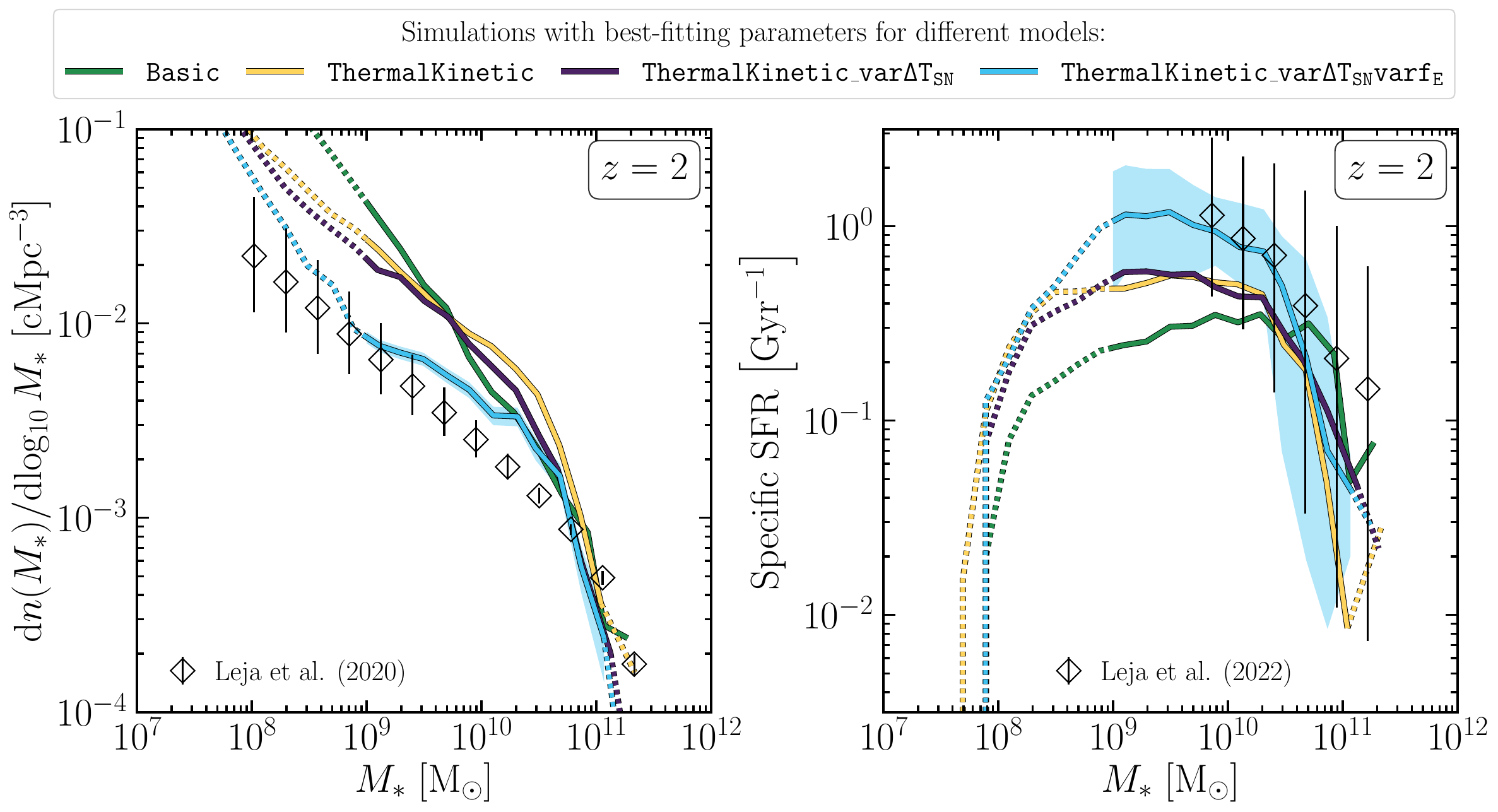}
    \caption{The $z=2$ galaxy stellar mass function (\gsmf; \textit{left}) and the $z=2$ median specific star formation rate of all galaxies (i.e. both star-forming and quenched; \textit{right}) as a function of stellar mass. The solid curves show results from simulations using the best-fitting parameters for the \texttt{Basic}, \texttt{ThermalKinetic}, \VardT{}, and \colibrefixedagntemp{} models. For comparison, we include the observed $z=2$ \gsmf{} from \citet{2020ApJ...893..111L} and the observed $z=2$ sSFR -- $M_*$ relation from \citet{2022ApJ...936..165L}. The shaded \colibrefixedagndtcolor{} region corresponds to the Poisson uncertainty in the left panel and the 16$^{\rm th}$-84$^{\rm th}$ percentile range in the right panel, both for the \colibrefixedagntemp{} model. Among the four models, \colibrefixedagntemp{} (in \colibrefixedagndtcolor) provides the best match to the observed $z=2$ \gsmf{} and sSFR  -- $M_*$ relation, although the simulated galaxies appear slightly overmassive -- by about $0.2$~dex -- relative to \citet{2020ApJ...893..111L}.}
    \label{fig: four_models_highz}
\end{figure*}

Fig.~\ref{fig: four_models_SFH} displays the redshift evolution of the cosmic star formation rate density (SFRD), using the same four models as in Fig.~\ref{fig: four_models_ssfr}. The SFRD is computed from the instantaneous SFRs of all star-forming gas particles in the simulations. For comparison, we include the $z \sim 0$ FUV-based SFRD estimate from \citet{2012MNRAS.427.3244D}, derived from the GAMA DR1, as well as radio-based SFRD estimates from the LOFAR Deep Fields at $0 < z < 4$ \citep{2023MNRAS.523.6082C}, and rest-frame FIR observations collected with ALMA at $z \approx 4.5-5.5$ \citep{2021A&A...649A.152K}. We also show results from the ALPINE multi-wavelength survey, based on 56 sub-mm continuum detections by ALMA in the ECDFS and COSMOS fields at $0.5 < z < 6$ \citep{2020A&A...643A...8G}, deep VLA COSMOS radio data at $0.3 < z < 5$ \citep{2017A&A...602A...5N}, and VLA radio measurements from a subsample of the GOODS-N survey at $0.1 < z < 3$ \citep{2022ApJ...927..204E}. In addition, we include the best-fitting analytic fit for the SFRD evolution from \citet[][black dashed curve]{2014ARA&A..52..415M}, derived from a compilation of IR and UV data across $0 < z < 8$, as well as predictions from the semi-empirical model \textsc{UniverseMachine} \citep{2019MNRAS.488.3143B}, showing both the ‘true’ and ‘observed’ SFRDs (grey hatched regions), where the latter accounts for systematic uncertainties in observationally inferred SFRs and the former is based on intrinsic SFR values predicted by \textsc{UniverseMachine}. Finally, we show a compilation of recent \textit{JWST} measurements at high redshifts ($4 < z < 18$) from \citet{2023MNRAS.523.1009B}, \citet{2023ApJS..265....5H}, \citet{2024MNRAS.533.3222D}, \citet{2025A&A...694A.178C}, and \citet{fu2025}, which are based on UV or H~$\upalpha$ observations and whose SFRD values are calculated down to an absolute UV magnitude limit of $\approx -17$~mag\footnote{Unlike the other four studies, \citet{2023MNRAS.523.1009B} used a limit of $-19$~mag. We re-normalize their results to $-17$~mag by shifting the SFRD values reported in \citet{2023MNRAS.523.1009B} upward by $0.5$~dex.}.

\begin{figure*}
        \centering
    \includegraphics[width=0.99\textwidth]{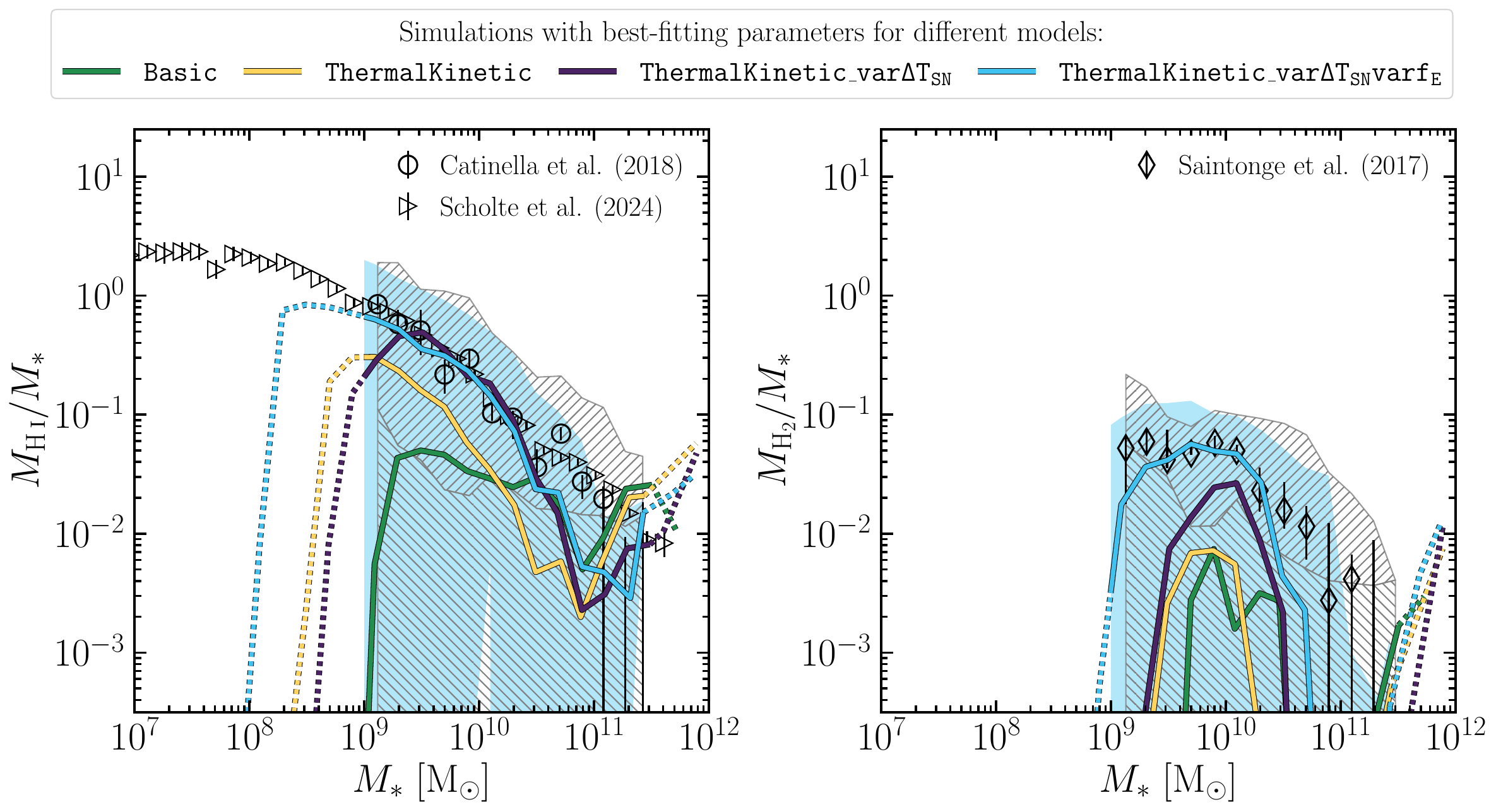}
    \caption{As in Fig.~\ref{fig: four_models_highz}, but showing the median $z=0$ H~\textsc{i}-to-stellar mass ratio (\textit{left}) and H$_2$-to-stellar mass ratio (\textit{right}) as functions of galaxy stellar mass. In both panels, the \colibrefixedagndtcolor{} shaded regions indicate the 16$^{\rm th}$ to 84$^{\rm th}$ percentile range for the \colibrefixedagntemp{} model. The shading extends to very low $y$ values because the lower percentile is influenced by objects with negligible cold gas masses. For comparison, we show the observed H~\textsc{i} mass fractions from \citet{2018MNRAS.476..875C} and \citet{2024MNRAS.535.2341S}, as well as H$_2$ measurements from \citet{2017ApJS..233...22S} (black symbols). We also include the scatter for the measurements from \citet{2018MNRAS.476..875C} and \citet{2017ApJS..233...22S}, shown as grey hatched regions: the upper hatched region, slanted toward the top right, represents the 16$^{\rm th}$–84$^{\rm th}$ percentile scatter when non-detections are treated as upper limits; the lower hatched region, slanted toward the bottom left, extends the 16$^{\rm th}$ percentile of the upper region to zero, corresponding to the case where non-detections are treated as zeroes. In both simulations and observations, molecular gas masses do not include a contribution from helium. Only the \colibrefixedagntemp{} model reproduces both the H~\textsc{i} and H$_2$ observed mass fractions for $M_* < 10^{10.5}~\mathrm{M_\odot}$ (including the observed scatter), while at $M_* > 10^{10.5}~\mathrm{M_\odot}$ all models undershoot the data.}
    \label{fig: four_models_cold}
\end{figure*}

The \texttt{Basic} and \texttt{ThermalKinetic} models predict significantly lower SFRD normalization than observed for $z<1$, underpredicting the observed SFRD by $\approx 0.5$~dex, which is consistent with both models having too low sSFR at $z=0$ in Fig.~\ref{fig: four_models_ssfr}. As is the case with the sSFR, the suppression in the SFRD at low redshifts is related to the high (constant) heating temperature used in SN thermal feedback: $\Delta T_{\rm SN}=10^{7.5}~\mathrm{K}$. Conversely, the SFRD in the other two models, \VardT{} and \colibrefixedagntemp, which incorporate the variable heating temperature, matches the $z\lesssim 1$ observed SFRD much better, with \colibrefixedagntemp{} predicting a $z=0$ SFRD of $\approx 10^{-2}~\mathrm{M_\odot \, yr^{-1} \, cMpc^{-3}}$ and showing the best agreement with the data.

At high redshifts ($z\gtrsim 1$), the SFRD in the \texttt{Basic}, \texttt{ThermalKinetic}, and \VardT{} models rises steeply with increasing redshift, peaking at $3<z<5$. The fact that these models predict a peak SFRD at higher redshifts than observed ($z\approx 2$) suggests that high$-z$ star formation in these models may be overly efficient. Since all four models were calibrated to match the observed \gsmf{} (and \ssm) at $z=0$, an excess in the early SFRD necessitates a suppressed SFRD at later times to ensure that the correct total stellar mass is formed by $z=0$.

The agreement with the data is noticeably improved in the \colibrefixedagntemp{} model, where the amount of stellar mass formed before $z=3$ is significantly reduced compared to the other models. The SFRD in the \colibrefixedagntemp{} model exhibits a broad peak between $z=4$ and $1$ and starts steeply declining with cosmic time only thereafter. This improvement is caused by the dependence of the SN energy on the stellar birth pressure (see equation~\ref{eq: stellar_birth_pressure_vs_SN_energy}), which is incorporated only into the \colibrefixedagntemp{} model. Specifically, the SN feedback in the \colibrefixedagntemp{} is more energetic at higher redshifts, as the star formation at high $z$ proceeds on average in higher gas-pressure environments (see Appendix \ref{appendix: redshift_evolution_of_sn_energy} for further details). Releasing more SN energy at high $z$ not only reduces the cosmic SFRD but also helps avoid runaway star formation in the centres of massive galaxies, which may lead to the formation of a pronounced stellar bulge component. The presence of a dominant bulge can be traced by the dip in the $z=0$ \ssm{} at $M_* \approx 2\times 10^{10}~\mathrm{M_\odot}$, which is absent only in the \colibrefixedagntemp{} model (see Figs. \ref{fig:basic_and_thermal_kinetic_model_calibration} and \ref{fig: models_three_and_four}).

\subsubsection{Galaxy stellar mass function and star formation rates at high redshift}

Fig.~\ref{fig: four_models_highz} shows the \gsmf{} and sSFR versus stellar mass at $z=2$ for the same four simulations as were shown in Figs. \ref{fig: four_models_ssfr} and \ref{fig: four_models_SFH}. At a given stellar mass, we show the median sSFR considering both passive and active galaxies. For comparison, we display the $z=2$ \gsmf{} and median sSFR derived by \citet{2020ApJ...893..111L,2022ApJ...936..165L} who applied the spectral energy distribution (SED) fitting code \textsc{Prospector} to measure SFRs and stellar masses of $\sim 10^5$ galaxies at $0.2<z<3$ from COSMOS2015 and 3D-\textit{HST} galaxy catalogues. The error bars in \citet{2020ApJ...893..111L,2022ApJ...936..165L} correspond to the 16$^{\rm th}$ and 84$^{\rm th}$ percentiles estimated by their best-fitting model.

At $z=2$, the \colibrefixedagntemp{} model shows significantly better agreement with the data than its three simpler counterparts. Despite being calibrated only to $z=0$ data, the model broadly reproduces the observed median sSFR, with discrepancies remaining $\approx 0.1$~dex at $M_* < 10^{10.5}~\mathrm{M_\odot}$. Additionally, it is systematically offset by just $\approx 0.2$~dex from the observed \gsmf, which is comparable to the systematic uncertainty in inferring $M_*$ using SED-fitting codes at these redshifts \citep[e.g.][]{2020MNRAS.492.5592K}. In contrast, the other three models perform less well: all systematically underpredict the observed sSFR by more than $\approx 0.3$~dex at $M_* < 10^{10.5}~\mathrm{M_\odot}$, and their \gsmf{} values exceed the observed data by at least $\approx 0.5$~dex at $M_* < 10^{10}~\mathrm{M_\odot}$.

\subsubsection{Cold gas properties}

We next investigate the $z=0$ properties of cold gas predicted by the simulations. The left and right panels of Fig.~\ref{fig: four_models_cold} show, respectively, the $z=0$ ratios of galaxy H~\textsc{i} mass to stellar mass and H$_{2}$ mass to stellar mass as functions of galaxy stellar mass. The solid curves correspond to the median mass fractions in the simulations, and the \colibrefixedagndtcolor{} shaded regions show the 16$^{\rm th}$ to 84$^{\rm th}$ percentile range for the \colibrefixedagntemp{} model. 

For reference, we show the $z \approx 0$ observed H~\textsc{i}-to-stellar mass fractions from \citet[][xGASS survey]{2018MNRAS.476..875C} and \citet[][ALFALFA survey]{2024MNRAS.535.2341S}, both derived from $21~\mathrm{cm}$ line observations. For H$_2$, we show mass fractions from xCOLD GASS \citep{2017ApJS..233...22S}, where H$_2$ masses were derived from CO(1–0) luminosity using the multi-variate $\alpha_{\rm CO}$ conversion factor from \citet{2017MNRAS.470.4750A}. To ensure fair comparison with simulations, we divided H$_2$ masses from \citet{2017ApJS..233...22S} by $1.36$ to remove a contribution from helium, which was included in the conversion factor $\alpha_{\rm CO}$ used by the authors. In both panels, black symbols indicate median observed gas fractions. Error bars for \citet{2018MNRAS.476..875C} and \citet{2017ApJS..233...22S} represent errors on the median, which we estimated via bootstrapping. Grey hatched regions show the 16$^{\rm th}$–84$^{\rm th}$ percentile scatter in the observations: in the left panel for \citet{2018MNRAS.476..875C}, and in the right panel for \citet{2017ApJS..233...22S}. In each panel, the upper hatched region (slanted top-right) shows the scatter when non-detections are treated as upper limits. The lower hatched region (slanted bottom-left) assumes non-detections are zeroes, resulting in a 16$^{\rm th}$ percentile extending to zero. Together, the lower percentiles of these two regions bracket the range for the true lower percentile.

We stress that the \colibre{} model does not impose an effective pressure and/or temperature floor and uses the non-equilibrium thermochemistry solver \textsc{chimes}, coupled to the \colibre{} dust grain model, to compute the abundances of primordial species. This allows us to take the (non-equilibrium) H$_{2}$ and H~\textsc{i} abundances directly from the simulations, which is in contrast to the previous generation of galaxy simulations of representative volumes, such as \textsc{eagle} and \textsc{IllustrisTNG}, where the atomic and molecular gas fractions would need to be estimated in post-processing \citep[see e.g.][]{2015MNRAS.452.3815L,2018ApJS..238...33D,2023MNRAS.523.2738M}. Such post-processing typically relies on (semi-)analytic models or fitting formulas calibrated using high-resolution radiative transfer simulations of smaller cosmological volumes \citep[e.g.][]{2011ApJ...728...88G,2013MNRAS.430.2427R,2013MNRAS.436.2747K}, and may include an intermediate step to obtain the neutral (H$_{2}$ + H~\textsc{i}) hydrogen fraction before the H$_{2}$ and H~\textsc{i} fractions can be estimated \citep[e.g.][]{2017MNRAS.464.4204C}.

Fig.~\ref{fig: four_models_cold} shows that while both the \VardT{} and \colibrefixedagntemp{} models are consistent with the observational data for H~\textsc{i} in the stellar mass range $10^9 < M_*/\mathrm{M_\odot} < 10^{10.5}$, only \colibrefixedagntemp{} also reproduces the observed H$_2$ mass fractions within this range, whereas \VardT{} systematically undershoots the H$_2$ data at all $M_*$. Moreover, comparing the \colibrefixedagndtcolor{} shaded regions with the grey hatched regions shows that \colibrefixedagntemp{} also reproduces the observed scatter for both H~\textsc{i} and H$_2$. By contrast, for the \texttt{Basic} and \texttt{ThermalKinetic} models, both the molecular and atomic gas fractions are, on average, too low compared to the data, consistent with these models underpredicting the observed sSFR at $z=0$ (Fig.~\ref{fig: four_models_ssfr}). The sharp downturn in the simulated molecular and atomic gas fractions in all four models below the stellar mass $M_* \sim 10^{9}~\mathrm{M_\odot}$ is driven by limited numerical resolution, with H$_2$ mass fractions starting to drop at slightly higher $M_*$ than H~\textsc{i}\footnote{Convergence tests show that higher resolution is required to robustly predict H$_2$ than to predict H~\textsc{i}.}.

At the high-mass end ($M_* > 10^{10.5}~\mathrm{M_\odot}$), all four models underpredict the observed gas fractions. In fact, there appears to be a tension in that the \colibrefixedagntemp{} model reproduces the quenched fraction and sSFRs in this mass range while undershooting the gas fractions. This discrepancy may highlight limitations in our relatively simple treatment of gas accretion onto SMBHs and/or AGN feedback, exacerbated by the relatively low resolution of the simulations ($m_{\rm gas}\sim 10^7~\mathrm{M_\odot}$), leading to overly efficient depletion of cold gas in massive galaxies. Indeed, \citet{schaye2025colibreproject}  demonstrate that agreement with the data at the high-mass end improves significantly when using the higher m6 resolution of \colibre{} (see their fig. 19). Finally, we note that the dip in the H~\textsc{i} mass fraction at $M_* \sim 5 \times 10^{10}~\mathrm{M_\odot}$ may also be mitigated by mock observing the simulated galaxies, rather than using intrinsic values predicted by the simulations. This would account for observational effects such as H~\textsc{i} emission blending, which can artificially boost the inferred neutral gas fractions at the high-mass end \citep[e.g.][]{2019MNRAS.483.5334S}.

\subsubsection{Metal content}

\begin{figure*}
        \centering
    \includegraphics[width=0.99\textwidth]{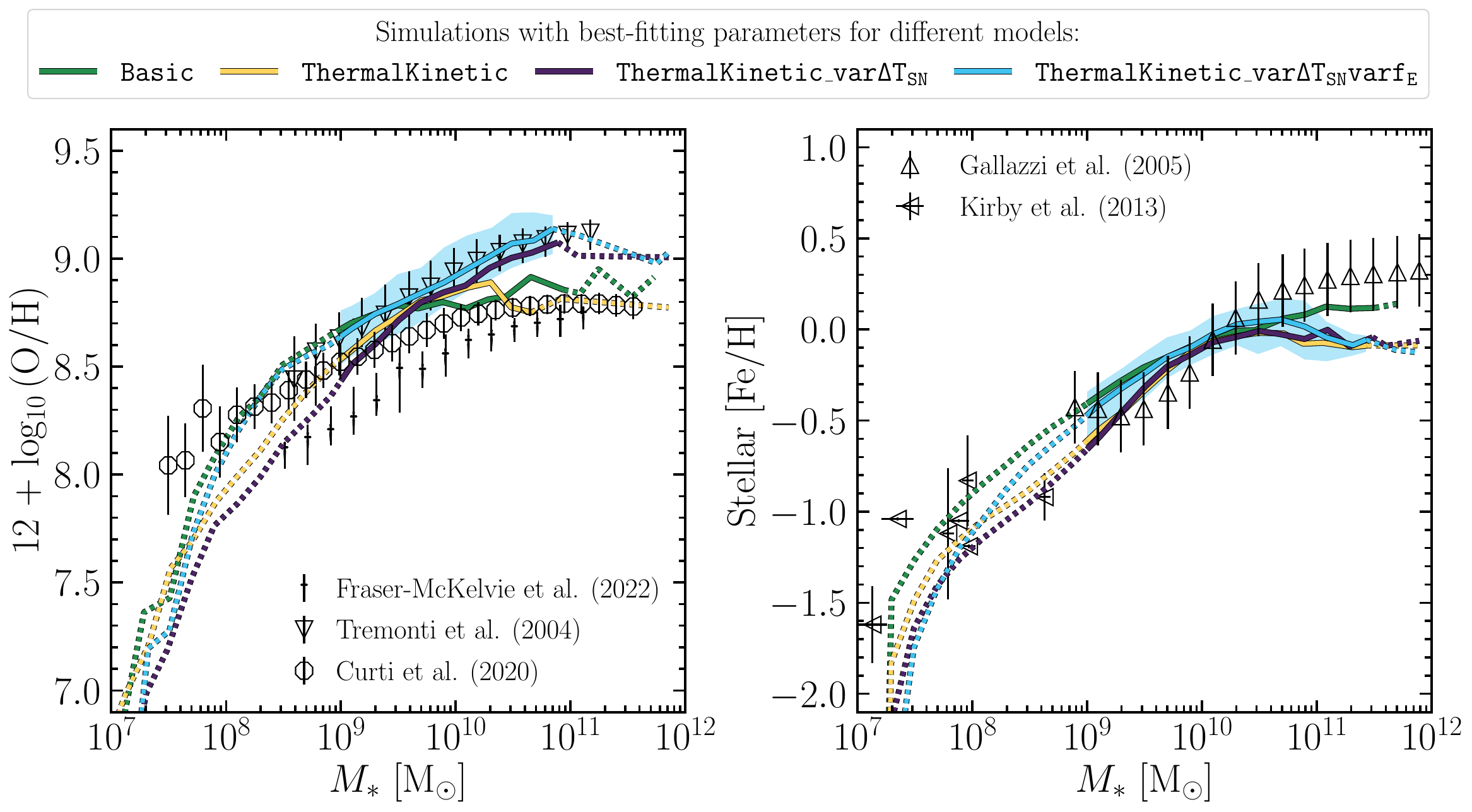}
    \caption{As Fig.~\ref{fig: four_models_highz}, but showing the median gas-phase metallicity (\textit{left}) and stellar metallicity (\textit{right}) versus galaxy stellar mass at $z=0$. The $z\approx 0$ comparison data are taken from \citet{2022MNRAS.510..320F}, \citet{ 2004ApJ...613..898T}, and \citet{2020MNRAS.491..944C} for the gas-phase metallicity, and from \citet{2005MNRAS.362...41G} and \citet{2013ApJ...779..102K} for the stellar metallicity. The gas-phase metallicity is computed in the gas that is sufficiently dense ($n_{\rm H} > 0.1~\mathrm{cm}^{-3}$) and cool ($T < 10^{4.5}~\mathrm{K}$) and only for star-forming galaxies ($\mathrm{sSFR} > 10^{-2}$~Gyr$^{-1}$), excluding metals that are present in dust. All four models are consistent with the observations for both the gas-phase and stellar metallicities at $M_* \lesssim 10^{11}~\mathrm{M_\odot}$.}
    \label{fig: four_models_met}
\end{figure*}

The left panel of Fig.~\ref{fig: four_models_met} shows the relationship between the metallicity of the gas phase, in units of $12 + \log_{10}(\rm O/H)$, and galaxy stellar mass at $z=0$. The gas metallicity in the simulations is derived directly from the oxygen abundance, as predicted by the chemistry model of \colibre. Each galaxy's ratio between the number of oxygen and hydrogen nuclei, $\rm O/H$, is calculated as
\begin{equation}
    \mathrm{O/H} = \frac{m_{\rm H}}{m_{\rm O}}\frac{\sum_i (X_{\rm O}/X_{\rm H})_{i} \, m_{{\rm gas}, i}}{\sum_i m_{\mathrm{gas},i}}\, ,
    \label{eq: O_over_H}
\end{equation}
where $(X_{\rm O}/X_{\rm H})_{i}$ is the ratio of the oxygen and hydrogen mass fractions carried by gas particle $i$, $m_{\mathrm{gas},i}$ is the mass of gas particle $i$, and $m_{\rm H}/m_{\rm O}$ is the ratio of the masses of hydrogen and oxygen nuclei. To aid the comparison with observational data, in equation (\ref{eq: O_over_H}) we consider only those gas particles that are dense ($n_{\rm H} > 0.1~\mathrm{cm}^{-3}$) and cool ($T < 10^{4.5}~\mathrm{K}$). Furthermore, we apply a spatial mask by requiring the selected particles to be within 50~pkpc apertures centred on the galaxies. We do not include metals that are present in dust. In a given stellar mass bin, we show the median value of $12 + \log_{10}(\rm O/H)$ considering only star-forming galaxies ($\mathrm{sSFR} > 10^{-2}~\mathrm{Gyr}^{-1}$).

For comparison, we display the gas-phase metallicities of 472 star-forming galaxies at $0.04 < z < 0.128$ from the SAMI Galaxy Survey \citep{2022MNRAS.510..320F}, as well as metallicities for $\sim 10^5$ local star-forming SDSS galaxies from \citet{2004ApJ...613..898T} and \citet{2020MNRAS.491..944C}. The data from \citet{2004ApJ...613..898T} and \citet{2022MNRAS.510..320F} follow two distinct metallicity tracks, offset by $\approx 0.3$~dex, while the measurements from \citet{2020MNRAS.491..944C} gradually transition from the upper to the lower track with increasing stellar mass. The $\approx 0.3$~dex systematic discrepancy between different observations arises from differences in the calibration of the methods used to infer gas-phase metallicities from galaxy spectra, as well as from differences in the methods themselves. In particular, the lower track typically corresponds to metallicities estimated using the so-called $T_{\rm e}$-method, while the upper track is based on calibrations using photoionization models \citep[see e.g.][]{2012MNRAS.426.2630L,2020MNRAS.491..944C}.

We find that all four models are consistent with the observational data, including the normalization, slope, and scatter of the mass–metallicity relation. At stellar masses $M_* > 10^{9.5}~\mathrm{M_\odot}$, \VardT{} and \colibrefixedagntemp{} closely follow the upper track of the observations, whereas the metallicities in \texttt{Basic} and \texttt{ThermalKinetic} saturate at values corresponding to the lower track.

The right panel of Fig.~\ref{fig: four_models_met} shows the relationship between galaxy stellar mass and stellar metallicity, [Fe/H], at $z=0$. The stellar metallicity in the simulations is derived directly from the galaxies' iron abundance. For each galaxy, we first calculate the ratio of the total numbers of iron and hydrogen nuclei, $\rm Fe/H$, where we employ the same expression as for the gas-phase O/H (equation~\ref{eq: O_over_H}) but apply it to stellar particle-carried fields and replace oxygen with iron. Stellar particles that contribute to Fe/H are selected within 50 pkpc apertures. The resulting ratio is subsequently normalized by the solar value of Fe/H, assuming a solar iron abundance of $12 + \log_{10} (\mathrm{Fe/H}) = 7.5$ \citep{doi:10.1146/annurev.astro.46.060407.145222}.  In a given stellar mass bin, we show the median value of [Fe/H].

For reference, we display the observed stellar metallicity-mass relation for a large ($\sim 10^5$) sample of $z\approx 0.1$ SDSS galaxies from \citet{2005MNRAS.362...41G} and dwarf irregular and spheroidal satellite galaxies of the Milky Way and M31 from \citet{2013ApJ...779..102K}. Where needed, the solar abundances used in the observations have been converted to the solar values reported by \citet{doi:10.1146/annurev.astro.46.060407.145222}.

Overall, all four models are consistent with the observations within the stellar mass range of $\sim 10^7$ to $10^{11}~\mathrm{M_\odot}$. At higher masses ($M_* \gtrsim 10^{11}~\mathrm{M_\odot}$), the \texttt{ThermalKinetic}, \VardT{}, and \colibrefixedagntemp{} models undershoot the \citet{2005MNRAS.362...41G} data by up to $\approx 0.35$~dex. This discrepancy is likely related to the fact that stellar metallicities in the simulations are derived from the Fe abundance, whereas the metallicities in \citet{2005MNRAS.362...41G} are estimated based on a combination of Mg and Fe absorption features in galaxy spectra, with the importance of Mg becoming higher in more massive galaxies due to $\alpha-$enhancement \citep[e.g.][]{2016MNRAS.461L.102S}. \citet{schaye2025colibreproject} show that the agreement between \colibre{} and the \citet{2005MNRAS.362...41G} data at the high-mass end improves if stellar metallicities in the simulations are derived from Mg instead of Fe, although a small discrepancy remains\footnote{Another potential source of discrepancy with the \citet{2005MNRAS.362...41G} data concerns our fiducial choice of aperture size, within which all galaxy properties are calculated: $50~\mathrm{pkpc}$. In contrast, \citet{2005MNRAS.362...41G} derived stellar metallicities from SDSS spectra taken with $3^{\prime\prime}$-diameter fibres, corresponding to a physical radius of $\approx 3~\mathrm{kpc}$ at $z=0.1$. We verified that reducing the aperture from $50$ to $3~\mathrm{kpc}$ increases the stellar metallicity at the high-mass end by no more than $\approx 0.1$~dex, which is insufficient to account for the full offset with respect to the \citet{2005MNRAS.362...41G} measurements ($\approx 0.35$~dex).}.

The stellar mass -- metallicity relation in the \texttt{Basic} model saturates at a metallicity that is $\approx 0.25$~dex higher than in the other three models. This results from the lower BH seed mass adopted in the \texttt{Basic} model, which weakens AGN feedback and consequently allows for more late-time star formation in massive galaxies, thereby increasing their present-day stellar metallicities.

\subsection{The COLIBRE fiducial model at m7, m6, and m5 resolutions}
\label{subsection: three resolutions}

In the previous sections, we showed that among the four models with the best-fitting parameter values identified by the emulators, \colibrefixedagntemp{} not only provides the closest match to the observed \gsmf{} and \ssm, but is also in overall better agreement with the observational data that were not used in the calibration. If the \colibre{} model were designed solely for the resolution at which the emulators were employed (m7; $m_{\rm gas} \approx m_{\rm dm} \sim 10^7~\mathrm{M_\odot}$), we could conclude our search for the best-fitting model here. However, since the \colibre{} suite also includes simulations at m6 ($m_{\rm gas} \approx m_{\rm dm} \sim 10^6~\mathrm{M_\odot}$) and m5 ($m_{\rm gas} \approx m_{\rm dm} \sim 10^5~\mathrm{M_\odot}$) resolutions, it is essential to verify that the \colibrefixedagntemp{} model performs well at these resolutions too before establishing it as the fiducial \colibre{} model.

\begin{figure*}
        \centering
    \includegraphics[width=0.99\textwidth]{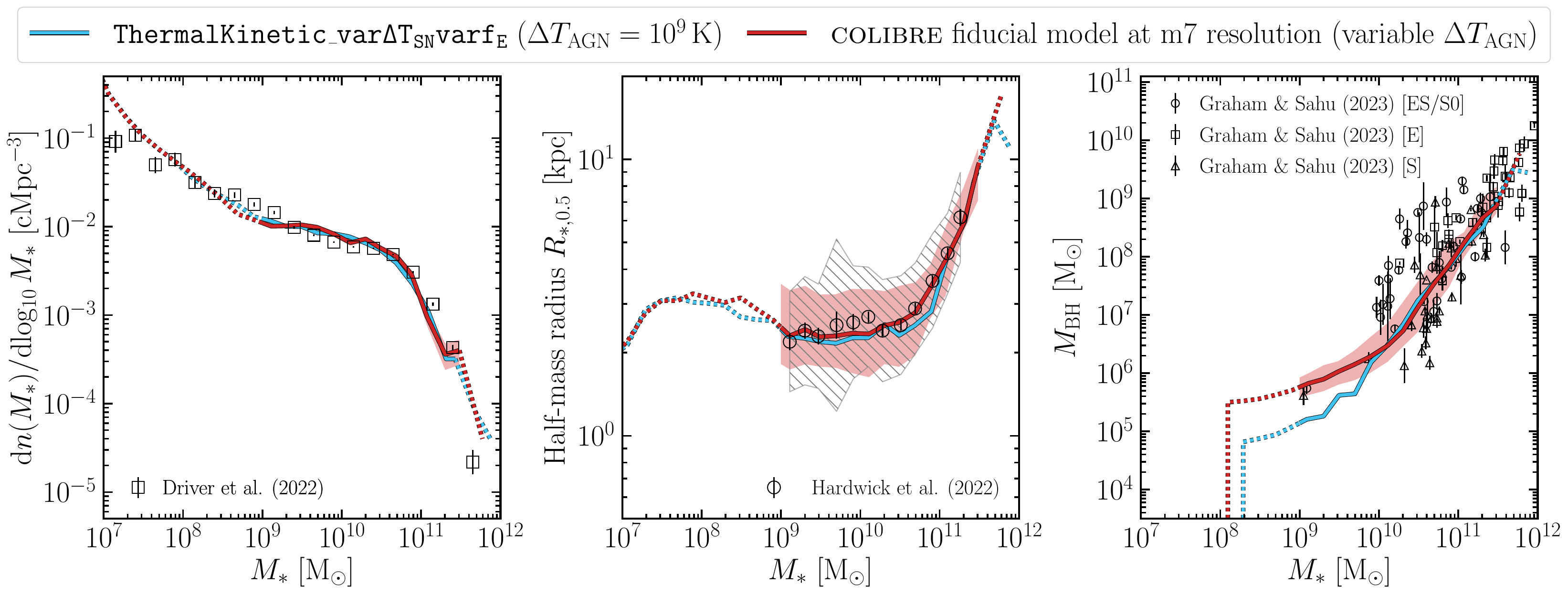}
    \caption{The galaxy stellar mass function (\gsmf; \textit{left}), the median size -- stellar mass relation (\ssm; \textit{middle}), and the median black hole mass -- stellar mass relation (\bhmsm; \textit{right}), all shown at $z=0$. The \colibrefixedagndtcolor{} and \colibrefiducialcolor{} solid curves are simulation predictions with the best-fitting \colibrefixedagntemp{} model (using $\Delta T_{\rm AGN} = 10^{9} \, \rm K$) and its modified version with the variable $\Delta T_{\rm AGN}$ (i.e. the fiducial \colibre{} model), respectively. Both simulations were run in a (50 cMpc)$^3$ volume at m7 resolution. The black symbols and grey hatched region indicate observational data. For reference, the uncertainty in the predictions of the \colibre{} fiducial model is indicated by the \colibrefiducialcolor{} shaded region, with boundaries corresponding to the Poisson uncertainty for the \gsmf{} and the 16$^{\rm th}$ to 84$^{\rm th}$ percentile range for the \ssm{} and \bhmsm{}. While the \gsmf{} and \ssm{} predicted by both models are nearly identical, each reproducing the observational data, the fiducial \colibre{} model predicts significantly higher BH masses in galaxies with $M_*\lesssim 10^{10}~\mathrm{M_\odot}$.}
    \label{fig: model_four_vs_final}
\end{figure*}

\subsubsection{Extension of the $\mathit{ThermalKinetic\_var\Delta T_{\mathit{SN}}varf_{\mathit{E}}}$ model with best-fitting parameter values to higher resolutions}
\label{subsection: extension_to_higher_resolutions}

By applying the best-fitting \colibrefixedagntemp{} model calibrated at m7 resolution to m6 and m5 resolutions, without adjusting any parameter values except for gravitational softening (which scales with $m_{\rm gas}$ as $\varepsilon_{\rm soft} \propto m_{\rm gas}^{1/3}$), we found that the fit to the observed \gsmf{} and \ssm{} at $z=0$ becomes progressively worse at higher resolutions. The primary factor reducing the accuracy of the fit is the increasingly earlier onset of AGN feedback. This result could be anticipated, as the tail of the gas density distribution in higher-resolution simulations extends to higher values, which can significantly boost BH accretion rates (see equation~\ref{eq: SMBH_acc_rate}). As a consequence, BH particles undergo faster growth and produce more aggressive AGN feedback, which ultimately results in galaxies with unrealistically low stellar masses and large sizes.

By manually adjusting the BH seed mass to suppress the overly efficient early BH growth and obtain a good fit to the observed \gsmf{} and \ssm, we found that at m6 resolution the BH seed mass needs to be reduced by a factor of $\sim 10$ compared to its best-fitting value at m7 resolution, and by another factor of $\sim 10$ when increasing the resolution from m6 to m5, giving $m_{\rm BH, seed} \sim 10^{3}~\mathrm{M_\odot}$ at m5 resolution. Because the BH accretion rate has a superlinear dependence on $m_{\rm BH}$ ($\dot{m}_{\rm BH} \propto m_{\rm BH}^2$), such a low value of $m_{\rm BH, seed}$ at m5 resolution yields two distinct populations of galaxies based on their stellar mass: (i) galaxies with $M_* \gtrsim 10^{10.5}~\mathrm{M_\odot}$ that have efficient BH growth, and (ii) galaxies with $M_* \lesssim 10^{10.5}~\mathrm{M_\odot}$ where BH accretion rates are very low. The transition from nearly no BH growth to efficient BH growth becomes a step-like function of $M_*$, which is undesirable. Fortunately, we found that a much smoother transition between the two regimes of BH growth, including an intermediate population of galaxies with moderately growing BHs, can be realised by adopting a much higher value of $m_{\rm BH, seed}$ alongside a \textit{variable} AGN heating temperature, which depends linearly on the BH mass, instead of the constant value $\Delta T_{\rm AGN} = 10^{9}~\mathrm{K}$. 

To ensure that the \colibre{} model remains consistent across all resolutions, we decided to implement the change from $\Delta T_{\rm AGN} = 10^{9}~\mathrm{K}$ to a variable $\Delta T_{\rm AGN}$ not only at m5 resolution but also at m6 and m7. In the following text, we detail how the change in $\Delta T_{\rm AGN}$ is implemented and how it affects the calibration at m7 resolution described in the previous sections. 

\subsubsection{The COLIBRE model with a variable $\Delta T_{\rm AGN}$}
\label{subsection: colibre_with_var_agn_heating_temperature}

\begin{figure*}
    \centering
     \includegraphics[width=0.99\textwidth]{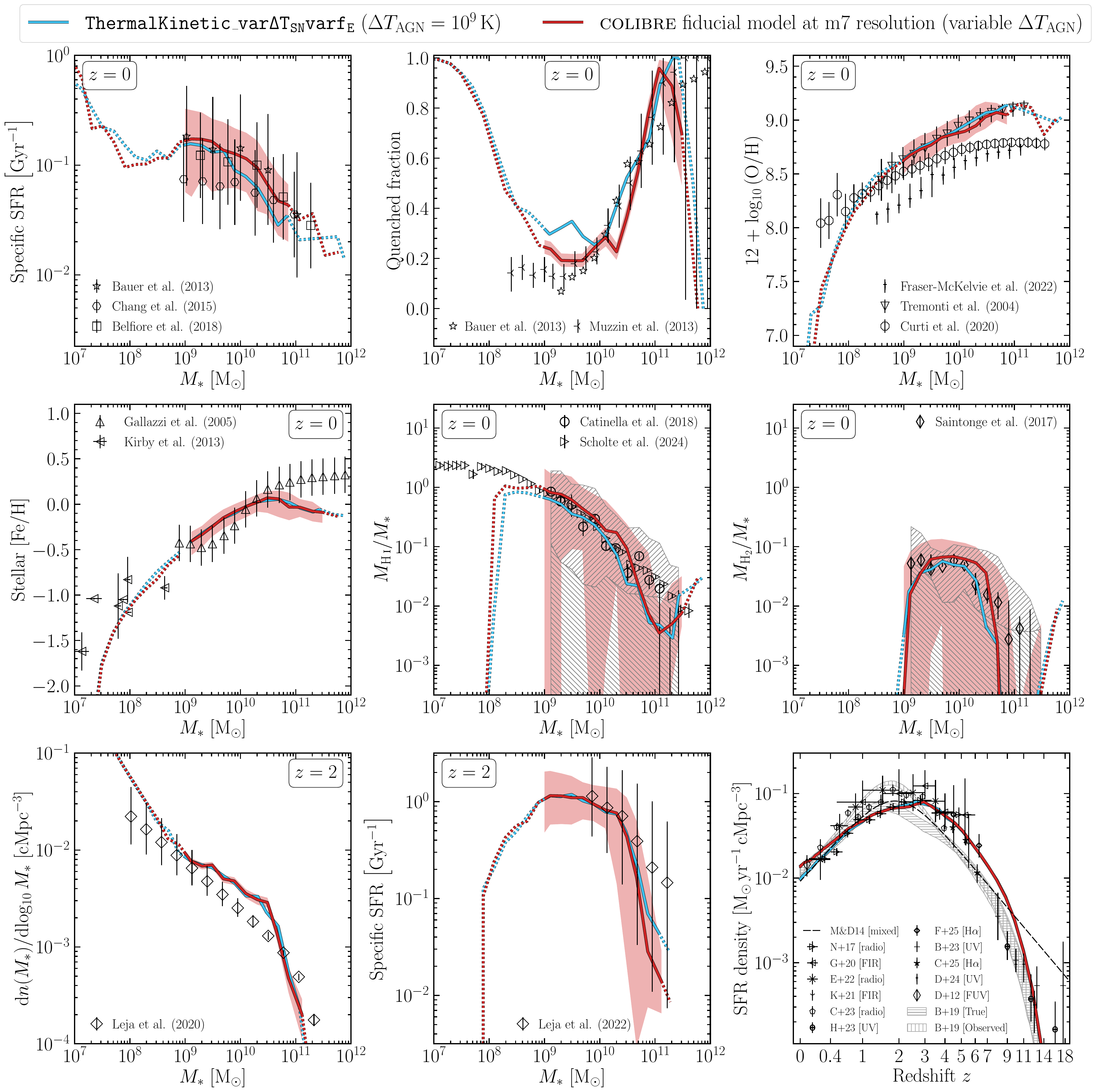}
    \caption{Comparison of the simulation predictions from the best-fitting \colibrefixedagntemp{} model (with $\Delta T_{\rm AGN} = 10^{9}~\mathrm{K}$; \colibrefixedagndtcolor) and its modified version with the variable $\Delta T_{\rm AGN}$ (i.e. the fiducial \colibre{} model; \colibrefiducialcolor) for various relations that were not used to calibrate the models. Both simulations were run in a (50 cMpc)$^3$ volume at m7 resolution. The panels, from left to right, top to bottom, display: the median sSFR versus stellar mass for active galaxies ($\textrm{sSFR} > 10^{-2}~\mathrm{Gyr^{-1}}$) at $z=0$, quenched fraction versus stellar mass at $z=0$, the gas-phase metallicity versus stellar mass for star-forming galaxies at $z=0$, the stellar metallicity versus stellar mass at $z=0$, the median H~\textsc{i} mass-to-stellar mass fraction versus stellar mass at $z=0$, the median H$_2$ mass-to-stellar mass fraction versus stellar mass at $z=0$, the \gsmf{} at $z=2$, the median sSFR versus stellar mass for all galaxies at $z=2$, and the cosmic star formation rate density versus redshift. Where present, the \colibrefiducialcolor{} shaded region indicates the 16$^{\rm th}$ to 84$^{\rm th}$ percentiles in the \colibre{} fiducial model, except for the quenched fraction, where it represents the $1\sigma$ confidence interval. The \colibrefixedagntemp{} and \colibre{} fiducial models perform similarly well in reproducing the observational data.}
    \label{fig: model_four_vs_final_2}
\end{figure*}

We take the \colibrefixedagntemp{} model and modify its prescription for AGN feedback by replacing a fixed $\Delta T_{\rm AGN} = 10^9~\mathrm{K}$ with a function that depends on the BH (subgrid) mass,
\begin{equation}
    \Delta T_{\rm AGN}(m_{\rm BH}) = \Delta T_{\rm AGN, pivot} \, \left( \frac{m_{\rm BH}}{m_{\rm BH,pivot}} \right) \, ,
\label{eq: variable_agn_temperature}
\end{equation}
where $\Delta T_{\rm AGN, pivot}$ and $m_{\rm BH,pivot}$ are (degenerate) parameters. The variable heating temperature is constrained within the bounds $\Delta T_{\rm AGN, min} < \Delta T_{\rm AGN}(m_{\rm BH}) < \Delta T_{\rm AGN, max}$, which at m7 resolution we set to $\Delta T_{\rm AGN, min} = 10^{6.5}~\mathrm{K}$ and $\Delta T_{\rm AGN, max} = 10^{9.5}~\mathrm{K}$. These limits ensure that AGN feedback remains both efficient and well-sampled across a wide range of halo masses, following the same rationale used to set the variable heating temperature in SN feedback (see $\S$\ref{subsubsection: dens_dep_dT}). The two remaining parameters, which are degenerate ($\Delta T_{\rm AGN, pivot} \propto m_{\rm BH, pivot}$), are set to $\Delta T_{\rm AGN, pivot} = 10^{9}~\mathrm{K}$ and $m_{\rm BH, pivot} = 10^{8}~\mathrm{M_\odot}$, such that a heating temperature of $10^{9}~\mathrm{K}$ is reached for a BH with $m_{\rm BH} = 10^{8}~\mathrm{M_\odot}$, the value that is typical for $M_* \sim 10^{11}~\mathrm{M_\odot}$ galaxies (see Fig.~\ref{fig: four_models_ssfr}). For BHs with masses below $10^{8}~\mathrm{M_\odot}$, the heating temperature is less than $10^{9}~\mathrm{K}$, leading to a better sampling of AGN feedback events compared to the case where $\Delta T_{\rm AGN}$ is fixed to $10^{9}~\mathrm{K}$. More generally, since the BH accretion rate scales as $\dot{m}_{\rm BH} \propto m_{\rm BH}^2$, while the accretion rate at the Eddington limit scales linearly with $m_{\rm BH}$, equation (\ref{eq: variable_agn_temperature}) ensures that the sampling of AGN feedback events is independent of the BH mass for BHs accreting at a fixed Eddington fraction.

Since the change from $\Delta T_{\rm AGN} = 10^9~\mathrm{K}$ to a variable $\Delta T_{\rm AGN}$ will impact the strength of AGN feedback, some of the best-fitting parameter values found for the \colibrefixedagntemp{} model with $\Delta T_{\rm AGN} = 10^9~\mathrm{K}$ may require adjustments to achieve similar goodness of fit to the observed \gsmf{} and \ssm. As an initial test, we ran several simulations in a (50 cMpc)$^3$ volume using the variation of the \colibrefixedagntemp{} model with the variable $\Delta T_{\rm AGN}$ given by equation (\ref{eq: variable_agn_temperature}), keeping the three SN parameters at their best-fitting values from Table \ref{table: best-fitting models} found for $\Delta T_{\rm AGN} = 10^9~\mathrm{K}$ while exploring different values of $m_{\rm BH, seed}$. These tests showed that the SN parameters can indeed remain unchanged, while $m_{\rm BH, seed}$ needs to be increased. Specifically, we found that if $m_{\rm BH, seed}$ is increased from its best-fitting value for $\Delta T_{\rm AGN}=10^{9}~\mathrm{K}$ of $10^{4.8}~\mathrm{M_\odot}$ to $10^{5.5}~\mathrm{M_\odot}$, the \colibrefixedagntemp{} model with the variable $\Delta T_{\rm AGN}$ matches the observed \gsmf{} and \ssm{} with a similar accuracy as the best-fitting \colibrefixedagntemp{} model with $\Delta T_{\rm AGN}=10^{9}~\mathrm{K}$. 

Fig.~\ref{fig: model_four_vs_final} shows the $z=0$ \gsmf{} (left panel), the $z=0$ \ssm{} (middle panel) and the $z=0$ \bhmsm{} (right panel) for the best-fitting \colibrefixedagntemp{} model with $\Delta T_{\rm AGN}=10^9~\mathrm{K}$ (\colibrefixedagndtcolor) and its modified version with the variable $\Delta T_{\rm AGN}$ (\colibrefiducialcolor). The only differences between the models are the treatment of $\Delta T_{\rm AGN}$ and the value of $m_{\rm BH, seed}$, which is set to $10^{4.8}~\mathrm{M_\odot}$ for $\Delta T_{\rm AGN}=10^{9}~\mathrm{K}$ and to $10^{5.5}~\mathrm{M_\odot}$ for the variable $\Delta T_{\rm AGN}$. 

The \gsmf{} and \ssm{} predicted by both models are virtually indistinguishable, each providing an excellent match to the \citet{2022MNRAS.513..439D} \gsmf{} and the \citet{2022MNRAS.509.3751H} \ssm. However, the \bhmsms{} show notable differences. While the predictions from both models converge for $M_* \gtrsim 10^{10}~\mathrm{M_\odot}$, reproducing the measurements of \citet{2023MNRAS.518.2177G}, the variable $\Delta T_{\rm AGN}$ model exhibits systematically more massive BHs at $M_* \lesssim 10^{10}~\mathrm{M_\odot}$ -- by up to $\approx 0.7$~dex -- due to its higher BH seed mass. As discussed in $\S$\ref{subsection: extension_to_higher_resolutions}, the ability to increase $m_{\rm BH, seed}$ without compromising the fit to the observed \gsmf{} and \ssm{} becomes crucial at m5 resolution, and is the main reason why the variable $\Delta T_{\rm AGN}$ model is preferred over its fixed $\Delta T_{\rm AGN}$ counterpart.

Fig.~\ref{fig: model_four_vs_final_2} presents a comparison between the predictions of the best-fitting \colibrefixedagntemp{} model (\colibrefixedagndtcolor) and its variation with the variable $\Delta T_{\rm AGN}$ (\colibrefiducialcolor) for nine different relations that were not used to calibrate the models. These relations were previously shown in $\S$\ref{subsection: other_galaxy_properties} in the context of comparing the four best-fitting models employing different SN feedback prescriptions. In Fig.~\ref{fig: model_four_vs_final_2}, the panels, from left to right and top to bottom, show: the median sSFR versus stellar mass at $z=0$ for active galaxies ($\textrm{sSFR} > 10^{-2} \, \rm Gyr^{-1}$), quenched fraction versus stellar mass at $z=0$, gas metallicity versus stellar mass at $z=0$ for star-forming galaxies, stellar metallicity versus stellar mass at $z=0$, H~\textsc{i} mass-to-stellar mass fraction versus stellar mass at $z=0$, H$_2$ mass-to-stellar mass fraction versus stellar mass at $z=0$, \gsmf{} at $z=2$, the median sSFR of all galaxies versus stellar mass at $z=2$, and the cosmic star formation rate density versus redshift. 

On average, the models with fixed and variable $\Delta T_{\rm AGN}$ perform similarly well in reproducing observational data at both $z=0$ and $z=2$. Small differences emerge in the $z=0$ sSFR, H~\textsc{i} fractions, H$_2$ fractions, and quenched fractions at $M_* \lesssim 10^{10.5}~\mathrm{M_\odot}$, where the variable $\Delta T_{\rm AGN}$ model exhibits slightly higher gas fractions and SFRs and slightly lower quenched fractions, leading to marginal or no improvement in the agreement with the data. Overall, Fig.~\ref{fig: model_four_vs_final_2} confirms that switching from the best-fitting \colibrefixedagntemp{} model with fixed $\Delta T_{\rm AGN}$ to its variation with variable $\Delta T_{\rm AGN}$ does not degrade the agreement with observations in any of the explored relations.

\begin{figure*}
        \centering
    \includegraphics[width=0.99\textwidth]{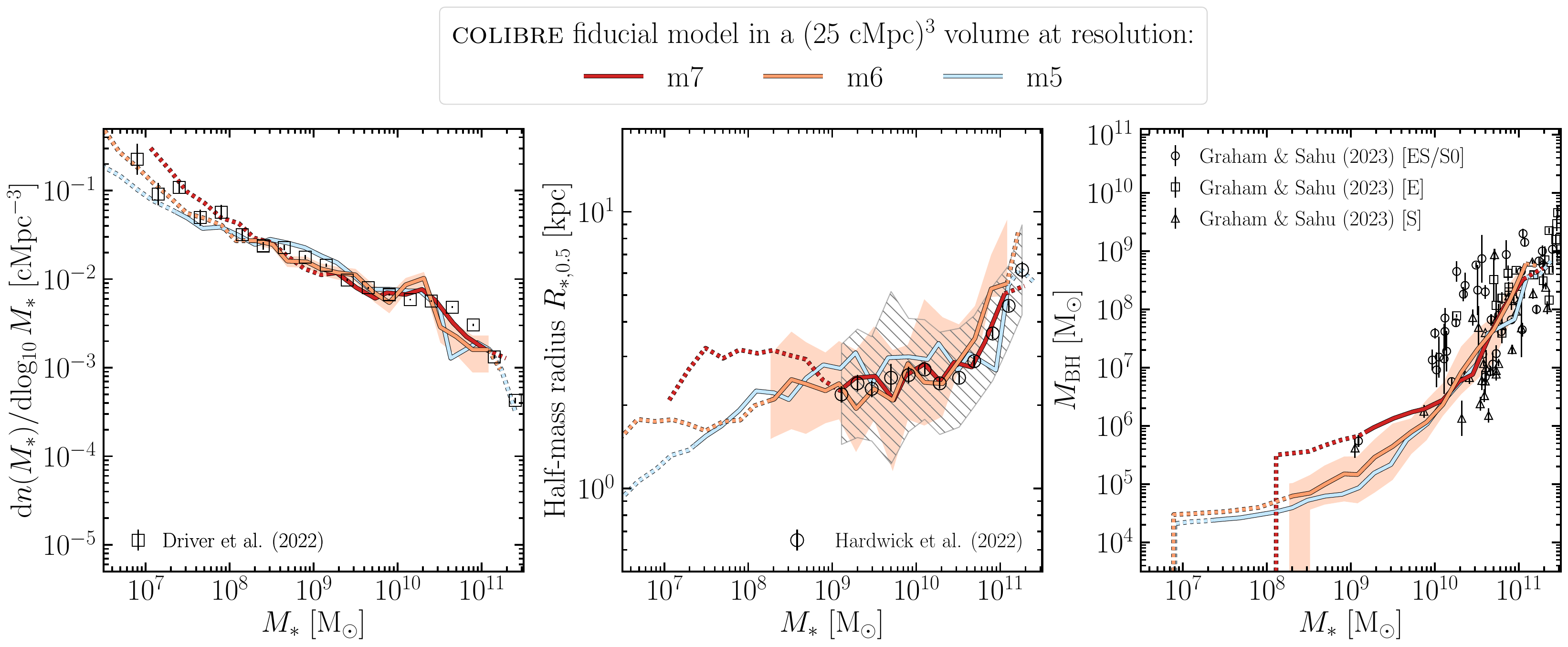}
    \caption{The $z=0$ galaxy stellar mass function (\gsmf; \textit{left}), the $z=0$ median size -- stellar mass relation (\ssm; \textit{middle}), and the $z=0$  median black hole mass -- stellar mass relation (\bhmsm; \textit{right}) in the simulations with the fiducial \colibre{} models (i.e. the best-fitting \colibre{} models with variable $\Delta T_{\rm AGN}$) at m7, m6, and m5 resolutions. All simulations were run in a (25 cMpc)$^3$ cosmological volume. Solid curves are converted to dotted curves below a stellar mass of $100 \, \times$ the baryonic particle mass of the simulations, to indicate that the corresponding galaxies are poorly resolved, as well as at the high-mass end, where there are fewer than five galaxies per bin. The \gsmf{} and \ssm{} predicted by \colibre{} exhibit very good convergence with resolution for $M_* \gtrsim 10^{9}~\mathrm{M_\odot}$ and reproduce the observed \gsmf{} from \citet{2022MNRAS.513..439D} and \ssm{} from \citet{2022MNRAS.509.3751H}. The \colibre{} \bhmsm{} agrees with the observational measurements from \citet{2023MNRAS.518.2177G} but converges only for $M_* \gtrsim 10^{10}~\mathrm{M_\odot}$ due to the different BH seed mass values used at different resolutions.}
    \label{fig: final_models_all_resolutions}
\end{figure*}

Based on the above findings, we opted to establish the \colibrefixedagntemp{} model with the variable $\Delta T_{\rm AGN}$ as the \textit{fiducial} \colibre{} model at all resolutions. At m7 resolution, the fiducial parameter values are therefore $m_{\rm BH, seed}=10^{5.5}~\mathrm{M_\odot}$, $f_{\rm kin}=0.1$, $P_{\rm E,pivot}/k_{\rm B}=8 \times 10^{3}~\mathrm{K \, cm^{-3}}$, and $n_{\rm H,pivot}=0.6~\mathrm{cm}^{-3}$. For m6 and m5 resolutions, several subgrid parameter values are adjusted from their m7 values to improve the fit to the observed $z=0$ \gsmf{} and \ssm{}. Specifically, $m_{\rm BH, seed}$ is decreased from  $m_{\rm BH, seed}=10^{5.5}~\mathrm{M_\odot}$ to $3\times 10^4~\mathrm{M_\odot}$, and $2\times 10^4~\mathrm{M_\odot}$, respectively, while $P_{\rm E,pivot}/k_{\rm B}$ is increased from $P_{\rm E,pivot}/k_{\rm B}=8 \times 10^{3}~\mathrm{K \, cm^{-3}}$ to $1 \times 10^4~\mathrm{K \, cm^{-3}}$ and $1.5 \times 10^4~\mathrm{K \, cm^{-3}}$. The parameter $f_{\rm kin}$ remains fixed at $0.1$ for all resolutions, while $n_{\rm H,pivot}$ is decreased from $0.6~\mathrm{cm}^{-3}$ to $0.5~\mathrm{cm}^{-3}$ at m6 resolution but increases to $1.0~\mathrm{cm}^{-3}$ at m5 resolution. Among the SN and AGN parameters not used in the emulation, $\Delta T_{\rm SN, min}$, $\Delta T_{\rm SN, max}$, $\Delta T_{\rm AGN, max}$, and $f_{\rm E,min}$ are set to $10^{6.75}~\mathrm{K}$, $10^{8}~\mathrm{K}$, $10^{10}~\mathrm{K}$, and $0.3$ at m6 resolution, and to $10^{7}~\mathrm{K}$, $10^{8}~\mathrm{K}$, $10^{10}~\mathrm{K}$, and $0.8$ at m5 resolution, respectively. Finally, the AGN feedback coupling efficiency, $\varepsilon_{\rm f}$, was reduced from $0.1$ at m7 resolution to $0.05$ at m6 and m5 resolutions to improve the agreement with the observed $z=0$ \bhmsm. Without this adjustment, the normalization of the \bhmsm{} predicted by the m6 and m5 models would have been too high by a factor of $\approx 2$ relative to the observations. A complete list of resolution-dependent subgrid parameter values is provided in table 1 of \citet{schaye2025colibreproject}.

The adjustments to the SN and AGN parameter values at m6 and m5 resolutions were determined through an iterative trial-and-error process, involving $\sim 100$ simulations in $25^3$ and $12.5^3$ cMpc$^3$ volumes, respectively. Our initial guess assumed the same parameter values as in the fiducial m7 model, and we then refined them iteratively to arrive at the final values reported above. The resolution dependence of most parameters is physically intuitive. For example, at higher resolution, BHs start growing earlier due to the presence of denser gas, requiring a lower $m_{\rm BH, seed}$ to counterbalance the enhanced BH growth. Conversely, thermal SN feedback becomes less efficient due to stronger radiative cooling losses in the denser gas and because a fixed temperature increase $\Delta T_{\rm SN}$ of a single gas particle corresponds to a smaller increase in the particle's internal energy, necessitating an increase in $\Delta T_{\rm SN, min}$ and $\Delta T_{\rm SN, max}$ at higher resolutions to maintain feedback strength comparable to m7 resolution. Star formation at higher resolution occurs in gas environments with higher pressures, so $P_{\rm E,pivot}$, which is representative of the median stellar birth pressure in the simulation (see Appendix \ref{appendix: redshift_evolution_of_sn_energy}), must increase. Finally, we note that the non-monotonic dependence of $n_{\rm H,pivot}$ on the resolution is due to its strong degeneracy with $P_{\rm E,pivot}$ and $\Delta T_{\rm SN, min}$, both of which change with resolution.

\begin{figure*}
    \centering
\includegraphics[width=0.99\textwidth]{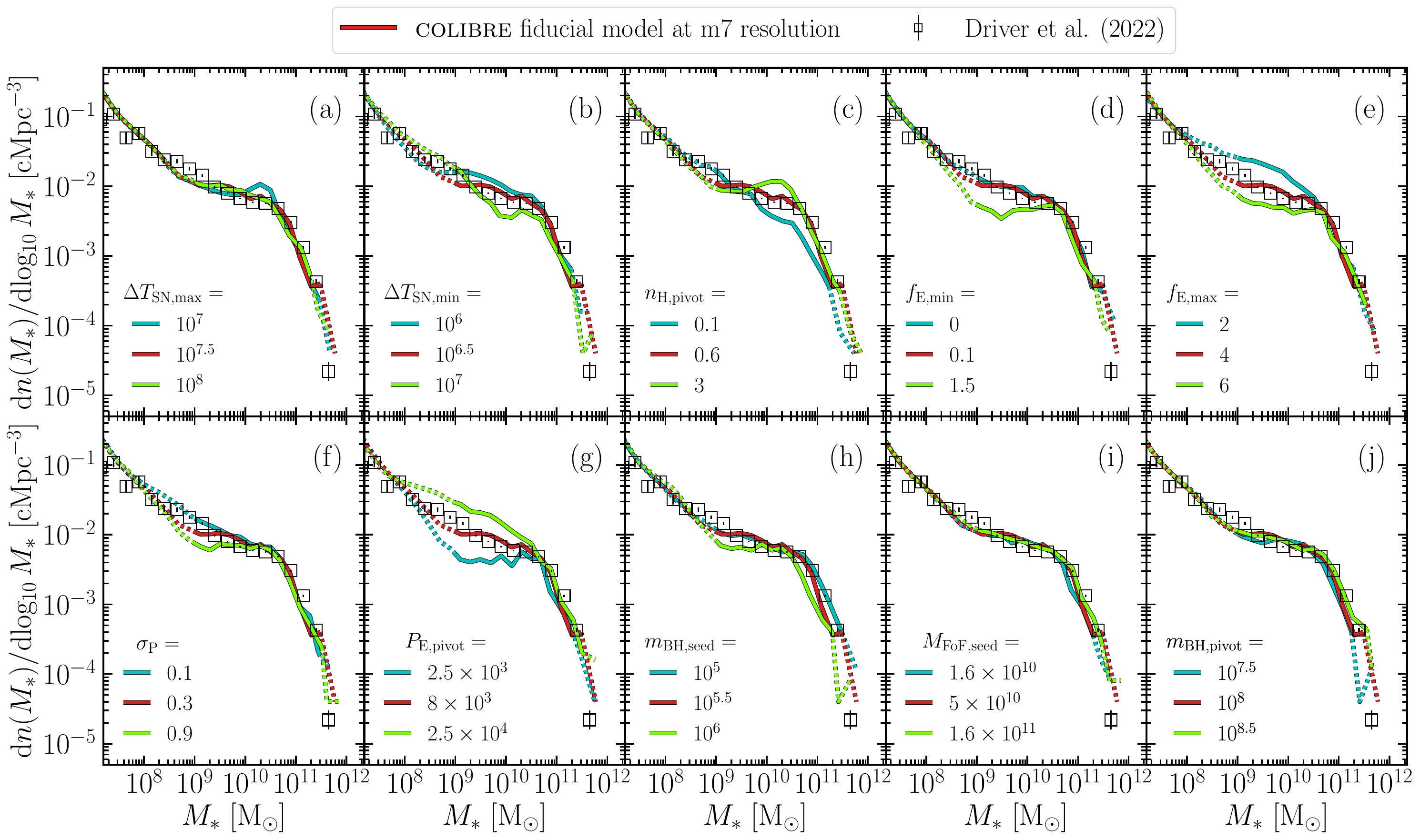}
\caption{The galaxy stellar mass function (\gsmf) at redshift $z=0$. Different panels show the effect of varying different SN and AGN feedback-related subgrid parameters of the \colibre{} fiducial model at m7 resolution. For each variation, we run a separate ($50~\mathrm{cMpc}$)$^3$ volume simulation. Only one parameter is varied per panel, while the others are held fixed to their best-fitting values. Starting from the top-left panel and going left to right, top to bottom, the varied parameters are $\Delta T_{\rm SN, max}$, $\Delta T_{\rm SN, min}$, $n_{\rm H, pivot}$, $f_{\rm E,min}$, $f_{\rm E,max}$, $\sigma_{\rm P}$, $P_{\rm E,pivot}$, $m_{\rm BH, seed}$, $M_{\rm FoF, seed}$, and $m_{\rm BH,pivot}$. The light-green (cyan) curve corresponds to the higher (lower) value of the parameter, relative to that in the fiducial m7 model, which is shown in \colibrefiducialcolor. The comparison data from \citet{2022MNRAS.513..439D} are displayed as black squares. Many parameters are degenerate with one another (i.e. they have a similar impact on the \gsmf), while others have little effect on the \gsmf.}
\label{fig: gsmf_one_param_var}
\end{figure*}

\begin{figure*}
        \centering
    \includegraphics[width=0.99\textwidth]{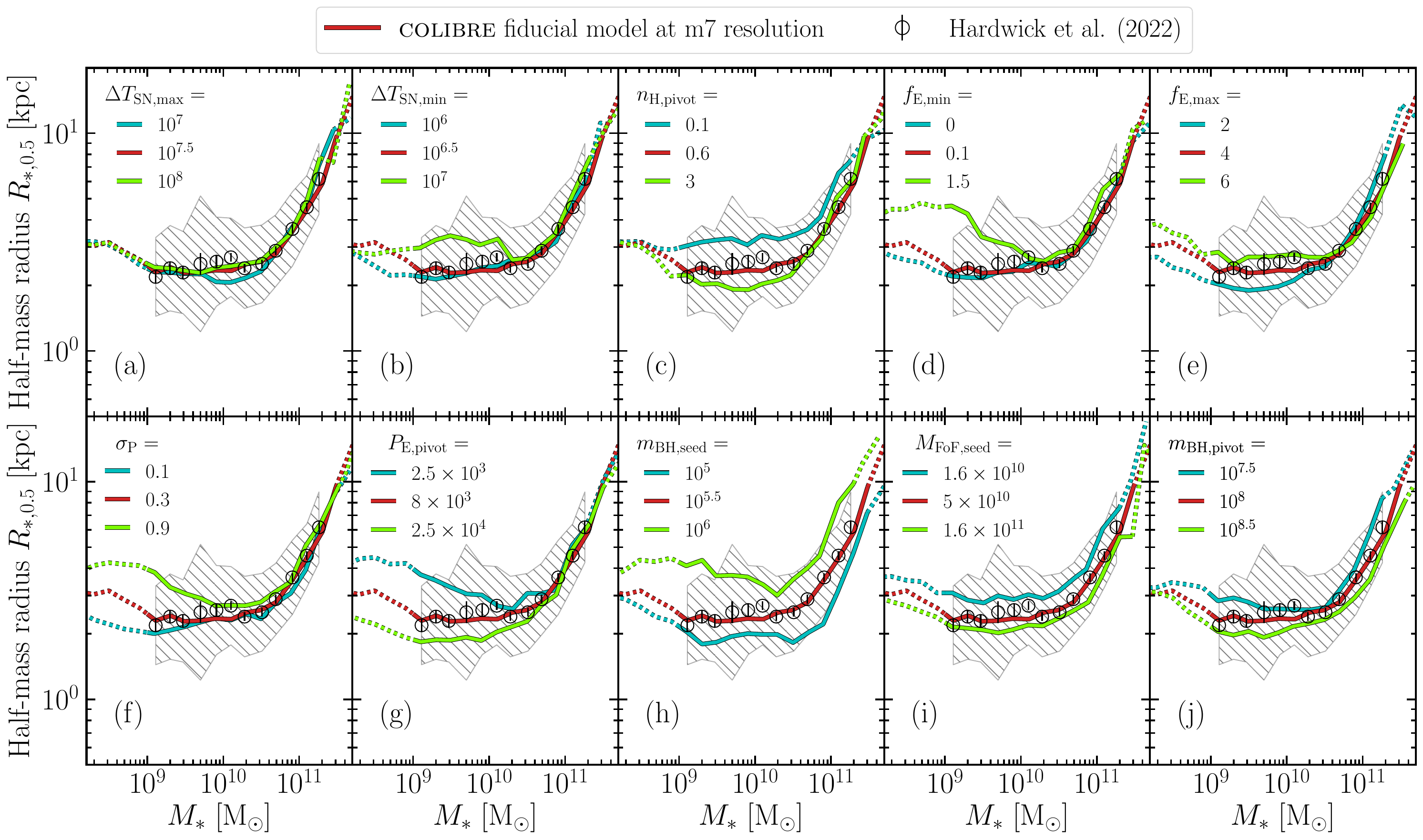}
    \caption{As Fig.~\ref{fig: gsmf_one_param_var}, but showing the $z=0$ median galaxy size -- stellar mass relation. The median observed relation from \citet{2022MNRAS.509.3751H} is displayed as black circles. The error bars show the $1\sigma$ error on the median observed sizes, while the grey hatched region indicates the galaxy population-wide scatter in \citet{2022MNRAS.509.3751H}.}
    \label{fig: sizes_one_param_var}
\end{figure*}

Fig.~\ref{fig: final_models_all_resolutions} shows the $z=0$ \gsmf{} (left panel), the $z=0$ \ssm{} (middle panel) and the $z=0$ \bhmsm{} (right panel) for the fiducial \colibre{} models at m7, m6 and m5 resolutions (differently coloured solid curves), all in a ($25$ cMpc)$^3$ cosmological volume. The observed \gsmf{} from \citet{2022MNRAS.513..439D}, the \ssm{} from \citet{2022MNRAS.509.3751H}, and the \bhmsm{} from \citet{2023MNRAS.518.2177G} are shown as black symbols.

The \colibre{} simulations reproduce the observed \gsmf{} and \ssm{} not only at m7 resolution, but also at m6 and m5, despite the best-fitting parameter values for m6 and m5 having been determined manually rather than via emulation. In particular, the \ssm{} predicted by \colibre{} at m6 and m5 resolutions remains within $\approx 0.1$~dex of the observed sizes across the full stellar mass range. In the range $10^9 \lesssim M_*/\mathrm{M_\odot} \lesssim 10^{10.5}$, the \ssm{} at m5 resolution appears slightly elevated compared to m6 and m7, but remains within $\approx0.1$~dex. At all three resolutions, the \gsmf{} agrees with the observational data to within $\approx 0.1$~dex at $M_* \lesssim 10^{10.5}~\mathrm{M_\odot}$, while at $M_* \gtrsim 10^{10.5}~\mathrm{M_\odot}$, the predictions are systematically offset from the observed \gsmf{}, owing to the small number of massive galaxies in the $(25~\mathrm{cMpc})^{3}$ volume. This is, however, not a concern, as we have verified that increasing the simulated volume brings the \gsmf{} predictions into much closer agreement with the data at the high-mass end (see Appendix \ref{appendix: boxsize_effect}). The agreement between the \colibre{} predictions for the \bhmsm{} and the observational measurements of \citet{2023MNRAS.518.2177G} is excellent at all resolutions, matching both the normalization and the slope of the observed relation.

The \colibre{} \gsmf{}, \ssm{}, and \bhmsm{} exhibit very good convergence\footnote{As discussed earlier in $\S$\ref{subsection: colibre_with_var_agn_heating_temperature}, some of the parameter values in the fiducial \colibre{} model are adjusted between different resolutions. Therefore, the resolution convergence should be interpreted as `weak convergence', in the language of \citet{2015MNRAS.446..521S}.} with resolution for $M_* \gtrsim 10^{8}~\mathrm{M_\odot}$, $M_* \gtrsim 10^{9}~\mathrm{M_\odot}$, and $M_* \gtrsim 10^{10}~\mathrm{M_\odot}$, respectively. The \gsmf{} converges down to a lower $M_*$ than the \ssm{} because the stellar mass is an integrated property, while the sizes depend on the spatial distribution of stellar particles within the galaxy and hence require more particles to be converged. At $M_* \lesssim 10^{9}~\mathrm{M_\odot}$, the limit corresponding to $\lesssim 100$ stellar particles at m7 resolution, galaxy half-mass radii decrease with increasing resolution. Meanwhile, the \bhmsm{} converges only for $M_* \gtrsim 10^{10}~\mathrm{M_\odot}$ due to differences in the BH seed mass used at different resolutions. Thanks to the variable $\Delta T_{\rm AGN}$, the lowest BH masses in the m5 model are $2 \times 10^4~\mathrm{M_\odot}$. The cutoff in the \bhmsm{} at $M_* \sim 10^7 - 10^8~\mathrm{M_\odot}$ is set by the minimum FoF mass, $M_{\rm FoF, seed}$, required for haloes to be seeded with a BH particle. $M_{\rm FoF, seed}$ is equal to $10^{10}~\mathrm{M_\odot}$ at m5 and m6 resolutions but to $5\times 10^{10}~\mathrm{M_\odot}$ at m7 resolution\footnote{Setting $M_{\rm FoF, seed}$ to $10^{10}~\mathrm{M_\odot}$ at m7 resolution, which is the value used at m6 and m5 resolutions, would result in massive BH particles having gravitational masses that exceed their subgrid masses, which is undesirable (see section 3.8.1 of \citealt{schaye2025colibreproject} for a discussion).}, shifting the cutoff of the \bhmsm{} to higher $M_*$ in the m7 model compared to m6 and m5.

\subsection{Parameter variations in the fiducial m7 COLIBRE model}
\label{subsection: parameter variations}

In the previous section, we showed that the fiducial \colibre{} models at m7, m6, and m5 resolutions reproduce the observational data to which they were calibrated: the $z=0$ \gsmf{} and \ssm{} using Gaussian process emulators and, independently, the $z=0$ \bhmsm{} by manually adjusting the coupling efficiency $\varepsilon_{\rm f}$ in equation~(\ref{eq: AGN_energy}). Because the m7 model was calibrated to the \gsmf{} and \ssm{} using emulators by optimizing the best-fitting values of a set of \textit{four} subgrid parameters, $\boldsymbol{\theta} = (f_{\rm kin}, n_{\rm H, pivot}, P_{\rm E, pivot}, m_{\rm BH, seed})$, an important question arises: why were these four parameters optimized, while other subgrid parameters related to stellar and AGN feedback were held fixed during emulation? This question was discussed in $\S$\ref{subsection: selection_theta}, though without presenting results supporting the choice of subgrid parameters. In this section, we further address this question by showing that the feedback-related subgrid parameters excluded from emulator-based calibration either have a negligible impact on the calibrated galaxy properties or are degenerate with parameters already included in $\boldsymbol{\theta}$. 

Figs.~\ref{fig: gsmf_one_param_var} and  \ref{fig: sizes_one_param_var} show how the $z=0$ \gsmf{} and \ssm{} in the \colibre{} fiducial m7 model respond to variations in $\Delta T_{\rm SN, max}$, $\Delta T_{\rm SN, min}$, $n_{\rm H, pivot}$, $f_{\rm E,min}$, $f_{\rm E,max}$, $\sigma_{\rm P}$, $P_{\rm E,pivot}$, $m_{\rm BH, seed}$, $M_{\rm FoF, seed}$, and $m_{\rm BH,pivot}$. Each figure contains 10 panels, where in each panel, we vary one of the 10 subgrid parameters, while the other parameters are fixed at their best-fitting values. For each variation, we run a separate ($50~\mathrm{cMpc}$)$^3$ volume simulation (that is, what is shown in the figures are the results from simulations, not from emulators). We show the results for three different values of each parameter: the value from the fiducial model (\colibrefiducialcolor), a lower value (cyan), and a higher value (light-green). The values of the varied parameters are indicated in each panel's legend.

First, we recall that the density-dependent heating temperature in the SN thermal feedback of the \colibre{} fiducial model depends on the subgrid parameters $\Delta T_{\rm SN, max}$, $\Delta T_{\rm SN, min}$, and $n_{\rm H, pivot}$ (see equation~\ref{eq: heating_temperature_vs_gas_density}). The effects of varying these parameters are shown in panels (a), (b), and (c), respectively. We observe that increasing $\Delta T_{\rm SN, max}$ by $0.5$~dex has no impact on either the \gsmf{} or galaxy sizes. In contrast, decreasing $\Delta T_{\rm SN, max}$~dex by $0.5$~dex introduces a bump in the \gsmf{} and a dip in galaxy sizes around $M_* = 10^{10.5}~\mathrm{M_\odot}$, indicating inefficient SN feedback at the mass scale just below where BHs take over. Similarly, changing $\Delta T_{\rm SN, min}$ by $\pm 0.5$~dex has a pronounced effect on both the \gsmf{} and \ssm. As expected, a higher (lower) $\Delta T_{\rm SN, min}$ enhances (weakens) SN thermal feedback, resulting in less (more) stellar mass formed and less (more) centrally concentrated galaxies, moving the low-mass end of the \gsmf{} down (up) and moving the low-mass end of the \ssm{} up (down). By comparing panels (a) and (b) with panel (c), where $n_{\rm H, pivot}$ is varied, we observe that most effects on the \gsmf{} and \ssm{} caused by changes in $\Delta T_{\rm SN, min}$ and/or $\Delta T_{\rm SN, max}$ can be captured by adjusting $n_{\rm H, pivot}$ alone. Furthermore, we recall that $\Delta T_{\rm SN, min}$ cannot be much lower than $10^{6.5}~\mathrm{K}$ (otherwise the SN thermal feedback would suffer from catastrophic overcooling) and $\Delta T_{\rm SN, max}$ cannot be much greater than $10^{7.5}~\mathrm{K}$ (otherwise the sampling of SN thermal injection events would become poor, see $\S$\ref{subsubsection: dens_dep_dT}). Based on these arguments, we decided to refrain from optimizing $\Delta T_{\rm SN, min}$ and $\Delta T_{\rm SN, max}$ and instead fix these parameters (at m7 resolution) to $10^{6.5}$ and $10^{7.5}$ K, respectively.

We next move to panels (d), (e), (f), and (g), which vary the parameters of the relation between the SN energy $f_{\rm E}$ and stellar birth pressure $P_{\rm birth}$ (equation~\ref{eq: stellar_birth_pressure_vs_SN_energy}). The parameters are: $f_{\rm E,min}$, $f_{\rm E,max}$, $\sigma_{\rm P}$ and $P_{\rm E,pivot}$. First, we observe that varying the parameters $f_{\rm E, min}$ and $f_{\rm E, max}$ -- which specify the energy injected by SN feedback from stellar particles formed in low- and high-pressure gas environments, respectively -- modulates the overall strength of SN feedback, predominantly affecting galaxies with $M_* \lesssim 10^{10.5}~\mathrm{M_\odot}$. Decreasing (increasing) the SN energy results in more (less) stellar mass formed and more (less) compact galaxies. Next, a nearly order-of-magnitude variation in the parameter $\sigma_{\rm P}$, which controls the width of the transition from $f_{\rm E, min}$ at low $P_{\rm birth}$ to $f_{\rm E, max}$ at high $P_{\rm birth}$, primarily impacts the low-mass end of the \gsmf{} and \ssm{} ($M_* \lesssim 10^{10}~\mathrm{M_\odot}$) in a manner similar to $f_{\rm E, min}$. Finally, the combined effect on the \gsmf{} and \ssm{} of varying $f_{\rm E, min}$, $f_{\rm E,max}$, and $\sigma_{\rm P}$ can be well captured by solely changing $P_{\rm E,pivot}$ (compare panels d, e, and f versus panel g), thereby justifying our choice to optimize $P_{\rm E,pivot}$ while keeping $f_{\rm E, min}$, $f_{\rm E,max}$, and $\sigma_{\rm P}$ fixed. Since lower values of $f_{\rm E, min}$ and $\sigma_{\rm P}$ yield a slightly better \ssm{} at low stellar masses (see panels d and f in Fig.~\ref{fig: sizes_one_param_var}), but $\sigma_{\rm P}$ must also not be too small to avoid an excessively sharp transition from $f_{\rm E, min}$ to $f_{\rm E, max}$, we set $f_{\rm E, min}$ to $0.1$ and $\sigma_{\rm P}$ to $0.3$. As for $f_{\rm E, max}$, we set its value to $4$, which gives an average energy per SN within the range $(1.5-2) \times 10^{51}~\mathrm{erg}$. Alternatively, we could choose a somewhat different value for $f_{\rm E, max}$ by adjusting $P_{\rm E,pivot}$ due to the degeneracy between these two parameters (compare panel e vs. panel g).

Lastly, the remaining three panels -- (h), (i), and (j) -- illustrate the effect of three BH-related parameters: $m_{\rm BH, seed}$, $M_{\rm FoF, seed}$, and $m_{\rm BH, pivot}$, respectively. Physically, increasing $m_{\rm BH, seed}$ promotes BH growth through gas accretion and mergers, leading to stronger AGN feedback, while decreasing $m_{\rm BH, seed}$ suppresses BH growth, resulting in weaker AGN feedback. Changing $M_{\rm FoF, seed}$ in the same direction as $m_{\rm BH, seed}$ has the opposite effect: larger values of $M_{\rm FoF, seed}$ delay BH seeding and reduce the overall number of seeded BHs, resulting in slower BH growth and weaker AGN feedback, while smaller values increase the rate of BH seeding, leading to faster BH growth and stronger AGN feedback. Finally, increasing (decreasing) the normalization of the relation between the BH mass and $\Delta T_{\rm AGN}$, $m_{\rm BH, pivot}$, leads to lower (higher) $\Delta T_{\rm AGN}$ at fixed BH mass, producing less (more) bursty -- and less (more) efficient -- AGN feedback\footnote{We verified that varying $\Delta T_{\rm AGN}$ by $0.5$~dex relative to $10^9~\mathrm{K}$ in the \colibrefixedagntemp{} model has a qualitatively similar effect on the \gsmf{} and \ssm{} as varying $m_{\rm BH, pivot}$ in the fiducial \colibre{} model, which uses a variable $\Delta T_{\rm AGN}$.}. Because our fiducial value for the BH seed mass is relatively large ($m_{\rm BH, seed}=10^{5.5}~\mathrm{M_\odot}$), varying these BH-related parameters affects all galaxies whose haloes are massive enough to have been endowed with a BH particle ($M_* \gtrsim 10^{8.5}~\mathrm{M_\odot}$). Galaxies experiencing stronger (weaker) AGN feedback on average form fewer (more) stars, and their stellar half-mass radii are larger (smaller). 

Overall, comparing panel (h) to panels (i) and (j), we find that varying $m_{\rm BH, seed}$ has a qualitatively similar effect on the \gsmf{} and \ssm{} as changing $M_{\rm FoF, seed}$ or $m_{\rm BH, pivot}$ in the opposite direction. This indicates that these three parameters are degenerate, meaning that optimizing one is sufficient. However, the \gsmf{} and \ssm{} are quantitatively more sensitive to variations in $m_{\rm BH, seed}$ than in $M_{\rm FoF, seed}$ or $m_{\rm BH, pivot}$. We remind that we chose to optimize $m_{\rm BH, seed}$ while fixing $M_{\rm FoF, seed}$ at $5\times 10^{10}~\mathrm{M_\odot}$ and $m_{\rm BH, pivot}$ at $10^{8}~\mathrm{M_\odot}$ in the \colibre{} fiducial m7 model; or equivalently, fixing $M_{\rm FoF, seed}$ at $5\times 10^{10}~\mathrm{M_\odot}$ and $\Delta T_{\rm AGN}$ at $10^9~\mathrm{K}$ in the \colibrefixedagntemp{} model. 

Lastly, we note that $m_{\rm BH, pivot}$ cannot be significantly smaller than $10^{8}~\mathrm{M_\odot}$, as this would correspond to $\Delta T_{\rm AGN} \ll 10^9~\mathrm{K}$ in massive objects according to equation (\ref{eq: variable_agn_temperature}), leading to implausibly high gas fractions therein, which are known to be sensitive to $\Delta T_{\rm AGN}$ \citep[e.g.][]{2014MNRAS.441.1270L,2017MNRAS.465.2936M,2023MNRAS.526.6103K}. Although we did not explicitly calibrate the models to cluster gas fractions, a few test simulations in a (100 cMpc)$^3$ volume confirmed that $\Delta T_{\rm AGN} = 10^9~\mathrm{K}$ in the \colibrefixedagntemp{} model and $m_{\rm BH, pivot} = 10^{8}~\mathrm{M_\odot}$ in the \colibre{} fiducial model yield plausible gas fractions in galaxy clusters. The analysis of the properties of galaxy clusters will be presented in future work.

\section{Conclusions}
\label{Section: Conclusions}

We have presented the calibration of the new \colibre{} subgrid model for cosmological hydrodynamical simulations of galaxy formation \citep{schaye2025colibreproject}. \colibre{} is available at three resolutions: $m_{\rm gas} \approx m_{\rm dm} \sim 10^7~\mathrm{M_\odot}$ (m7), $10^6~\mathrm{M_\odot}$ (m6), and $10^5~\mathrm{M_\odot}$ (m5). It has evolved from the \textsc{owls} \citep{2010MNRAS.402.1536S} and \textsc{eagle} \citep{2015MNRAS.446..521S} galaxy formation models with a large number of improvements and modifications. The most significant ones are: (i) the presence of a cold interstellar gas phase; (ii) the suppression of spurious energy transfer from DM to baryons (by using four times more DM particles than baryonic particles); (iii) a model for the formation and evolution of dust grains coupled to the chemistry; (iv) the use of a non-equilibrium network for the calculation of radiative cooling rates and ion and molecular fractions of hydrogen and helium; and (v) improved prescriptions for the modelling of all subgrid physics processes, including the prescriptions for radiative cooling, star formation, stellar mass loss, BHs, and feedback from stars and SMBHs.

We used Gaussian process emulators to calibrate SN and AGN feedback in the \colibre{} model. The emulators were trained on $\sim 200$ simulations at m7 resolution in a (50 cMpc)$^3$ volume. Each simulation was run with a unique combination of subgrid parameter values governing the strengths of SN and AGN feedback, enabling the emulators to learn how galaxy properties vary as functions of these parameters. These four parameters are: (i) the fraction of SN energy injected in kinetic form, $f_{\rm kin}$, (ii) the pivot density in the thermal channel of SN feedback with a variable heating temperature, $n_{\rm H, pivot}$; (iii) the pivot stellar birth pressure in the relation between the SN energy and stellar birth pressure, $P_{\rm E,pivot}$; and (iv) the BH seed mass, $m_{\rm BH, seed}$. By fitting the trained emulators to the observed $z=0$ galaxy stellar mass function (\gsmf) and to the observed $z=0$ galaxy size -- stellar mass relation (\ssm) in the stellar mass range $10^9 < M_*/\mathrm{M_\odot} < 10^{11.3}$, we found the values of the subgrid parameters that result in the best agreement with the target observational data. 

The emulator-based calibration used a fixed $\Delta T_{\rm AGN} = 10^9~\mathrm{K}$. After the emulation was completed, we updated the model to use an AGN heating temperature that increases linearly with BH mass (equation~\ref{eq: variable_agn_temperature}), as this proved important for higher resolution simulations. Paired with an increase in the BH seed mass, this change has a negligible effect on any of the calibration diagnostics. In the following, we first summarize our conclusions concerning the calibration with emulators at m7 resolution, using the \colibre{} model with $\Delta T_{\rm AGN} = 10^9~\mathrm{K}$ (referred to as \colibrefixedagntemp), and then discuss the transition to the fiducial \colibre{} model, which uses a variable $\Delta T_{\rm AGN}$. 

In the prescription for SN feedback in the \colibre{} model, (i) stellar particles inject their SN energy into surrounding gas in both thermal and kinetic forms, (ii) the heating temperature in the thermal channel, $\Delta T_{\rm SN}$, is an increasing function of the gas density (equation~\ref{eq: heating_temperature_vs_gas_density}), and (iii) the energy per SN in units of $10^{51}$ erg, $f_{\rm E}$, is an increasing function of stellar birth gas pressure, $P_{\rm birth}$ (equation~\ref{eq: stellar_birth_pressure_vs_SN_energy}). In order to demonstrate that these model ingredients are all necessary to reproduce the observed \gsmf{} and \ssm, we explored three variations of the \colibre{} model in which the modelling of SN feedback was significantly simplified:
\begin{enumerate}
    \item We first considered the \texttt{Basic} model, in which the energy in SN feedback is constant (i.e. $f_{\rm E}$ is independent of $P_{\rm birth}$) and is only injected thermally, stochastically heating the gas by a constant value of $\Delta T_{\rm SN}=10^{7.5}~\mathrm{K}$. 
    
    \item Our second simplified model was \texttt{ThermalKinetic}, which allows some fraction of the SN energy, $f_{\rm kin}$, to be injected kinetically via low-energy kicks with a target kick velocity of 50 km s$^{-1}$, while the remainder is injected thermally as in the \texttt{Basic} model.
    
    \item Finally, in the third simplified model, \VardT{}, the heating temperature $\Delta T_{\rm SN}$ increases with the density of the gas surrounding the SNe. Compared to the \colibrefixedagntemp{} model, this model uses a constant energy per SN, as opposed to the SN energy increasing with the stellar birth pressure as adopted in the \colibrefixedagntemp{} model.
\end{enumerate}

These three simplified models were fit to the observed $z=0$ \gsmf{} and \ssm{} using emulators in the same manner as the \colibrefixedagntemp{} model, and for each model, the best-fitting subgrid parameter values were found. For the \texttt{Basic} model, we optimized the parameters $m_{\rm BH, seed}$ and $f_{\rm E}$; for \texttt{ThermalKinetic}, the parameters $m_{\rm BH, seed}$, $f_{\rm E}$, and $f_{\rm kin}$; and for \VardT, $m_{\rm BH, seed}$, $f_{\rm E}$, $f_{\rm kin}$, and $n_{\rm H,pivot}$. In total, we ran $\approx 200$ simulations for various combinations of the subgrid parameters and models. Our main results with regard to the calibration are as follows:

\begin{itemize}
    \item The \texttt{Basic} model fails to produce a good fit to the observed $z=0$ \gsmf{} (Fig.~\ref{fig:basic_and_thermal_kinetic_model_calibration}). The \gsmf{} exhibits a power-law shape as opposed to the observed \citet{1976ApJ...203..297S} shape. Increasing or decreasing the SN energy, which is described by the subgrid parameter $f_{\rm E}$, cannot resolve this discrepancy (middle panel of Fig.~\ref{fig:smhm_emulator_thermalkinetic}).
    
    \item The \texttt{ThermalKinetic} model can successfully reproduce the observed $z=0$ \gsmf{} or the observed \ssm{} separately but cannot fit both relations simultaneously (Fig.~\ref{fig: thermal_kinetic_model_with_var_constraints}). The fact that \texttt{ThermalKinetic} can provide a good match to the observed \gsmf{} is a consequence of the ability to combine the large energy injections of the thermal channel of SN feedback with the low-energy kicks of the kinetic channel (Fig.~\ref{fig:smhm_emulator_thermalkinetic}). The relative strengths of the two channels are optimized by the emulators via the parameter $f_{\rm kin}$: the model fit to the observed \gsmf{} (\ssm) prefers $f_{\rm kin}\approx 0.6$ ($f_{\rm kin}\approx 0$), while fitting to both constraints gives an intermediate value of  $f_{\rm kin}\approx 0.3$ (Fig.~\ref{fig:thermalkinetic_model_posterior}).

    \item Adopting a density-dependent heating temperature $\Delta T_{\rm SN}$ (the \VardT{} model) improves the combined fit to the \gsmf{} and \ssm, while additionally introducing a stellar birth pressure dependence of the SN energy (the \colibrefixedagntemp{} model) results in excellent agreement with the observed \gsmf{} and \ssm{} (Fig.~\ref{fig: models_three_and_four}).
\end{itemize}

Having calibrated each model to the observed \gsmf{} and \ssm, we proceeded to compare the best-fitting versions of each model to a number of observables that were not considered in the emulator-based calibration:

\begin{itemize}

    \item The observed $z=0$ sSFR and the galaxy quenched fractions are broadly matched by the \VardT{} and \colibrefixedagntemp{} models, but not by \texttt{Basic} and \texttt{ThermalKinetic} (Fig.~\ref{fig: four_models_ssfr}). The SN feedback with a constant $\Delta T_{\rm SN}$ of $10^{7.5}$ K, which is employed in the latter two models, is overly powerful, leading to a lack of star-forming gas by $z=0$ in low- and intermediate-mass galaxies.

    \item The observed $z=0$ cold gas fractions in the stellar mass range $10^{9} < M_*/\mathrm{M_\odot} < 10^{10.5}$ are reproduced by both the \VardT{} and \colibrefixedagntemp{} models for H~\textsc{i}, but only by \colibrefixedagntemp{} for H$_2$ (Fig.~\ref{fig: four_models_cold}). In addition, \colibrefixedagntemp{} reproduces the observed scatter for both H~\textsc{i} and H$_2$. At higher stellar masses, $M_* > 10^{10.5}~\mathrm{M_\odot}$, all four models underestimate the gas fractions (Fig.~\ref{fig: four_models_ssfr}), though, as shown by \citet{schaye2025colibreproject}, the agreement with the data at the high-mass end can be improved by using higher resolution.
    
    \item Owing to its stellar birth pressure dependence of the energy in SN feedback, the \colibrefixedagntemp{} model is the only model that provides a reasonably good match to the observed \gsmf{} and sSFR at $z=2$ (Fig.~\ref{fig: four_models_highz}). 
    
    \item Similarly, due to the pressure dependence of its SN feedback, the cosmic star formation rate density (SFRD) in the \colibrefixedagntemp{} model is suppressed at high $z$ relative to the other three models (Fig.~\ref{fig: four_models_SFH}). As a result, the SFRD in the \colibrefixedagntemp{} model has a broad peak between $1<z<4$ and only begins to decline steeply below $z\approx 1$, which agrees with observations. In contrast, in the other models, the SFRD is a steeply declining function of cosmic time already after $z\approx 3$.
    
    \item The observed $z=0$ relations between galaxy stellar mass and stellar and gas-phase metallicities are reproduced at $M_* \lesssim 10^{11}~\mathrm{M_\odot}$ in all four models (Fig.~\ref{fig: four_models_met}).

    \item The $z=0$ masses of SMBHs, as well as their scaling with the stellar mass of the host galaxy, are consistent with observations for all four models (Fig.~\ref{fig: four_models_ssfr}). This agreement is expected as the value of the AGN feedback coupling efficiency, $\varepsilon_{\rm f}$, was chosen to match the $z=0$ observed \bhmsm{} at the high galaxy masses for which dynamical BH mass measurements are possible. Because $\varepsilon_{\rm f}$ primarily affects the normalization of the \bhmsm{} but has less impact on the \gsmf{} and \ssm, its value was set independently of the other subgrid parameters and without using emulators.
\end{itemize}

Having demonstrated that the best-fitting \colibrefixedagntemp{} model outperforms its three counterparts with simplified SN feedback, we applied it to simulations at two higher \colibre{} resolutions: m6 and m5. Using the best-fitting parameter values at m7 resolution as an initial guess, we manually adjusted the subgrid parameter values to achieve a fit to the $z=0$ \gsmf, \ssm, and \bhmsm{} at m6 and m5 resolutions that is similarly good as at m7. At m5 resolution, we found that $\Delta T_{\rm AGN} = 10^9~\mathrm{K}$ -- the heating temperature used in the AGN feedback of the \colibrefixedagntemp{} model -- requires a low BH seed mass ($m_{\rm BH, seed} \sim 10^3~\mathrm{M_\odot}$), yielding a probably unrealistically steep relation between galaxy stellar mass and BH mass; otherwise, AGN feedback is excessively strong. To allow for a higher $m_{\rm BH, seed}$ without compromising the fit to the observed \gsmf{} and \ssm{}, we replaced the fixed $\Delta T_{\rm AGN} = 10^9~\mathrm{K}$ with a variable $\Delta T_{\rm AGN}$ (equation~\ref{eq: variable_agn_temperature}). This change was applied at all three \colibre{} resolutions to maintain consistency. The updated model (i.e. the modified version of the best-fitting \colibrefixedagntemp{} model that uses the variable $\Delta T_{\rm AGN}$) was established as the fiducial \colibre{} model. At m7 resolution, the fiducial values of the four calibrated subgrid parameters of SN and AGN feedback are: $m_{\rm BH, seed}=10^{5.5}~\mathrm{M_\odot}$, $f_{\rm kin}=0.1$, $P_{\rm E,pivot}/k_{\rm B} =8 \times 10^{3}~\mathrm{K \, cm^{-3}}$, and $n_{\rm H,pivot}=0.6~\mathrm{cm}^{-3}$. In relation to both the change in $\Delta T_{\rm AGN}$ and the extension of the calibrated m7 model to higher resolutions, we demonstrated that:
\begin{itemize}
    \item The fit of the \colibre{} fiducial m7 model, which uses a variable $\Delta T_{\rm AGN}$, to the observed \gsmf, \ssm, and \bhmsm{} remains as good as that of \colibrefixedagntemp{}, which employs $\Delta T_{\rm AGN} = 10^9~\mathrm{K}$ and was calibrated using emulators (Fig.~\ref{fig: model_four_vs_final}). Additionally, the fiducial model demonstrates a similar level of agreement with observational data that were not used in the calibration (Fig.~\ref{fig: model_four_vs_final_2}).

    \item The fiducial m6 and m5 \colibre{} models, both of which also use a variable $\Delta T_{\rm AGN}$ but were calibrated manually, exhibit a similar level of agreement with the observed \gsmf, \ssm, and \bhmsm{} as the fiducial m7 model (Fig.~\ref{fig: final_models_all_resolutions}).

    \item At m7 resolution, switching from a fixed to a variable $\Delta T_{\rm AGN}$ allows for an increase in the BH seed mass, leading to more massive BHs in low-mass galaxies by up to $0.7$~dex (right-hand panel of Fig.~\ref{fig: model_four_vs_final}). At m5 resolution, the BH mass never falls below $2\times 10^4~\mathrm{M_\odot}$, ensuring that the median BH mass grows smoothly with increasing galaxy stellar mass (right-hand panel of Fig.~\ref{fig: final_models_all_resolutions}).
\end{itemize}

Having calibrated the \colibre{} fiducial model at all resolutions, we proceeded to investigate the effect of changing individual subgrid parameters in the fiducial m7 model, including the parameters that were not optimized by the emulators. We confirmed that the latter parameters are either degenerate with the parameters that were optimized and/or have little impact on the $z=0$ \gsmf{} and \ssm{} (Figs. \ref{fig: gsmf_one_param_var} and  \ref{fig: sizes_one_param_var}).

In closing, we stress that calibrating a galaxy formation model is a numerically demanding process with no guarantee of success. The fact that the \colibre{} fiducial m7 model fit to the $z=0$ \gsmf{} and \ssm{} reproduces many observed relations that were not considered during the calibration (Fig.~\ref{fig: model_four_vs_final_2}) is an encouraging result. Part of this agreement can be attributed to the choices made during the development and testing of various physical prescriptions implemented in \colibre, including the model for radiative cooling \citep{2025arXiv250615773P}, the model for dust grains \citep{2025arXiv250513056T}, the star formation prescription \citep{2024MNRAS.532.3299N}, the model for metal enrichment and turbulent diffusion of element mass fractions (Correa et al., submitted), the prescription for early stellar feedback \citep{2025arXiv250925309B}, and the model for SN feedback \citep{2023MNRAS.523.3709C}. For a detailed discussion of the performance of the \colibre{} simulations at higher resolutions and in larger cosmological volumes, we refer the reader to \citet{schaye2025colibreproject}, who present results from the fiducial \colibre{} models at m7, m6, and m5 resolutions in the largest available cosmological volumes at $z=0$: $400^3~\mathrm{cMpc}^3$, $200^3~\mathrm{cMpc}^3$, and $25^3~\mathrm{cMpc}^3$, respectively.

\section*{Acknowledgements}

The authors of this work acknowledge the pioneering impact that the late Richard Bower had on the use of emulation techniques in galaxy formation simulations. As a founding member of the COLIBRE team, his influence is felt throughout the project from both his personal contributions and role as an advisor, mentor, and colleague. Of particular note was his multi-year effort to recalibrate the \textsc{eagle} model programmatically using Gaussian process emulation, which, through many twists and turns, led to the highly successful emulation campaign for \textsc{flamingo}, and now for the \colibre{} model. His overwhelmingly positive demeanor and sharp physical insight are sorely missed by the \colibre{} team and across the astronomical community. We thank the referee for a constructive report. This work used the DiRAC@Durham facility managed by the Institute for Computational Cosmology on behalf of the STFC DiRAC HPC Facility (www.dirac.ac.uk). The equipment was funded by BEIS capital funding via STFC capital grants ST/K00042X/1, ST/P002293/1, ST/R002371/1 and ST/S002502/1, Durham University and STFC operations grant ST/R000832/1. DiRAC is part of the National e-Infrastructure. This project has received funding from the Netherlands Organization for Scientific Research (NWO) through research programme Athena 184.034.002. ABL acknowledges support by the Italian Ministry for Universities and Research (MUR), program `Dipartimenti di Eccellenza 2023-2027' within the Centro Bicocca di Cosmologia Quantitativa (BiCoQ), and support by UNIMIB's Fondo di Ateneo Quota Competitiva (project 2024-ATEQC-0050). CGL acknowledges support from STFC grants ST/T000244/1 and ST/X001075/1. CSF acknowledges support from European Research Council (ERC) Advanced Grant DMIDAS (GA 786910). EC was supported by the funding from the European Union's Horizon 2020 research and innovation programme under the Marie Skłodowska-Curie grant agreement No 860744 (BiD4BESt). JT acknowledges support of STFC Grant ST/X004651/1. RAC acknowledges support from STFC grants ST/Y002482/1 and ST/Y001907/1. SP acknowledges support from the Austrian Science Fund (FWF) through project V 982-N. VJFM acknowledges support by NWO through the Dark Universe Science Collaboration (OCENW.XL21.XL21.025). YMB acknowledges support from UK Research and Innovation through a Future Leaders Fellowship (grant agreement MR/X035166/1) and financial support from the Swiss National Science Foundation (SNSF) under project 200021\_213076. The research in this paper made use of the \textsc{Swift} open-source simulation code (\url{http://www.swiftsim.com}, \citealt{2018ascl.soft05020S}) version 1.0.0. The data analysis was carried out with the use of \textsc{swiftsimio} \citep{Borrow2020simio,2021arXiv210605281B}, \textsc{numpy} \citep{2020Natur.585..357H}, \textsc{scipy} \citep{2020SciPy-NMeth}, \textsc{matplotlib} \citep{2007CSE.....9...90H}, and \textsc{seaborn} \citep{Waskom2021}. 

\section*{Data Availability}

The data underlying this article will be shared on reasonable request to the corresponding author. The public version of the \textsc{Swift} simulation code can be found on \href{http://www.swiftsim.com}{www.swiftsim.com}. The \textsc{Swift} modules related to the \colibre{} galaxy formation model will be integrated into the public version after the public release of \colibre. The \textsc{chimes} astrochemistry code is publicly available at \href{https://richings.bitbucket.io/chimes/home.html}{https://richings.bitbucket.io/chimes/home.html}.



\bibliographystyle{mnras}
\bibliography{main} 

\begin{thebibliography}{}
\makeatletter
\relax
\def\mn@urlcharsother{\let\do\@makeother \do\$\do\&\do\#\do\^\do\_\do\%\do\~}
\def\mn@doi{\begingroup\mn@urlcharsother \@ifnextchar [ {\mn@doi@} {\mn@doi@[]}}
\def\mn@doi@[#1]#2{\def\@tempa{#1}\ifx\@tempa\@empty \href {http://dx.doi.org/#2} {doi:#2}\else \href {http://dx.doi.org/#2} {#1}\fi \endgroup}
\def\mn@eprint#1#2{\mn@eprint@#1:#2::\@nil}
\def\mn@eprint@arXiv#1{\href {http://arxiv.org/abs/#1} {{\tt arXiv:#1}}}
\def\mn@eprint@dblp#1{\href {http://dblp.uni-trier.de/rec/bibtex/#1.xml} {dblp:#1}}
\def\mn@eprint@#1:#2:#3:#4\@nil{\def\@tempa {#1}\def\@tempb {#2}\def\@tempc {#3}\ifx \@tempc \@empty \let \@tempc \@tempb \let \@tempb \@tempa \fi \ifx \@tempb \@empty \def\@tempb {arXiv}\fi \@ifundefined {mn@eprint@\@tempb}{\@tempb:\@tempc}{\expandafter \expandafter \csname mn@eprint@\@tempb\endcsname \expandafter{\@tempc}}}

\bibitem[\protect\citeauthoryear{{Abbott} et~al.,}{{Abbott} et~al.}{2022}]{2022PhRvD.105b3520A}
{Abbott} T.~M.~C.,  et~al., 2022, \mn@doi [\prd] {10.1103/PhysRevD.105.023520}, \href {https://ui.adsabs.harvard.edu/abs/2022PhRvD.105b3520A} {105, 023520}

\bibitem[\protect\citeauthoryear{{Accurso} et~al.,}{{Accurso} et~al.}{2017}]{2017MNRAS.470.4750A}
{Accurso} G.,  et~al., 2017, \mn@doi [\mnras] {10.1093/mnras/stx1556}, \href {https://ui.adsabs.harvard.edu/abs/2017MNRAS.470.4750A} {470, 4750}

\bibitem[\protect\citeauthoryear{Asplund, Grevesse, Sauval  \& Scott}{Asplund et~al.}{2009}]{doi:10.1146/annurev.astro.46.060407.145222}
Asplund M.,  Grevesse N.,  Sauval A.~J.,   Scott P.,  2009, \mn@doi [Annual Review of Astronomy and Astrophysics] {10.1146/annurev.astro.46.060407.145222}, 47, 481

\bibitem[\protect\citeauthoryear{{Bah{\'e}} et~al.,}{{Bah{\'e}} et~al.}{2016}]{2016MNRAS.456.1115B}
{Bah{\'e}} Y.~M.,  et~al., 2016, \mn@doi [\mnras] {10.1093/mnras/stv2674}, \href {https://ui.adsabs.harvard.edu/abs/2016MNRAS.456.1115B} {456, 1115}

\bibitem[\protect\citeauthoryear{{Bah{\'e}} et~al.,}{{Bah{\'e}} et~al.}{2022}]{2022MNRAS.516..167B}
{Bah{\'e}} Y.~M.,  et~al., 2022, \mn@doi [\mnras] {10.1093/mnras/stac1339}, \href {https://ui.adsabs.harvard.edu/abs/2022MNRAS.516..167B} {516, 167}

\bibitem[\protect\citeauthoryear{Baldwin, Phillips  \& Terlevich}{Baldwin et~al.}{1981}]{Baldwin1981}
Baldwin J.~A.,  Phillips M.~M.,   Terlevich R.,  1981, \mn@doi [Publications of the Astronomical Society of the Pacific] {10.1086/130766}, 93, 5

\bibitem[\protect\citeauthoryear{{Barber}, {Crain}  \& {Schaye}}{{Barber} et~al.}{2018}]{2018MNRAS.479.5448B}
{Barber} C.,  {Crain} R.~A.,   {Schaye} J.,  2018, \mn@doi [\mnras] {10.1093/mnras/sty1826}, \href {https://ui.adsabs.harvard.edu/abs/2018MNRAS.479.5448B} {479, 5448}

\bibitem[\protect\citeauthoryear{{Barber}, {Schaye}  \& {Crain}}{{Barber} et~al.}{2019}]{2019MNRAS.483..985B}
{Barber} C.,  {Schaye} J.,   {Crain} R.~A.,  2019, \mn@doi [\mnras] {10.1093/mnras/sty3011}, \href {https://ui.adsabs.harvard.edu/abs/2019MNRAS.483..985B} {483, 985}

\bibitem[\protect\citeauthoryear{{Bate} \& {Burkert}}{{Bate} \& {Burkert}}{1997}]{1997MNRAS.288.1060B}
{Bate} M.~R.,  {Burkert} A.,  1997, \mn@doi [\mnras] {10.1093/mnras/288.4.1060}, \href {https://ui.adsabs.harvard.edu/abs/1997MNRAS.288.1060B} {288, 1060}

\bibitem[\protect\citeauthoryear{{Bauer} et~al.,}{{Bauer} et~al.}{2013}]{2013MNRAS.434..209B}
{Bauer} A.~E.,  et~al., 2013, \mn@doi [\mnras] {10.1093/mnras/stt1011}, \href {https://ui.adsabs.harvard.edu/abs/2013MNRAS.434..209B} {434, 209}

\bibitem[\protect\citeauthoryear{{Behroozi}, {Wechsler}, {Hearin}  \& {Conroy}}{{Behroozi} et~al.}{2019}]{2019MNRAS.488.3143B}
{Behroozi} P.,  {Wechsler} R.~H.,  {Hearin} A.~P.,   {Conroy} C.,  2019, \mn@doi [\mnras] {10.1093/mnras/stz1182}, \href {https://ui.adsabs.harvard.edu/abs/2019MNRAS.488.3143B} {488, 3143}

\bibitem[\protect\citeauthoryear{{Belfiore} et~al.,}{{Belfiore} et~al.}{2018}]{2018MNRAS.477.3014B}
{Belfiore} F.,  et~al., 2018, \mn@doi [\mnras] {10.1093/mnras/sty768}, \href {https://ui.adsabs.harvard.edu/abs/2018MNRAS.477.3014B} {477, 3014}

\bibitem[\protect\citeauthoryear{{Ben{\'\i}tez-Llambay} et~al.,}{{Ben{\'\i}tez-Llambay} et~al.}{2025}]{2025arXiv250925309B}
{Ben{\'\i}tez-Llambay} A.,  et~al., 2025, \mn@doi [arXiv e-prints] {10.48550/arXiv.2509.25309}, \href {https://ui.adsabs.harvard.edu/abs/2025arXiv250925309B} {p. arXiv:2509.25309}

\bibitem[\protect\citeauthoryear{{B{\'e}thermin} et~al.,}{{B{\'e}thermin} et~al.}{2023}]{2023A&A...680L...8B}
{B{\'e}thermin} M.,  et~al., 2023, \mn@doi [\aap] {10.1051/0004-6361/202348115}, \href {https://ui.adsabs.harvard.edu/abs/2023A&A...680L...8B} {680, L8}

\bibitem[\protect\citeauthoryear{{Bird}, {Ni}, {Di Matteo}, {Croft}, {Feng}  \& {Chen}}{{Bird} et~al.}{2022}]{2022MNRAS.512.3703B}
{Bird} S.,  {Ni} Y.,  {Di Matteo} T.,  {Croft} R.,  {Feng} Y.,   {Chen} N.,  2022, \mn@doi [\mnras] {10.1093/mnras/stac648}, \href {https://ui.adsabs.harvard.edu/abs/2022MNRAS.512.3703B} {512, 3703}

\bibitem[\protect\citeauthoryear{{Black}}{{Black}}{1987}]{1987ASSL..134..731B}
{Black} J.~H.,  1987, {Heating and Cooling of the Interstellar Gas}, \mn@doi{10.1007/978-94-009-3861-8_27}

\bibitem[\protect\citeauthoryear{{Booth} \& {Schaye}}{{Booth} \& {Schaye}}{2009}]{2009MNRAS.398...53B}
{Booth} C.~M.,  {Schaye} J.,  2009, \mn@doi [\mnras] {10.1111/j.1365-2966.2009.15043.x}, \href {https://ui.adsabs.harvard.edu/abs/2009MNRAS.398...53B} {398, 53}

\bibitem[\protect\citeauthoryear{{Booth} \& {Schaye}}{{Booth} \& {Schaye}}{2010}]{2010MNRAS.405L...1B}
{Booth} C.~M.,  {Schaye} J.,  2010, \mn@doi [\mnras] {10.1111/j.1745-3933.2010.00832.x}, \href {https://ui.adsabs.harvard.edu/abs/2010MNRAS.405L...1B} {405, L1}

\bibitem[\protect\citeauthoryear{Borrow \& Borrisov}{Borrow \& Borrisov}{2020}]{Borrow2020simio}
Borrow J.,  Borrisov A.,  2020, \mn@doi [Journal of Open Source Software] {10.21105/joss.02430}, p.~2430

\bibitem[\protect\citeauthoryear{{Borrow} \& {Kelly}}{{Borrow} \& {Kelly}}{2021}]{2021arXiv210605281B}
{Borrow} J.,  {Kelly} A.~J.,  2021, arXiv e-prints, \href {https://ui.adsabs.harvard.edu/abs/2021arXiv210605281B} {p. arXiv:2106.05281}

\bibitem[\protect\citeauthoryear{{Borrow}, {Schaller}, {Bower}  \& {Schaye}}{{Borrow} et~al.}{2022}]{2022MNRAS.511.2367B}
{Borrow} J.,  {Schaller} M.,  {Bower} R.~G.,   {Schaye} J.,  2022, \mn@doi [\mnras] {10.1093/mnras/stab3166}, \href {https://ui.adsabs.harvard.edu/abs/2022MNRAS.511.2367B} {511, 2367}

\bibitem[\protect\citeauthoryear{{Borrow}, {Schaller}, {Bah{\'e}}, {Schaye}, {Ludlow}, {Ploeckinger}, {Nobels}  \& {Altamura}}{{Borrow} et~al.}{2023}]{2023MNRAS.tmp.2803B}
{Borrow} J.,  {Schaller} M.,  {Bah{\'e}} Y.~M.,  {Schaye} J.,  {Ludlow} A.~D.,  {Ploeckinger} S.,  {Nobels} F. S.~J.,   {Altamura} E.,  2023, \mn@doi [\mnras] {10.1093/mnras/stad2928}, \href {https://ui.adsabs.harvard.edu/abs/2023MNRAS.tmp.2803B} {}

\bibitem[\protect\citeauthoryear{{Bouwens}, {Illingworth}, {Oesch}, {Stefanon}, {Naidu}, {van Leeuwen}  \& {Magee}}{{Bouwens} et~al.}{2023}]{2023MNRAS.523.1009B}
{Bouwens} R.,  {Illingworth} G.,  {Oesch} P.,  {Stefanon} M.,  {Naidu} R.,  {van Leeuwen} I.,   {Magee} D.,  2023, \mn@doi [\mnras] {10.1093/mnras/stad1014}, \href {https://ui.adsabs.harvard.edu/abs/2023MNRAS.523.1009B} {523, 1009}

\bibitem[\protect\citeauthoryear{{Bower}, {Vernon}, {Goldstein}, {Benson}, {Lacey}, {Baugh}, {Cole}  \& {Frenk}}{{Bower} et~al.}{2010}]{2010MNRAS.407.2017B}
{Bower} R.~G.,  {Vernon} I.,  {Goldstein} M.,  {Benson} A.~J.,  {Lacey} C.~G.,  {Baugh} C.~M.,  {Cole} S.,   {Frenk} C.~S.,  2010, \mn@doi [\mnras] {10.1111/j.1365-2966.2010.16991.x}, \href {https://ui.adsabs.harvard.edu/abs/2010MNRAS.407.2017B} {407, 2017}

\bibitem[\protect\citeauthoryear{{Bregman} \& {Harrington}}{{Bregman} \& {Harrington}}{1986}]{Bregman1986}
{Bregman} J.~N.,  {Harrington} J.~P.,  1986, \mn@doi [\apj] {10.1086/164652}, \href {https://ui.adsabs.harvard.edu/abs/1986ApJ...309..833B} {309, 833}

\bibitem[\protect\citeauthoryear{{Bryan} \& {Norman}}{{Bryan} \& {Norman}}{1998}]{1998ApJ...495...80B}
{Bryan} G.~L.,  {Norman} M.~L.,  1998, \mn@doi [\apj] {10.1086/305262}, \href {https://ui.adsabs.harvard.edu/abs/1998ApJ...495...80B} {495, 80}

\bibitem[\protect\citeauthoryear{Bundy et~al.,}{Bundy et~al.}{2014}]{Bundy2015}
Bundy K.,  et~al., 2014, \mn@doi [The Astrophysical Journal] {10.1088/0004-637X/798/1/7}, 798, 7

\bibitem[\protect\citeauthoryear{{Cappellari} et~al.,}{{Cappellari} et~al.}{2012}]{2012Natur.484..485C}
{Cappellari} M.,  et~al., 2012, \mn@doi [\nat] {10.1038/nature10972}, \href {https://ui.adsabs.harvard.edu/abs/2012Natur.484..485C} {484, 485}

\bibitem[\protect\citeauthoryear{{Catinella} et~al.,}{{Catinella} et~al.}{2018}]{2018MNRAS.476..875C}
{Catinella} B.,  et~al., 2018, \mn@doi [\mnras] {10.1093/mnras/sty089}, \href {https://ui.adsabs.harvard.edu/abs/2018MNRAS.476..875C} {476, 875}

\bibitem[\protect\citeauthoryear{{Chabrier}}{{Chabrier}}{2003}]{Chabrier2003}
{Chabrier} G.,  2003, \mn@doi [\pasp] {10.1086/376392}, \href {https://ui.adsabs.harvard.edu/abs/2003PASP..115..763C} {115, 763}

\bibitem[\protect\citeauthoryear{{Chaikin}, {Schaye}, {Schaller}, {Bah{\'e}}, {Nobels}  \& {Ploeckinger}}{{Chaikin} et~al.}{2022}]{2022MNRAS.514..249C}
{Chaikin} E.,  {Schaye} J.,  {Schaller} M.,  {Bah{\'e}} Y.~M.,  {Nobels} F. S.~J.,   {Ploeckinger} S.,  2022, \mn@doi [\mnras] {10.1093/mnras/stac1132}, \href {https://ui.adsabs.harvard.edu/abs/2022MNRAS.514..249C} {514, 249}

\bibitem[\protect\citeauthoryear{{Chaikin}, {Schaye}, {Schaller}, {Ben{\'\i}tez-Llambay}, {Nobels}  \& {Ploeckinger}}{{Chaikin} et~al.}{2023}]{2023MNRAS.523.3709C}
{Chaikin} E.,  {Schaye} J.,  {Schaller} M.,  {Ben{\'\i}tez-Llambay} A.,  {Nobels} F. S.~J.,   {Ploeckinger} S.,  2023, \mn@doi [\mnras] {10.1093/mnras/stad1626}, \href {https://ui.adsabs.harvard.edu/abs/2023MNRAS.523.3709C} {523, 3709}

\bibitem[\protect\citeauthoryear{{Chang}, {van der Wel}, {da Cunha}  \& {Rix}}{{Chang} et~al.}{2015}]{2015ApJS..219....8C}
{Chang} Y.-Y.,  {van der Wel} A.,  {da Cunha} E.,   {Rix} H.-W.,  2015, \mn@doi [\apjs] {10.1088/0067-0049/219/1/8}, \href {https://ui.adsabs.harvard.edu/abs/2015ApJS..219....8C} {219, 8}

\bibitem[\protect\citeauthoryear{{Chen}, {Stark}, {Endsley}, {Topping}, {Whitler}  \& {Charlot}}{{Chen} et~al.}{2023}]{2023MNRAS.518.5607C}
{Chen} Z.,  {Stark} D.~P.,  {Endsley} R.,  {Topping} M.,  {Whitler} L.,   {Charlot} S.,  2023, \mn@doi [\mnras] {10.1093/mnras/stac3476}, \href {https://ui.adsabs.harvard.edu/abs/2023MNRAS.518.5607C} {518, 5607}

\bibitem[\protect\citeauthoryear{{Cochrane} et~al.,}{{Cochrane} et~al.}{2023}]{2023MNRAS.523.6082C}
{Cochrane} R.~K.,  et~al., 2023, \mn@doi [\mnras] {10.1093/mnras/stad1602}, \href {https://ui.adsabs.harvard.edu/abs/2023MNRAS.523.6082C} {523, 6082}

\bibitem[\protect\citeauthoryear{{Cook}, {Cortese}, {Catinella}  \& {Robotham}}{{Cook} et~al.}{2019}]{2019MNRAS.490.4060C}
{Cook} R. H.~W.,  {Cortese} L.,  {Catinella} B.,   {Robotham} A.,  2019, \mn@doi [\mnras] {10.1093/mnras/stz2789}, \href {https://ui.adsabs.harvard.edu/abs/2019MNRAS.490.4060C} {490, 4060}

\bibitem[\protect\citeauthoryear{{Covelo-Paz} et~al.,}{{Covelo-Paz} et~al.}{2025}]{2025A&A...694A.178C}
{Covelo-Paz} A.,  et~al., 2025, \mn@doi [\aap] {10.1051/0004-6361/202452363}, \href {https://ui.adsabs.harvard.edu/abs/2025A&A...694A.178C} {694, A178}

\bibitem[\protect\citeauthoryear{{Crain} \& {van de Voort}}{{Crain} \& {van de Voort}}{2023}]{2023ARA&A..61..473C}
{Crain} R.~A.,  {van de Voort} F.,  2023, \mn@doi [\araa] {10.1146/annurev-astro-041923-043618}, \href {https://ui.adsabs.harvard.edu/abs/2023ARA&A..61..473C} {61, 473}

\bibitem[\protect\citeauthoryear{{Crain} et~al.,}{{Crain} et~al.}{2015}]{2015MNRAS.450.1937C}
{Crain} R.~A.,  et~al., 2015, \mn@doi [\mnras] {10.1093/mnras/stv725}, \href {https://ui.adsabs.harvard.edu/abs/2015MNRAS.450.1937C} {450, 1937}

\bibitem[\protect\citeauthoryear{{Crain} et~al.,}{{Crain} et~al.}{2017}]{2017MNRAS.464.4204C}
{Crain} R.~A.,  et~al., 2017, \mn@doi [\mnras] {10.1093/mnras/stw2586}, \href {https://ui.adsabs.harvard.edu/abs/2017MNRAS.464.4204C} {464, 4204}

\bibitem[\protect\citeauthoryear{{Curti}, {Mannucci}, {Cresci}  \& {Maiolino}}{{Curti} et~al.}{2020}]{2020MNRAS.491..944C}
{Curti} M.,  {Mannucci} F.,  {Cresci} G.,   {Maiolino} R.,  2020, \mn@doi [\mnras] {10.1093/mnras/stz2910}, \href {https://ui.adsabs.harvard.edu/abs/2020MNRAS.491..944C} {491, 944}

\bibitem[\protect\citeauthoryear{{Dalgarno} \& {McCray}}{{Dalgarno} \& {McCray}}{1972}]{1972ARA&A..10..375D}
{Dalgarno} A.,  {McCray} R.~A.,  1972, \mn@doi [\araa] {10.1146/annurev.aa.10.090172.002111}, \href {https://ui.adsabs.harvard.edu/abs/1972ARA&A..10..375D} {10, 375}

\bibitem[\protect\citeauthoryear{{Dalla Vecchia} \& {Schaye}}{{Dalla Vecchia} \& {Schaye}}{2008}]{2008MNRAS.387.1431D}
{Dalla Vecchia} C.,  {Schaye} J.,  2008, \mn@doi [\mnras] {10.1111/j.1365-2966.2008.13322.x}, \href {https://ui.adsabs.harvard.edu/abs/2008MNRAS.387.1431D} {387, 1431}

\bibitem[\protect\citeauthoryear{{Dalla Vecchia} \& {Schaye}}{{Dalla Vecchia} \& {Schaye}}{2012}]{2012MNRAS.426..140D}
{Dalla Vecchia} C.,  {Schaye} J.,  2012, \mn@doi [\mnras] {10.1111/j.1365-2966.2012.21704.x}, \href {https://ui.adsabs.harvard.edu/abs/2012MNRAS.426..140D} {426, 140}

\bibitem[\protect\citeauthoryear{{Dav{\'e}}, {Thompson}  \& {Hopkins}}{{Dav{\'e}} et~al.}{2016}]{2016MNRAS.462.3265D}
{Dav{\'e}} R.,  {Thompson} R.,   {Hopkins} P.~F.,  2016, \mn@doi [\mnras] {10.1093/mnras/stw1862}, \href {https://ui.adsabs.harvard.edu/abs/2016MNRAS.462.3265D} {462, 3265}

\bibitem[\protect\citeauthoryear{{Dav{\'e}}, {Angl{\'e}s-Alc{\'a}zar}, {Narayanan}, {Li}, {Rafieferantsoa}  \& {Appleby}}{{Dav{\'e}} et~al.}{2019}]{2019MNRAS.486.2827D}
{Dav{\'e}} R.,  {Angl{\'e}s-Alc{\'a}zar} D.,  {Narayanan} D.,  {Li} Q.,  {Rafieferantsoa} M.~H.,   {Appleby} S.,  2019, \mn@doi [\mnras] {10.1093/mnras/stz937}, \href {https://ui.adsabs.harvard.edu/abs/2019MNRAS.486.2827D} {486, 2827}

\bibitem[\protect\citeauthoryear{{Dehnen} \& {Aly}}{{Dehnen} \& {Aly}}{2012}]{2012MNRAS.425.1068D}
{Dehnen} W.,  {Aly} H.,  2012, \mn@doi [\mnras] {10.1111/j.1365-2966.2012.21439.x}, \href {https://ui.adsabs.harvard.edu/abs/2012MNRAS.425.1068D} {425, 1068}

\bibitem[\protect\citeauthoryear{{Di Matteo}, {Colberg}, {Springel}, {Hernquist}  \& {Sijacki}}{{Di Matteo} et~al.}{2008}]{2008ApJ...676...33D}
{Di Matteo} T.,  {Colberg} J.,  {Springel} V.,  {Hernquist} L.,   {Sijacki} D.,  2008, \mn@doi [\apj] {10.1086/524921}, \href {https://ui.adsabs.harvard.edu/abs/2008ApJ...676...33D} {676, 33}

\bibitem[\protect\citeauthoryear{{Diemer} et~al.,}{{Diemer} et~al.}{2018}]{2018ApJS..238...33D}
{Diemer} B.,  et~al., 2018, \mn@doi [\apjs] {10.3847/1538-4365/aae387}, \href {https://ui.adsabs.harvard.edu/abs/2018ApJS..238...33D} {238, 33}

\bibitem[\protect\citeauthoryear{{Dolag} et~al.,}{{Dolag} et~al.}{2025}]{2025arXiv250401061D}
{Dolag} K.,  et~al., 2025, \mn@doi [arXiv e-prints] {10.48550/arXiv.2504.01061}, \href {https://ui.adsabs.harvard.edu/abs/2025arXiv250401061D} {p. arXiv:2504.01061}

\bibitem[\protect\citeauthoryear{{Donnan} et~al.,}{{Donnan} et~al.}{2024}]{2024MNRAS.533.3222D}
{Donnan} C.~T.,  et~al., 2024, \mn@doi [\mnras] {10.1093/mnras/stae2037}, \href {https://ui.adsabs.harvard.edu/abs/2024MNRAS.533.3222D} {533, 3222}

\bibitem[\protect\citeauthoryear{{Driver} et~al.,}{{Driver} et~al.}{2011}]{2011MNRAS.413..971D}
{Driver} S.~P.,  et~al., 2011, \mn@doi [\mnras] {10.1111/j.1365-2966.2010.18188.x}, \href {https://ui.adsabs.harvard.edu/abs/2011MNRAS.413..971D} {413, 971}

\bibitem[\protect\citeauthoryear{{Driver} et~al.,}{{Driver} et~al.}{2012}]{2012MNRAS.427.3244D}
{Driver} S.~P.,  et~al., 2012, \mn@doi [\mnras] {10.1111/j.1365-2966.2012.22036.x}, \href {https://ui.adsabs.harvard.edu/abs/2012MNRAS.427.3244D} {427, 3244}

\bibitem[\protect\citeauthoryear{{Driver} et~al.,}{{Driver} et~al.}{2022}]{2022MNRAS.513..439D}
{Driver} S.~P.,  et~al., 2022, \mn@doi [\mnras] {10.1093/mnras/stac472}, \href {https://ui.adsabs.harvard.edu/abs/2022MNRAS.513..439D} {513, 439}

\bibitem[\protect\citeauthoryear{{Dubois} et~al.,}{{Dubois} et~al.}{2014}]{2014MNRAS.444.1453D}
{Dubois} Y.,  et~al., 2014, \mn@doi [\mnras] {10.1093/mnras/stu1227}, \href {https://ui.adsabs.harvard.edu/abs/2014MNRAS.444.1453D} {444, 1453}

\bibitem[\protect\citeauthoryear{{Dubois} et~al.,}{{Dubois} et~al.}{2021}]{2021A&A...651A.109D}
{Dubois} Y.,  et~al., 2021, \mn@doi [\aap] {10.1051/0004-6361/202039429}, \href {https://ui.adsabs.harvard.edu/abs/2021A&A...651A.109D} {651, A109}

\bibitem[\protect\citeauthoryear{{Eddington}}{{Eddington}}{1913}]{1913MNRAS..73..359E}
{Eddington} A.~S.,  1913, \mn@doi [\mnras] {10.1093/mnras/73.5.359}, \href {https://ui.adsabs.harvard.edu/abs/1913MNRAS..73..359E} {73, 359}

\bibitem[\protect\citeauthoryear{{Eke}, {Cole}  \& {Frenk}}{{Eke} et~al.}{1996}]{1996MNRAS.282..263E}
{Eke} V.~R.,  {Cole} S.,   {Frenk} C.~S.,  1996, \mn@doi [\mnras] {10.1093/mnras/282.1.263}, \href {https://ui.adsabs.harvard.edu/abs/1996MNRAS.282..263E} {282, 263}

\bibitem[\protect\citeauthoryear{{Eldridge}, {Stanway}, {Xiao}, {McClelland }, {Taylor}, {Ng}, {Greis}  \& {Bray}}{{Eldridge} et~al.}{2017}]{BPASS2017}
{Eldridge} J.~J.,  {Stanway} E.~R.,  {Xiao} L.,  {McClelland } L.~A.~S.,  {Taylor} G.,  {Ng} M.,  {Greis} S.~M.~L.,   {Bray} J.~C.,  2017, \mn@doi [\pasa] {10.1017/pasa.2017.51}, \href {https://ui.adsabs.harvard.edu/abs/2017PASA...34...58E} {34, e058}

\bibitem[\protect\citeauthoryear{{Enia} et~al.,}{{Enia} et~al.}{2022}]{2022ApJ...927..204E}
{Enia} A.,  et~al., 2022, \mn@doi [\apj] {10.3847/1538-4357/ac51ca}, \href {https://ui.adsabs.harvard.edu/abs/2022ApJ...927..204E} {927, 204}

\bibitem[\protect\citeauthoryear{{Faucher-Gigu{\`e}re}}{{Faucher-Gigu{\`e}re}}{2020}]{2020MNRAS.493.1614F}
{Faucher-Gigu{\`e}re} C.-A.,  2020, \mn@doi [\mnras] {10.1093/mnras/staa302}, \href {https://ui.adsabs.harvard.edu/abs/2020MNRAS.493.1614F} {493, 1614}

\bibitem[\protect\citeauthoryear{{Feldmann} et~al.,}{{Feldmann} et~al.}{2023}]{2023MNRAS.522.3831F}
{Feldmann} R.,  et~al., 2023, \mn@doi [\mnras] {10.1093/mnras/stad1205}, \href {https://ui.adsabs.harvard.edu/abs/2023MNRAS.522.3831F} {522, 3831}

\bibitem[\protect\citeauthoryear{{Ferland}, {Korista}, {Verner}, {Ferguson}, {Kingdon}  \& {Verner}}{{Ferland} et~al.}{1998}]{1998PASP..110..761F}
{Ferland} G.~J.,  {Korista} K.~T.,  {Verner} D.~A.,  {Ferguson} J.~W.,  {Kingdon} J.~B.,   {Verner} E.~M.,  1998, \mn@doi [\pasp] {10.1086/316190}, \href {https://ui.adsabs.harvard.edu/abs/1998PASP..110..761F} {110, 761}

\bibitem[\protect\citeauthoryear{{Foreman-Mackey}, {Hogg}, {Lang}  \& {Goodman}}{{Foreman-Mackey} et~al.}{2013}]{2013PASP..125..306F}
{Foreman-Mackey} D.,  {Hogg} D.~W.,  {Lang} D.,   {Goodman} J.,  2013, \mn@doi [\pasp] {10.1086/670067}, \href {https://ui.adsabs.harvard.edu/abs/2013PASP..125..306F} {125, 306}

\bibitem[\protect\citeauthoryear{{Forouhar Moreno}, {Helly}, {McGibbon}, {Schaye}, {Schaller}, {Han}  \& {Kugel}}{{Forouhar Moreno} et~al.}{2025}]{2025arXiv250206932F}
{Forouhar Moreno} V.~J.,  {Helly} J.,  {McGibbon} R.,  {Schaye} J.,  {Schaller} M.,  {Han} J.,   {Kugel} R.,  2025, \mn@doi [arXiv e-prints] {10.48550/arXiv.2502.06932}, \href {https://ui.adsabs.harvard.edu/abs/2025arXiv250206932F} {p. arXiv:2502.06932}

\bibitem[\protect\citeauthoryear{{Fraser-McKelvie} et~al.,}{{Fraser-McKelvie} et~al.}{2022}]{2022MNRAS.510..320F}
{Fraser-McKelvie} A.,  et~al., 2022, \mn@doi [\mnras] {10.1093/mnras/stab3430}, \href {https://ui.adsabs.harvard.edu/abs/2022MNRAS.510..320F} {510, 320}

\bibitem[\protect\citeauthoryear{Fu et~al.,}{Fu et~al.}{2025}]{fu2025}
Fu S.,  et~al., 2025, \apj

\bibitem[\protect\citeauthoryear{{Gallazzi}, {Charlot}, {Brinchmann}, {White}  \& {Tremonti}}{{Gallazzi} et~al.}{2005}]{2005MNRAS.362...41G}
{Gallazzi} A.,  {Charlot} S.,  {Brinchmann} J.,  {White} S. D.~M.,   {Tremonti} C.~A.,  2005, \mn@doi [\mnras] {10.1111/j.1365-2966.2005.09321.x}, \href {https://ui.adsabs.harvard.edu/abs/2005MNRAS.362...41G} {362, 41}

\bibitem[\protect\citeauthoryear{Gardner et~al.,}{Gardner et~al.}{2006}]{JWST63183b28f893464d8ea2c9680f466d66}
Gardner J.,  et~al., 2006, \mn@doi [Space Science Reviews] {10.1007/s11214-006-8315-7}, 123, 485

\bibitem[\protect\citeauthoryear{{Gim{\'e}nez-Arteaga} et~al.,}{{Gim{\'e}nez-Arteaga} et~al.}{2023}]{2023ApJ...948..126G}
{Gim{\'e}nez-Arteaga} C.,  et~al., 2023, \mn@doi [\apj] {10.3847/1538-4357/acc5ea}, \href {https://ui.adsabs.harvard.edu/abs/2023ApJ...948..126G} {948, 126}

\bibitem[\protect\citeauthoryear{{Gnedin} \& {Kravtsov}}{{Gnedin} \& {Kravtsov}}{2011}]{2011ApJ...728...88G}
{Gnedin} N.~Y.,  {Kravtsov} A.~V.,  2011, \mn@doi [\apj] {10.1088/0004-637X/728/2/88}, \href {https://ui.adsabs.harvard.edu/abs/2011ApJ...728...88G} {728, 88}

\bibitem[\protect\citeauthoryear{{Goodman} \& {Weare}}{{Goodman} \& {Weare}}{2010}]{2010CAMCS...5...65G}
{Goodman} J.,  {Weare} J.,  2010, \mn@doi [Communications in Applied Mathematics and Computational Science] {10.2140/camcos.2010.5.65}, \href {https://ui.adsabs.harvard.edu/abs/2010CAMCS...5...65G} {5, 65}

\bibitem[\protect\citeauthoryear{{\VAN{Graaff}{de Graaff}{de Graaff}}, {Trayford}, {Franx}, {Schaller}, {Schaye}  \& {van der Wel}}{{\VAN{Graaff}{de Graaff}{de Graaff}} et~al.}{2022}]{2022MNRAS.511.2544D}
{\VAN{Graaff}{de Graaff}{de Graaff}} A.,  {Trayford} J.,  {Franx} M.,  {Schaller} M.,  {Schaye} J.,   {van der Wel} A.,  2022, \mn@doi [\mnras] {10.1093/mnras/stab3510}, \href {https://ui.adsabs.harvard.edu/abs/2022MNRAS.511.2544D} {511, 2544}

\bibitem[\protect\citeauthoryear{{Graham} \& {Sahu}}{{Graham} \& {Sahu}}{2023}]{2023MNRAS.518.2177G}
{Graham} A.~W.,  {Sahu} N.,  2023, \mn@doi [\mnras] {10.1093/mnras/stac2019}, \href {https://ui.adsabs.harvard.edu/abs/2023MNRAS.518.2177G} {518, 2177}

\bibitem[\protect\citeauthoryear{{Graham} \& {Sahu}}{{Graham} \& {Sahu}}{2024}]{2024MNRAS.530.3429G}
{Graham} A.~W.,  {Sahu} N.,  2024, \mn@doi [\mnras] {10.1093/mnras/stae1079}, \href {https://ui.adsabs.harvard.edu/abs/2024MNRAS.530.3429G} {530, 3429}

\bibitem[\protect\citeauthoryear{{Greengard} \& {Rokhlin}}{{Greengard} \& {Rokhlin}}{1987}]{1987JCoPh..73..325G}
{Greengard} L.,  {Rokhlin} V.,  1987, \mn@doi [Journal of Computational Physics] {10.1016/0021-9991(87)90140-9}, \href {https://ui.adsabs.harvard.edu/abs/1987JCoPh..73..325G} {73, 325}

\bibitem[\protect\citeauthoryear{{Gruppioni} et~al.,}{{Gruppioni} et~al.}{2020}]{2020A&A...643A...8G}
{Gruppioni} C.,  et~al., 2020, \mn@doi [\aap] {10.1051/0004-6361/202038487}, \href {https://ui.adsabs.harvard.edu/abs/2020A&A...643A...8G} {643, A8}

\bibitem[\protect\citeauthoryear{{Hahn}, {Michaux}, {Rampf}, {Uhlemann}  \& {Angulo}}{{Hahn} et~al.}{2020}]{2020ascl.soft08024H}
{Hahn} O.,  {Michaux} M.,  {Rampf} C.,  {Uhlemann} C.,   {Angulo} R.~E.,  2020, {MUSIC2-monofonIC: 3LPT initial condition generator}, Astrophysics Source Code Library, record ascl:2008.024 (\mn@eprint {ascl} {2008.024})

\bibitem[\protect\citeauthoryear{{Hahn}, {Rampf}  \& {Uhlemann}}{{Hahn} et~al.}{2021}]{2021MNRAS.503..426H}
{Hahn} O.,  {Rampf} C.,   {Uhlemann} C.,  2021, \mn@doi [\mnras] {10.1093/mnras/staa3773}, \href {https://ui.adsabs.harvard.edu/abs/2021MNRAS.503..426H} {503, 426}

\bibitem[\protect\citeauthoryear{{Han}, {Cole}, {Frenk}, {Benitez-Llambay}  \& {Helly}}{{Han} et~al.}{2018}]{2018MNRAS.474..604H}
{Han} J.,  {Cole} S.,  {Frenk} C.~S.,  {Benitez-Llambay} A.,   {Helly} J.,  2018, \mn@doi [\mnras] {10.1093/mnras/stx2792}, \href {https://ui.adsabs.harvard.edu/abs/2018MNRAS.474..604H} {474, 604}

\bibitem[\protect\citeauthoryear{{Hardwick}, {Cortese}, {Obreschkow}, {Catinella}  \& {Cook}}{{Hardwick} et~al.}{2022}]{2022MNRAS.509.3751H}
{Hardwick} J.~A.,  {Cortese} L.,  {Obreschkow} D.,  {Catinella} B.,   {Cook} R. H.~W.,  2022, \mn@doi [\mnras] {10.1093/mnras/stab3261}, \href {https://ui.adsabs.harvard.edu/abs/2022MNRAS.509.3751H} {509, 3751}

\bibitem[\protect\citeauthoryear{{Harikane} et~al.,}{{Harikane} et~al.}{2023}]{2023ApJS..265....5H}
{Harikane} Y.,  et~al., 2023, \mn@doi [\apjs] {10.3847/1538-4365/acaaa9}, \href {https://ui.adsabs.harvard.edu/abs/2023ApJS..265....5H} {265, 5}

\bibitem[\protect\citeauthoryear{{Harris} et~al.,}{{Harris} et~al.}{2020}]{2020Natur.585..357H}
{Harris} C.~R.,  et~al., 2020, \mn@doi [\nat] {10.1038/s41586-020-2649-2}, \href {https://ui.adsabs.harvard.edu/abs/2020Natur.585..357H} {585, 357}

\bibitem[\protect\citeauthoryear{{Henden}, {Puchwein}, {Shen}  \& {Sijacki}}{{Henden} et~al.}{2018}]{2018MNRAS.479.5385H}
{Henden} N.~A.,  {Puchwein} E.,  {Shen} S.,   {Sijacki} D.,  2018, \mn@doi [\mnras] {10.1093/mnras/sty1780}, \href {https://ui.adsabs.harvard.edu/abs/2018MNRAS.479.5385H} {479, 5385}

\bibitem[\protect\citeauthoryear{{Hopkins}}{{Hopkins}}{2015}]{2015MNRAS.450...53H}
{Hopkins} P.~F.,  2015, \mn@doi [\mnras] {10.1093/mnras/stv195}, \href {https://ui.adsabs.harvard.edu/abs/2015MNRAS.450...53H} {450, 53}

\bibitem[\protect\citeauthoryear{{Hopkins} et~al.,}{{Hopkins} et~al.}{2018}]{2018MNRAS.480..800H}
{Hopkins} P.~F.,  et~al., 2018, \mn@doi [\mnras] {10.1093/mnras/sty1690}, \href {https://ui.adsabs.harvard.edu/abs/2018MNRAS.480..800H} {480, 800}

\bibitem[\protect\citeauthoryear{{Hunter}}{{Hunter}}{2007}]{2007CSE.....9...90H}
{Hunter} J.~D.,  2007, \mn@doi [Computing in Science and Engineering] {10.1109/MCSE.2007.55}, \href {https://ui.adsabs.harvard.edu/abs/2007CSE.....9...90H} {9, 90}

\bibitem[\protect\citeauthoryear{{Hu{\v{s}}ko} et~al.,}{{Hu{\v{s}}ko} et~al.}{2025}]{2025arXiv250905179H}
{Hu{\v{s}}ko} F.,  et~al., 2025, \mn@doi [arXiv e-prints] {10.48550/arXiv.2509.05179}, \href {https://ui.adsabs.harvard.edu/abs/2025arXiv250905179H} {p. arXiv:2509.05179}

\bibitem[\protect\citeauthoryear{{Ikeda} et~al.,}{{Ikeda} et~al.}{2022}]{2022ApJ...933...11I}
{Ikeda} R.,  et~al., 2022, \mn@doi [\apj] {10.3847/1538-4357/ac6cdc}, \href {https://ui.adsabs.harvard.edu/abs/2022ApJ...933...11I} {933, 11}

\bibitem[\protect\citeauthoryear{{Katsianis} et~al.,}{{Katsianis} et~al.}{2020}]{2020MNRAS.492.5592K}
{Katsianis} A.,  et~al., 2020, \mn@doi [\mnras] {10.1093/mnras/staa157}, \href {https://ui.adsabs.harvard.edu/abs/2020MNRAS.492.5592K} {492, 5592}

\bibitem[\protect\citeauthoryear{{Kaviraj}, {Lazar}, {Watkins}, {Laigle}, {Martin}  \& {Jackson}}{{Kaviraj} et~al.}{2025}]{2025MNRAS.538..153K}
{Kaviraj} S.,  {Lazar} I.,  {Watkins} A.~E.,  {Laigle} C.,  {Martin} G.,   {Jackson} R.~A.,  2025, \mn@doi [\mnras] {10.1093/mnras/staf233}, \href {https://ui.adsabs.harvard.edu/abs/2025MNRAS.538..153K} {538, 153}

\bibitem[\protect\citeauthoryear{{Kennicutt}}{{Kennicutt}}{1998}]{1998ApJ...498..541K}
{Kennicutt} Robert~C. J.,  1998, \mn@doi [\apj] {10.1086/305588}, \href {https://ui.adsabs.harvard.edu/abs/1998ApJ...498..541K} {498, 541}

\bibitem[\protect\citeauthoryear{{Kennicutt} Robert~C. et~al.,}{{Kennicutt} et~al.}{2007}]{2007ApJ...671..333K}
{Kennicutt} Robert~C. J.,  et~al., 2007, \mn@doi [\apj] {10.1086/522300}, \href {https://ui.adsabs.harvard.edu/abs/2007ApJ...671..333K} {671, 333}

\bibitem[\protect\citeauthoryear{{Khusanova} et~al.,}{{Khusanova} et~al.}{2021}]{2021A&A...649A.152K}
{Khusanova} Y.,  et~al., 2021, \mn@doi [\aap] {10.1051/0004-6361/202038944}, \href {https://ui.adsabs.harvard.edu/abs/2021A&A...649A.152K} {649, A152}

\bibitem[\protect\citeauthoryear{{Kirby}, {Cohen}, {Guhathakurta}, {Cheng}, {Bullock}  \& {Gallazzi}}{{Kirby} et~al.}{2013}]{2013ApJ...779..102K}
{Kirby} E.~N.,  {Cohen} J.~G.,  {Guhathakurta} P.,  {Cheng} L.,  {Bullock} J.~S.,   {Gallazzi} A.,  2013, \mn@doi [\apj] {10.1088/0004-637X/779/2/102}, \href {https://ui.adsabs.harvard.edu/abs/2013ApJ...779..102K} {779, 102}

\bibitem[\protect\citeauthoryear{{Krumholz}}{{Krumholz}}{2013}]{2013MNRAS.436.2747K}
{Krumholz} M.~R.,  2013, \mn@doi [\mnras] {10.1093/mnras/stt1780}, \href {https://ui.adsabs.harvard.edu/abs/2013MNRAS.436.2747K} {436, 2747}

\bibitem[\protect\citeauthoryear{{Krumholz} \& {Gnedin}}{{Krumholz} \& {Gnedin}}{2011}]{2011ApJ...729...36K}
{Krumholz} M.~R.,  {Gnedin} N.~Y.,  2011, \mn@doi [\apj] {10.1088/0004-637X/729/1/36}, \href {https://ui.adsabs.harvard.edu/abs/2011ApJ...729...36K} {729, 36}

\bibitem[\protect\citeauthoryear{{Krumholz}, {McKee}  \& {Klein}}{{Krumholz} et~al.}{2006}]{Krumholz_et_al_2006}
{Krumholz} M.~R.,  {McKee} C.~F.,   {Klein} R.~I.,  2006, \mn@doi [\apj] {10.1086/498844}, \href {https://ui.adsabs.harvard.edu/abs/2006ApJ...638..369K} {638, 369}

\bibitem[\protect\citeauthoryear{Kugel \& Borrow}{Kugel \& Borrow}{2022}]{Kugel2022}
Kugel R.,  Borrow J.,  2022, \mn@doi [Journal of Open Source Software] {10.21105/joss.04240}, 7, 4240

\bibitem[\protect\citeauthoryear{{Kugel} et~al.,}{{Kugel} et~al.}{2023}]{2023MNRAS.526.6103K}
{Kugel} R.,  et~al., 2023, \mn@doi [\mnras] {10.1093/mnras/stad2540}, \href {https://ui.adsabs.harvard.edu/abs/2023MNRAS.526.6103K} {526, 6103}

\bibitem[\protect\citeauthoryear{{Lagos} et~al.,}{{Lagos} et~al.}{2015}]{2015MNRAS.452.3815L}
{Lagos} C. d.~P.,  et~al., 2015, \mn@doi [\mnras] {10.1093/mnras/stv1488}, \href {https://ui.adsabs.harvard.edu/abs/2015MNRAS.452.3815L} {452, 3815}

\bibitem[\protect\citeauthoryear{{Le Brun}, {McCarthy}, {Schaye}  \& {Ponman}}{{Le Brun} et~al.}{2014}]{2014MNRAS.441.1270L}
{Le Brun} A. M.~C.,  {McCarthy} I.~G.,  {Schaye} J.,   {Ponman} T.~J.,  2014, \mn@doi [\mnras] {10.1093/mnras/stu608}, \href {https://ui.adsabs.harvard.edu/abs/2014MNRAS.441.1270L} {441, 1270}

\bibitem[\protect\citeauthoryear{{Le F{\`e}vre} et~al.,}{{Le F{\`e}vre} et~al.}{2020}]{2020A&A...643A...1L}
{Le F{\`e}vre} O.,  et~al., 2020, \mn@doi [\aap] {10.1051/0004-6361/201936965}, \href {https://ui.adsabs.harvard.edu/abs/2020A&A...643A...1L} {643, A1}

\bibitem[\protect\citeauthoryear{{Leja}, {Speagle}, {Johnson}, {Conroy}, {van Dokkum}  \& {Franx}}{{Leja} et~al.}{2020}]{2020ApJ...893..111L}
{Leja} J.,  {Speagle} J.~S.,  {Johnson} B.~D.,  {Conroy} C.,  {van Dokkum} P.,   {Franx} M.,  2020, \mn@doi [\apj] {10.3847/1538-4357/ab7e27}, \href {https://ui.adsabs.harvard.edu/abs/2020ApJ...893..111L} {893, 111}

\bibitem[\protect\citeauthoryear{{Leja} et~al.,}{{Leja} et~al.}{2022}]{2022ApJ...936..165L}
{Leja} J.,  et~al., 2022, \mn@doi [\apj] {10.3847/1538-4357/ac887d}, \href {https://ui.adsabs.harvard.edu/abs/2022ApJ...936..165L} {936, 165}

\bibitem[\protect\citeauthoryear{{Li} et~al.,}{{Li} et~al.}{2017}]{2017ApJ...838...77L}
{Li} H.,  et~al., 2017, \mn@doi [\apj] {10.3847/1538-4357/aa662a}, \href {https://ui.adsabs.harvard.edu/abs/2017ApJ...838...77L} {838, 77}

\bibitem[\protect\citeauthoryear{{L{\'o}pez-S{\'a}nchez}, {Dopita}, {Kewley}, {Zahid}, {Nicholls}  \& {Scharw{\"a}chter}}{{L{\'o}pez-S{\'a}nchez} et~al.}{2012}]{2012MNRAS.426.2630L}
{L{\'o}pez-S{\'a}nchez} {\'A}.~R.,  {Dopita} M.~A.,  {Kewley} L.~J.,  {Zahid} H.~J.,  {Nicholls} D.~C.,   {Scharw{\"a}chter} J.,  2012, \mn@doi [\mnras] {10.1111/j.1365-2966.2012.21145.x}, \href {https://ui.adsabs.harvard.edu/abs/2012MNRAS.426.2630L} {426, 2630}

\bibitem[\protect\citeauthoryear{{Ludlow}, {Schaye}  \& {Bower}}{{Ludlow} et~al.}{2019}]{2019MNRAS.488.3663L}
{Ludlow} A.~D.,  {Schaye} J.,   {Bower} R.,  2019, \mn@doi [\mnras] {10.1093/mnras/stz1821}, \href {https://ui.adsabs.harvard.edu/abs/2019MNRAS.488.3663L} {488, 3663}

\bibitem[\protect\citeauthoryear{{Ludlow}, {Fall}, {Schaye}  \& {Obreschkow}}{{Ludlow} et~al.}{2021}]{2021MNRAS.508.5114L}
{Ludlow} A.~D.,  {Fall} S.~M.,  {Schaye} J.,   {Obreschkow} D.,  2021, \mn@doi [\mnras] {10.1093/mnras/stab2770}, \href {https://ui.adsabs.harvard.edu/abs/2021MNRAS.508.5114L} {508, 5114}

\bibitem[\protect\citeauthoryear{{Ludlow}, {Fall}, {Wilkinson}, {Schaye}  \& {Obreschkow}}{{Ludlow} et~al.}{2023}]{2023MNRAS.525.5614L}
{Ludlow} A.~D.,  {Fall} S.~M.,  {Wilkinson} M.~J.,  {Schaye} J.,   {Obreschkow} D.,  2023, \mn@doi [\mnras] {10.1093/mnras/stad2615}, \href {https://ui.adsabs.harvard.edu/abs/2023MNRAS.525.5614L} {525, 5614}

\bibitem[\protect\citeauthoryear{{Madau} \& {Dickinson}}{{Madau} \& {Dickinson}}{2014}]{2014ARA&A..52..415M}
{Madau} P.,  {Dickinson} M.,  2014, \mn@doi [\araa] {10.1146/annurev-astro-081811-125615}, \href {https://ui.adsabs.harvard.edu/abs/2014ARA&A..52..415M} {52, 415}

\bibitem[\protect\citeauthoryear{{Manuwal} \& {Stevens}}{{Manuwal} \& {Stevens}}{2023}]{2023MNRAS.523.2738M}
{Manuwal} A.,  {Stevens} A. R.~H.,  2023, \mn@doi [\mnras] {10.1093/mnras/stad1587}, \href {https://ui.adsabs.harvard.edu/abs/2023MNRAS.523.2738M} {523, 2738}

\bibitem[\protect\citeauthoryear{{Mart{\'\i}n-Navarro}, {La Barbera}, {Vazdekis}, {Falc{\'o}n-Barroso}  \& {Ferreras}}{{Mart{\'\i}n-Navarro} et~al.}{2015}]{2015MNRAS.447.1033M}
{Mart{\'\i}n-Navarro} I.,  {La Barbera} F.,  {Vazdekis} A.,  {Falc{\'o}n-Barroso} J.,   {Ferreras} I.,  2015, \mn@doi [\mnras] {10.1093/mnras/stu2480}, \href {https://ui.adsabs.harvard.edu/abs/2015MNRAS.447.1033M} {447, 1033}

\bibitem[\protect\citeauthoryear{{Mathis}, {Mezger}  \& {Panagia}}{{Mathis} et~al.}{1983}]{Mathis1983}
{Mathis} J.~S.,  {Mezger} P.~G.,   {Panagia} N.,  1983, \aap, \href {https://ui.adsabs.harvard.edu/abs/1983A&A...128..212M} {128, 212}

\bibitem[\protect\citeauthoryear{{McCarthy}, {Schaye}, {Bird}  \& {Le Brun}}{{McCarthy} et~al.}{2017}]{2017MNRAS.465.2936M}
{McCarthy} I.~G.,  {Schaye} J.,  {Bird} S.,   {Le Brun} A. M.~C.,  2017, \mn@doi [\mnras] {10.1093/mnras/stw2792}, \href {https://ui.adsabs.harvard.edu/abs/2017MNRAS.465.2936M} {465, 2936}

\bibitem[\protect\citeauthoryear{{McGibbon}, {Helly}, {Schaye}, {Schaller}  \& {Vandenbroucke}}{{McGibbon} et~al.}{2025}]{2025JOSS...10.8252M}
{McGibbon} R.,  {Helly} J.,  {Schaye} J.,  {Schaller} M.,   {Vandenbroucke} B.,  2025, \mn@doi [The Journal of Open Source Software] {10.21105/joss.08252}, \href {https://ui.adsabs.harvard.edu/abs/2025JOSS...10.8252M} {10, 8252}

\bibitem[\protect\citeauthoryear{McKay, Beckman  \& Conover}{McKay et~al.}{1979}]{ef76b040-2f28-37ba-b0c4-02ed99573416}
McKay M.~D.,  Beckman R.~J.,   Conover W.~J.,  1979, Technometrics, 21, 239

\bibitem[\protect\citeauthoryear{{Michaux}, {Hahn}, {Rampf}  \& {Angulo}}{{Michaux} et~al.}{2021}]{2021MNRAS.500..663M}
{Michaux} M.,  {Hahn} O.,  {Rampf} C.,   {Angulo} R.~E.,  2021, \mn@doi [\mnras] {10.1093/mnras/staa3149}, \href {https://ui.adsabs.harvard.edu/abs/2021MNRAS.500..663M} {500, 663}

\bibitem[\protect\citeauthoryear{{Moster}, {Naab}  \& {White}}{{Moster} et~al.}{2018}]{2018MNRAS.477.1822M}
{Moster} B.~P.,  {Naab} T.,   {White} S. D.~M.,  2018, \mn@doi [\mnras] {10.1093/mnras/sty655}, \href {https://ui.adsabs.harvard.edu/abs/2018MNRAS.477.1822M} {477, 1822}

\bibitem[\protect\citeauthoryear{{Muzzin} et~al.,}{{Muzzin} et~al.}{2013}]{2013ApJ...777...18M}
{Muzzin} A.,  et~al., 2013, \mn@doi [\apj] {10.1088/0004-637X/777/1/18}, \href {https://ui.adsabs.harvard.edu/abs/2013ApJ...777...18M} {777, 18}

\bibitem[\protect\citeauthoryear{{Nobels}, {Schaye}, {Schaller}, {Ploeckinger}, {Chaikin}  \& {Richings}}{{Nobels} et~al.}{2024}]{2024MNRAS.532.3299N}
{Nobels} F. S.~J.,  {Schaye} J.,  {Schaller} M.,  {Ploeckinger} S.,  {Chaikin} E.,   {Richings} A.~J.,  2024, \mn@doi [\mnras] {10.1093/mnras/stae1390}, \href {https://ui.adsabs.harvard.edu/abs/2024MNRAS.532.3299N} {532, 3299}

\bibitem[\protect\citeauthoryear{{Novak} et~al.,}{{Novak} et~al.}{2017}]{2017A&A...602A...5N}
{Novak} M.,  et~al., 2017, \mn@doi [\aap] {10.1051/0004-6361/201629436}, \href {https://ui.adsabs.harvard.edu/abs/2017A&A...602A...5N} {602, A5}

\bibitem[\protect\citeauthoryear{{Ostriker}}{{Ostriker}}{1999}]{1999ApJ...513..252O}
{Ostriker} E.~C.,  1999, \mn@doi [\apj] {10.1086/306858}, \href {https://ui.adsabs.harvard.edu/abs/1999ApJ...513..252O} {513, 252}

\bibitem[\protect\citeauthoryear{{Pakmor} et~al.,}{{Pakmor} et~al.}{2023}]{2023MNRAS.524.2539P}
{Pakmor} R.,  et~al., 2023, \mn@doi [\mnras] {10.1093/mnras/stac3620}, \href {https://ui.adsabs.harvard.edu/abs/2023MNRAS.524.2539P} {524, 2539}

\bibitem[\protect\citeauthoryear{{P{\'e}roux} \& {Howk}}{{P{\'e}roux} \& {Howk}}{2020}]{2020ARA&A..58..363P}
{P{\'e}roux} C.,  {Howk} J.~C.,  2020, \mn@doi [\araa] {10.1146/annurev-astro-021820-120014}, \href {https://ui.adsabs.harvard.edu/abs/2020ARA&A..58..363P} {58, 363}

\bibitem[\protect\citeauthoryear{{Pillepich} et~al.,}{{Pillepich} et~al.}{2018}]{2018MNRAS.473.4077P}
{Pillepich} A.,  et~al., 2018, \mn@doi [\mnras] {10.1093/mnras/stx2656}, \href {https://ui.adsabs.harvard.edu/abs/2018MNRAS.473.4077P} {473, 4077}

\bibitem[\protect\citeauthoryear{{Ploeckinger} \& {Schaye}}{{Ploeckinger} \& {Schaye}}{2020}]{2020MNRAS.497.4857P}
{Ploeckinger} S.,  {Schaye} J.,  2020, \mn@doi [\mnras] {10.1093/mnras/staa2172}, \href {https://ui.adsabs.harvard.edu/abs/2020MNRAS.497.4857P} {497, 4857}

\bibitem[\protect\citeauthoryear{{Ploeckinger}, {Nobels}, {Schaller}  \& {Schaye}}{{Ploeckinger} et~al.}{2024}]{2024MNRAS.528.2930P}
{Ploeckinger} S.,  {Nobels} F. S.~J.,  {Schaller} M.,   {Schaye} J.,  2024, \mn@doi [\mnras] {10.1093/mnras/stad3935}, \href {https://ui.adsabs.harvard.edu/abs/2024MNRAS.528.2930P} {528, 2930}

\bibitem[\protect\citeauthoryear{{Ploeckinger}, {Richings}, {Schaye}, {Trayford}, {Schaller}  \& {Chaikin}}{{Ploeckinger} et~al.}{2025}]{2025arXiv250615773P}
{Ploeckinger} S.,  {Richings} A.~J.,  {Schaye} J.,  {Trayford} J.~W.,  {Schaller} M.,   {Chaikin} E.,  2025, arXiv e-prints, \href {https://ui.adsabs.harvard.edu/abs/2025arXiv250615773P} {p. arXiv:2506.15773}

\bibitem[\protect\citeauthoryear{{Portinari}, {Chiosi}  \& {Bressan}}{{Portinari} et~al.}{1998}]{Portinari1998}
{Portinari} L.,  {Chiosi} C.,   {Bressan} A.,  1998, \aap, \href {https://ui.adsabs.harvard.edu/abs/1998A&A...334..505P} {334, 505}

\bibitem[\protect\citeauthoryear{{Rahmati}, {Pawlik}, {Rai{\v{c}}evi{\'c}}  \& {Schaye}}{{Rahmati} et~al.}{2013}]{2013MNRAS.430.2427R}
{Rahmati} A.,  {Pawlik} A.~H.,  {Rai{\v{c}}evi{\'c}} M.,   {Schaye} J.,  2013, \mn@doi [\mnras] {10.1093/mnras/stt066}, \href {https://ui.adsabs.harvard.edu/abs/2013MNRAS.430.2427R} {430, 2427}

\bibitem[\protect\citeauthoryear{{Ramos Almeida} et~al.,}{{Ramos Almeida} et~al.}{2022}]{2022A&A...658A.155R}
{Ramos Almeida} C.,  et~al., 2022, \mn@doi [\aap] {10.1051/0004-6361/202141906}, \href {https://ui.adsabs.harvard.edu/abs/2022A&A...658A.155R} {658, A155}

\bibitem[\protect\citeauthoryear{Rasmussen \& Williams}{Rasmussen \& Williams}{2006}]{books/lib/RasmussenW06}
Rasmussen C.~E.,  Williams C. K.~I.,  2006, Gaussian processes for machine learning..
Adaptive computation and machine learning, MIT Press

\bibitem[\protect\citeauthoryear{{Richings}, {Schaye}  \& {Oppenheimer}}{{Richings} et~al.}{2014a}]{2014MNRAS.440.3349R}
{Richings} A.~J.,  {Schaye} J.,   {Oppenheimer} B.~D.,  2014a, \mn@doi [\mnras] {10.1093/mnras/stu525}, \href {https://ui.adsabs.harvard.edu/abs/2014MNRAS.440.3349R} {440, 3349}

\bibitem[\protect\citeauthoryear{{Richings}, {Schaye}  \& {Oppenheimer}}{{Richings} et~al.}{2014b}]{2014MNRAS.442.2780R}
{Richings} A.~J.,  {Schaye} J.,   {Oppenheimer} B.~D.,  2014b, \mn@doi [\mnras] {10.1093/mnras/stu1046}, \href {https://ui.adsabs.harvard.edu/abs/2014MNRAS.442.2780R} {442, 2780}

\bibitem[\protect\citeauthoryear{{Robertson} \& {Kravtsov}}{{Robertson} \& {Kravtsov}}{2008}]{2008ApJ...680.1083R}
{Robertson} B.~E.,  {Kravtsov} A.~V.,  2008, \mn@doi [\apj] {10.1086/587796}, \href {https://ui.adsabs.harvard.edu/abs/2008ApJ...680.1083R} {680, 1083}

\bibitem[\protect\citeauthoryear{{Robotham}, {Bellstedt}, {Lagos}, {Thorne}, {Davies}, {Driver}  \& {Bravo}}{{Robotham} et~al.}{2020}]{2020MNRAS.495..905R}
{Robotham} A.~S.~G.,  {Bellstedt} S.,  {Lagos} C. d.~P.,  {Thorne} J.~E.,  {Davies} L.~J.,  {Driver} S.~P.,   {Bravo} M.,  2020, \mn@doi [\mnras] {10.1093/mnras/staa1116}, \href {https://ui.adsabs.harvard.edu/abs/2020MNRAS.495..905R} {495, 905}

\bibitem[\protect\citeauthoryear{{Saintonge} et~al.,}{{Saintonge} et~al.}{2017}]{2017ApJS..233...22S}
{Saintonge} A.,  et~al., 2017, \mn@doi [\apjs] {10.3847/1538-4365/aa97e0}, \href {https://ui.adsabs.harvard.edu/abs/2017ApJS..233...22S} {233, 22}

\bibitem[\protect\citeauthoryear{{Schaller}, {Dalla Vecchia}, {Schaye}, {Bower}, {Theuns}, {Crain}, {Furlong}  \& {McCarthy}}{{Schaller} et~al.}{2015}]{2015MNRAS.454.2277S}
{Schaller} M.,  {Dalla Vecchia} C.,  {Schaye} J.,  {Bower} R.~G.,  {Theuns} T.,  {Crain} R.~A.,  {Furlong} M.,   {McCarthy} I.~G.,  2015, \mn@doi [\mnras] {10.1093/mnras/stv2169}, \href {https://ui.adsabs.harvard.edu/abs/2015MNRAS.454.2277S} {454, 2277}

\bibitem[\protect\citeauthoryear{{Schaller} et~al.}{{Schaller} et~al.}{2018}]{2018ascl.soft05020S}
{Schaller} M.,  et~al., 2018, {SWIFT: SPH With Inter-dependent Fine-grained Tasking}, Astrophysics Source Code Library (\mn@eprint {ascl} {1805.020})

\bibitem[\protect\citeauthoryear{{Schaller} et~al.,}{{Schaller} et~al.}{2024}]{2024MNRAS.530.2378S}
{Schaller} M.,  et~al., 2024, \mn@doi [\mnras] {10.1093/mnras/stae922}, \href {https://ui.adsabs.harvard.edu/abs/2024MNRAS.530.2378S} {530, 2378}

\bibitem[\protect\citeauthoryear{{Schaye} \& {Dalla Vecchia}}{{Schaye} \& {Dalla Vecchia}}{2008}]{2008MNRAS.383.1210S}
{Schaye} J.,  {Dalla Vecchia} C.,  2008, \mn@doi [\mnras] {10.1111/j.1365-2966.2007.12639.x}, \href {https://ui.adsabs.harvard.edu/abs/2008MNRAS.383.1210S} {383, 1210}

\bibitem[\protect\citeauthoryear{{Schaye} et~al.,}{{Schaye} et~al.}{2010}]{2010MNRAS.402.1536S}
{Schaye} J.,  et~al., 2010, \mn@doi [\mnras] {10.1111/j.1365-2966.2009.16029.x}, \href {https://ui.adsabs.harvard.edu/abs/2010MNRAS.402.1536S} {402, 1536}

\bibitem[\protect\citeauthoryear{{Schaye} et~al.,}{{Schaye} et~al.}{2015}]{2015MNRAS.446..521S}
{Schaye} J.,  et~al., 2015, \mn@doi [\mnras] {10.1093/mnras/stu2058}, \href {https://ui.adsabs.harvard.edu/abs/2015MNRAS.446..521S} {446, 521}

\bibitem[\protect\citeauthoryear{{Schaye} et~al.,}{{Schaye} et~al.}{2023}]{2023MNRAS.526.4978S}
{Schaye} J.,  et~al., 2023, \mn@doi [\mnras] {10.1093/mnras/stad2419}, \href {https://ui.adsabs.harvard.edu/abs/2023MNRAS.526.4978S} {526, 4978}

\bibitem[\protect\citeauthoryear{{Schaye} et~al.,}{{Schaye} et~al.}{2025}]{schaye2025colibreproject}
{Schaye} J.,  et~al., 2025, \mn@doi [arXiv e-prints] {10.48550/arXiv.2508.21126}, \href {https://ui.adsabs.harvard.edu/abs/2025arXiv250821126S} {p. arXiv:2508.21126}

\bibitem[\protect\citeauthoryear{{Schechter}}{{Schechter}}{1976}]{1976ApJ...203..297S}
{Schechter} P.,  1976, \mn@doi [\apj] {10.1086/154079}, \href {https://ui.adsabs.harvard.edu/abs/1976ApJ...203..297S} {203, 297}

\bibitem[\protect\citeauthoryear{{Schmidt}}{{Schmidt}}{1959}]{1959ApJ...129..243S}
{Schmidt} M.,  1959, \mn@doi [\apj] {10.1086/146614}, \href {https://ui.adsabs.harvard.edu/abs/1959ApJ...129..243S} {129, 243}

\bibitem[\protect\citeauthoryear{{Scholte} et~al.,}{{Scholte} et~al.}{2024}]{2024MNRAS.535.2341S}
{Scholte} D.,  et~al., 2024, \mn@doi [\mnras] {10.1093/mnras/stae2477}, \href {https://ui.adsabs.harvard.edu/abs/2024MNRAS.535.2341S} {535, 2341}

\bibitem[\protect\citeauthoryear{{Segers}, {Schaye}, {Bower}, {Crain}, {Schaller}  \& {Theuns}}{{Segers} et~al.}{2016}]{2016MNRAS.461L.102S}
{Segers} M.~C.,  {Schaye} J.,  {Bower} R.~G.,  {Crain} R.~A.,  {Schaller} M.,   {Theuns} T.,  2016, \mn@doi [\mnras] {10.1093/mnrasl/slw111}, \href {https://ui.adsabs.harvard.edu/abs/2016MNRAS.461L.102S} {461, L102}

\bibitem[\protect\citeauthoryear{{Shakura} \& {Sunyaev}}{{Shakura} \& {Sunyaev}}{1973}]{1973A&A....24..337S}
{Shakura} N.~I.,  {Sunyaev} R.~A.,  1973, \aap, \href {https://ui.adsabs.harvard.edu/abs/1973A&A....24..337S} {24, 337}

\bibitem[\protect\citeauthoryear{{Smith} et~al.,}{{Smith} et~al.}{2024}]{2024MNRAS.527.1216S}
{Smith} M.~C.,  et~al., 2024, \mn@doi [\mnras] {10.1093/mnras/stad3168}, \href {https://ui.adsabs.harvard.edu/abs/2024MNRAS.527.1216S} {527, 1216}

\bibitem[\protect\citeauthoryear{{Springel} \& {Hernquist}}{{Springel} \& {Hernquist}}{2003}]{2003MNRAS.339..289S}
{Springel} V.,  {Hernquist} L.,  2003, \mn@doi [\mnras] {10.1046/j.1365-8711.2003.06206.x}, \href {https://ui.adsabs.harvard.edu/abs/2003MNRAS.339..289S} {339, 289}

\bibitem[\protect\citeauthoryear{{Springel}, {Di Matteo}  \& {Hernquist}}{{Springel} et~al.}{2005}]{2005MNRAS.361..776S}
{Springel} V.,  {Di Matteo} T.,   {Hernquist} L.,  2005, \mn@doi [\mnras] {10.1111/j.1365-2966.2005.09238.x}, \href {https://ui.adsabs.harvard.edu/abs/2005MNRAS.361..776S} {361, 776}

\bibitem[\protect\citeauthoryear{{Springel}, {Pakmor}, {Zier}  \& {Reinecke}}{{Springel} et~al.}{2021}]{2021MNRAS.506.2871S}
{Springel} V.,  {Pakmor} R.,  {Zier} O.,   {Reinecke} M.,  2021, \mn@doi [\mnras] {10.1093/mnras/stab1855}, \href {https://ui.adsabs.harvard.edu/abs/2021MNRAS.506.2871S} {506, 2871}

\bibitem[\protect\citeauthoryear{{Stanway} \& {Eldridge}}{{Stanway} \& {Eldridge}}{2018}]{BPASS2018}
{Stanway} E.~R.,  {Eldridge} J.~J.,  2018, \mn@doi [\mnras] {10.1093/mnras/sty1353}, \href {https://ui.adsabs.harvard.edu/abs/2018MNRAS.479...75S} {479, 75}

\bibitem[\protect\citeauthoryear{{Stevens} et~al.,}{{Stevens} et~al.}{2019}]{2019MNRAS.483.5334S}
{Stevens} A. R.~H.,  et~al., 2019, \mn@doi [\mnras] {10.1093/mnras/sty3451}, \href {https://ui.adsabs.harvard.edu/abs/2019MNRAS.483.5334S} {483, 5334}

\bibitem[\protect\citeauthoryear{{Stinson}, {Seth}, {Katz}, {Wadsley}, {Governato}  \& {Quinn}}{{Stinson} et~al.}{2006}]{2006MNRAS.373.1074S}
{Stinson} G.,  {Seth} A.,  {Katz} N.,  {Wadsley} J.,  {Governato} F.,   {Quinn} T.,  2006, \mn@doi [\mnras] {10.1111/j.1365-2966.2006.11097.x}, \href {https://ui.adsabs.harvard.edu/abs/2006MNRAS.373.1074S} {373, 1074}

\bibitem[\protect\citeauthoryear{{Sutherland} \& {Dopita}}{{Sutherland} \& {Dopita}}{1993}]{1993ApJS...88..253S}
{Sutherland} R.~S.,  {Dopita} M.~A.,  1993, \mn@doi [\apjs] {10.1086/191823}, \href {https://ui.adsabs.harvard.edu/abs/1993ApJS...88..253S} {88, 253}

\bibitem[\protect\citeauthoryear{Tacconi, Genzel  \& Sternberg}{Tacconi et~al.}{2020}]{doi:10.1146/annurev-astro-082812-141034}
Tacconi L.~J.,  Genzel R.,   Sternberg A.,  2020, \mn@doi [Annual Review of Astronomy and Astrophysics] {10.1146/annurev-astro-082812-141034}, 58, 157

\bibitem[\protect\citeauthoryear{{Teyssier}}{{Teyssier}}{2002}]{2002A&A...385..337T}
{Teyssier} R.,  2002, \mn@doi [\aap] {10.1051/0004-6361:20011817}, \href {https://ui.adsabs.harvard.edu/abs/2002A&A...385..337T} {385, 337}

\bibitem[\protect\citeauthoryear{{Thomas} et~al.,}{{Thomas} et~al.}{2011}]{2011MNRAS.415..545T}
{Thomas} J.,  et~al., 2011, \mn@doi [\mnras] {10.1111/j.1365-2966.2011.18725.x}, \href {https://ui.adsabs.harvard.edu/abs/2011MNRAS.415..545T} {415, 545}

\bibitem[\protect\citeauthoryear{{Trayford} et~al.,}{{Trayford} et~al.}{2025}]{2025arXiv250513056T}
{Trayford} J.~W.,  et~al., 2025, \mn@doi [arXiv e-prints] {10.48550/arXiv.2505.13056}, \href {https://ui.adsabs.harvard.edu/abs/2025arXiv250513056T} {p. arXiv:2505.13056}

\bibitem[\protect\citeauthoryear{{Tremmel}, {Karcher}, {Governato}, {Volonteri}, {Quinn}, {Pontzen}, {Anderson}  \& {Bellovary}}{{Tremmel} et~al.}{2017}]{2017MNRAS.470.1121T}
{Tremmel} M.,  {Karcher} M.,  {Governato} F.,  {Volonteri} M.,  {Quinn} T.~R.,  {Pontzen} A.,  {Anderson} L.,   {Bellovary} J.,  2017, \mn@doi [\mnras] {10.1093/mnras/stx1160}, \href {https://ui.adsabs.harvard.edu/abs/2017MNRAS.470.1121T} {470, 1121}

\bibitem[\protect\citeauthoryear{{Tremonti} et~al.,}{{Tremonti} et~al.}{2004}]{2004ApJ...613..898T}
{Tremonti} C.~A.,  et~al., 2004, \mn@doi [\apj] {10.1086/423264}, \href {https://ui.adsabs.harvard.edu/abs/2004ApJ...613..898T} {613, 898}

\bibitem[\protect\citeauthoryear{{Truelove}, {Klein}, {McKee}, {Holliman}, {Howell}  \& {Greenough}}{{Truelove} et~al.}{1997}]{1997ApJ...489L.179T}
{Truelove} J.~K.,  {Klein} R.~I.,  {McKee} C.~F.,  {Holliman} II J.~H.,  {Howell} L.~H.,   {Greenough} J.~A.,  1997, \mn@doi [\apjl] {10.1086/310975}, \href {https://ui.adsabs.harvard.edu/abs/1997ApJ...489L.179T} {489, L179}

\bibitem[\protect\citeauthoryear{Virtanen et~al.,}{Virtanen et~al.}{2020}]{2020SciPy-NMeth}
Virtanen P.,  et~al., 2020, \mn@doi [Nature Methods] {10.1038/s41592-019-0686-2}, \href {https://rdcu.be/b08Wh} {17, 261}

\bibitem[\protect\citeauthoryear{{Vogelsberger}, {Genel}, {Sijacki}, {Torrey}, {Springel}  \& {Hernquist}}{{Vogelsberger} et~al.}{2013}]{2013MNRAS.436.3031V}
{Vogelsberger} M.,  {Genel} S.,  {Sijacki} D.,  {Torrey} P.,  {Springel} V.,   {Hernquist} L.,  2013, \mn@doi [\mnras] {10.1093/mnras/stt1789}, \href {https://ui.adsabs.harvard.edu/abs/2013MNRAS.436.3031V} {436, 3031}

\bibitem[\protect\citeauthoryear{Waskom}{Waskom}{2021}]{Waskom2021}
Waskom M.~L.,  2021, \mn@doi [Journal of Open Source Software] {10.21105/joss.03021}, 6, 3021

\bibitem[\protect\citeauthoryear{{Wetzel}, {Tinker}  \& {Conroy}}{{Wetzel} et~al.}{2012}]{2012MNRAS.424..232W}
{Wetzel} A.~R.,  {Tinker} J.~L.,   {Conroy} C.,  2012, \mn@doi [\mnras] {10.1111/j.1365-2966.2012.21188.x}, \href {https://ui.adsabs.harvard.edu/abs/2012MNRAS.424..232W} {424, 232}

\bibitem[\protect\citeauthoryear{{Wiersma}, {Schaye}, {Theuns}, {Dalla Vecchia}  \& {Tornatore}}{{Wiersma} et~al.}{2009}]{2009MNRAS.399..574W}
{Wiersma} R. P.~C.,  {Schaye} J.,  {Theuns} T.,  {Dalla Vecchia} C.,   {Tornatore} L.,  2009, \mn@doi [\mnras] {10.1111/j.1365-2966.2009.15331.x}, \href {https://ui.adsabs.harvard.edu/abs/2009MNRAS.399..574W} {399, 574}

\bibitem[\protect\citeauthoryear{{Wilkinson}, {Ludlow}, {Lagos}, {Fall}, {Schaye}  \& {Obreschkow}}{{Wilkinson} et~al.}{2023}]{2023MNRAS.519.5942W}
{Wilkinson} M.~J.,  {Ludlow} A.~D.,  {Lagos} C. d.~P.,  {Fall} S.~M.,  {Schaye} J.,   {Obreschkow} D.,  2023, \mn@doi [\mnras] {10.1093/mnras/stad055}, \href {https://ui.adsabs.harvard.edu/abs/2023MNRAS.519.5942W} {519, 5942}

\bibitem[\protect\citeauthoryear{{Woosley}, {Eastman}  \& {Schmidt}}{{Woosley} et~al.}{1999}]{1999ApJ...516..788W}
{Woosley} S.~E.,  {Eastman} R.~G.,   {Schmidt} B.~P.,  1999, \mn@doi [\apj] {10.1086/307131}, \href {https://ui.adsabs.harvard.edu/abs/1999ApJ...516..788W} {516, 788}

\bibitem[\protect\citeauthoryear{Wootten \& Thompson}{Wootten \& Thompson}{2009}]{ALMA5136193}
Wootten A.,  Thompson A.~R.,  2009, \mn@doi [Proceedings of the IEEE] {10.1109/JPROC.2009.2020572}, 97, 1463

\bibitem[\protect\citeauthoryear{{Wright} et~al.,}{{Wright} et~al.}{2010}]{2010AJ....140.1868W}
{Wright} E.~L.,  et~al., 2010, \mn@doi [\aj] {10.1088/0004-6256/140/6/1868}, \href {https://ui.adsabs.harvard.edu/abs/2010AJ....140.1868W} {140, 1868}

\bibitem[\protect\citeauthoryear{{Zibetti}, {Charlot}  \& {Rix}}{{Zibetti} et~al.}{2009}]{2009MNRAS.400.1181Z}
{Zibetti} S.,  {Charlot} S.,   {Rix} H.-W.,  2009, \mn@doi [\mnras] {10.1111/j.1365-2966.2009.15528.x}, \href {https://ui.adsabs.harvard.edu/abs/2009MNRAS.400.1181Z} {400, 1181}

\makeatother
\end{thebibliography}




\appendix

\section{The effect of the simulation box size}
\label{appendix: boxsize_effect}

\begin{figure*}
    \centering
    \includegraphics[width=0.99\textwidth]{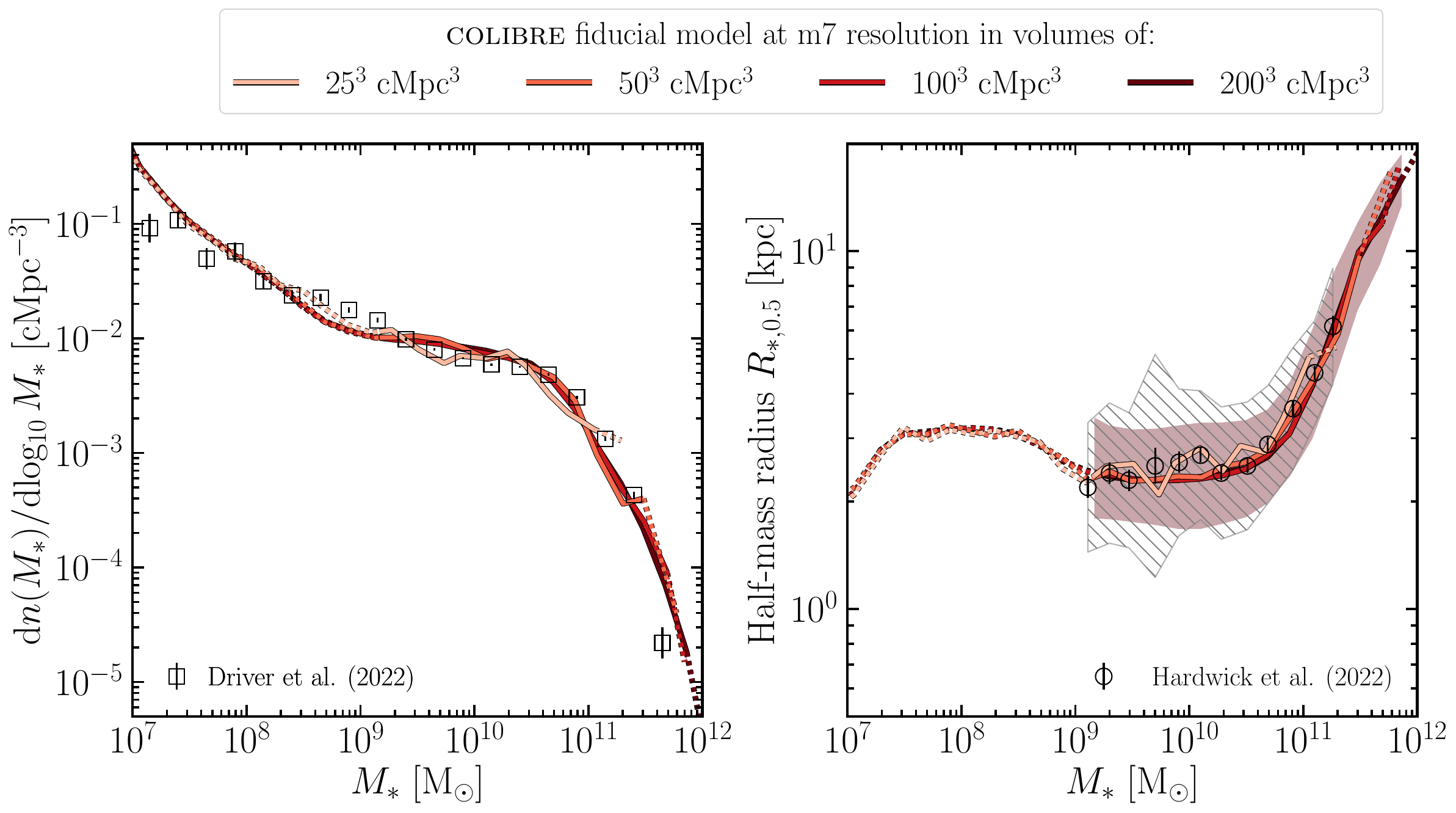}
    \caption{The $z=0$ galaxy stellar mass function (\gsmf; \textit{left}) and the $z=0$ median size -- stellar mass relation (\ssm; \textit{right}) for the \colibre{} fiducial \colibre{} m7 model in different cosmological volumes: $25^3$, $50^3$, $100^3$, and $200^3$ cMpc$^3$ (shown in progressively darker shades of red). The dark-red shaded region represents Poisson uncertainty for the \gsmf{} and the 16$^{\rm th}$ to 84$^{\rm th}$ percentile scatter for the \ssm{} in the largest simulation. Observational data are shown as black squares in the left panel (\gsmf{} from \citealt{2022MNRAS.513..439D}) and black circles in the right panel (\ssm{} from \citealt{2022MNRAS.509.3751H}), with the grey hatched region indicating the galaxy population-wide scatter from \citet{2022MNRAS.509.3751H}. Despite being calibrated using emulators trained on $50^3$ cMpc$^3$ volume simulations, the \colibre{} model shows excellent agreement with the observed $z=0$ \gsmf{} and \ssm{} also in $100^3$ and $200^3$ cMpc$^3$ cosmological volumes.}
\label{fig: box_size_effect}
\end{figure*}

Fig.~\ref{fig: box_size_effect} shows the $z=0$ \gsmf{} (left panel) and the $z=0$ median \ssm{} (right panel) from simulations using the fiducial \colibre{} m7 model in different cosmological volumes: $25^3$, $50^3$, $100^3$, and $200^3~\mathrm{cMpc}^3$ (shown in progressively darker shades of red). The dark-red shaded region indicates the Poisson uncertainty for the \gsmf{} and the 16$^{\rm th}$–84$^{\rm th}$ percentile range for the \ssm{} in the $(200~\mathrm{cMpc})^3$ simulation.

The fiducial m7 model in $100^3$ and $200^3$ cMpc$^3$ volumes exhibits excellent agreement with the observed $z=0$ \gsmf{} and \ssm{}, despite being calibrated to these data using emulators trained on (50 cMpc)$^3$ volume simulations. In contrast, the simulation in the smallest volume, (25 cMpc)$^3$, shows noticeable deviations from the observations, particularly at the high-mass end, due to the limited number of massive haloes. In other words, this comparison demonstrates that, although not ideal, a (50 cMpc)$^3$ volume is sufficient for performing calibration to the observed $z=0$ \gsmf{} and \ssm, while (25 cMpc)$^3$ would have been too small.

\section{Redshift evolution of the energy in SN feedback}
\label{appendix: redshift_evolution_of_sn_energy}

In Section \ref{Section: Results}, we showed that among the four models with best-fitting parameter values identified by the emulators, \colibrefixedagntemp{} provides the closest match to the observational data. A major reason for its superior performance over the other three models is the stellar birth pressure-dependent energy in SN feedback, employed only in \colibrefixedagntemp{} (the other three models use a fixed energy per SN). In this section, we provide additional details on the pressure dependence in the \colibrefixedagntemp{} model and show its impact on the energy released by SNe at different redshifts.

\begin{figure}
    \centering
     \includegraphics[width=0.47\textwidth]{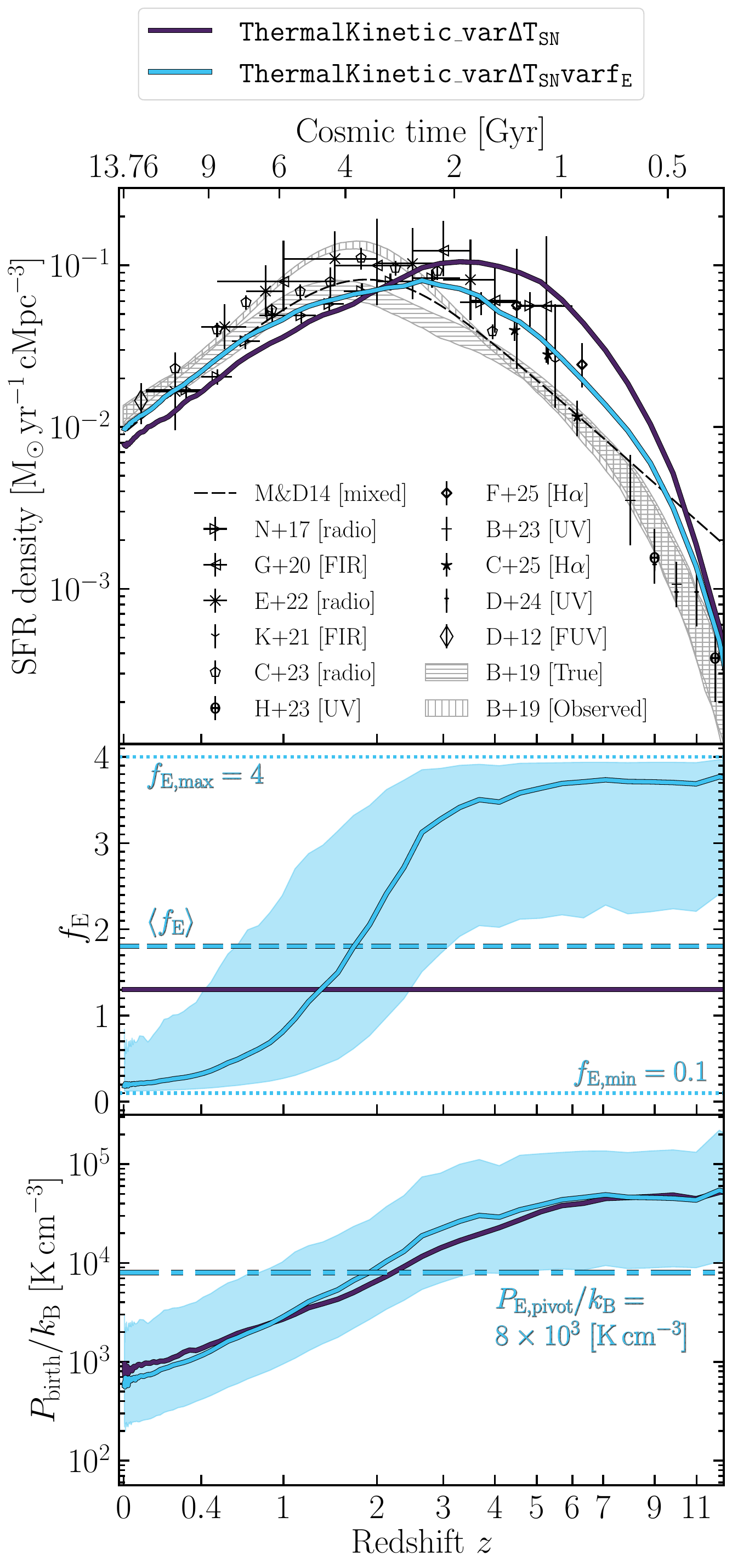}
    \caption{Comparison of the best-fitting \VardT{} (\modelvardtcolor) and \colibrefixedagntemp{} (\colibrefixedagndtcolor) models at m7 resolution in a (50 cMpc)$^{3}$ volume. The top, middle, and bottom panels show, respectively, the cosmic star formation rate density (SFRD), the median energy per CC SN in units of $10^{51}$ erg, and the median stellar birth pressure versus redshift. In the middle and bottom panels, the shaded regions represent the 16$^{\rm th}$ to 84$^{\rm th}$ percentile scatter in the \colibrefixedagntemp{} model. For reference, in the middle panel, the horizontal \colibrefixedagndtcolor{} dotted lines mark the minimum and maximum allowed CC SN energy values in \colibrefixedagntemp{}, while the horizontal long-dashed line indicates the average SN energy over the entire simulation, all in units of $10^{51}$~erg. The dash-dotted line in the bottom panel indicates the value of the subgrid parameter $P_{\rm E,pivot}/k_{\rm B}$ used in \colibrefixedagntemp{}. The median SN energy in the \colibrefixedagntemp{} model increases monotonically with redshift, causing it to better match the observed SFRD at high $z$ compared to \VardT, where the CC SN energy is constant.}
    \label{fig: f_E_vs_redshift}
\end{figure}

Fig.~\ref{fig: f_E_vs_redshift} compares the best-fitting \VardT{} (\modelvardtcolor) and \colibrefixedagntemp{} (\colibrefixedagndtcolor) models at m7 resolution in a (50 cMpc)$^3$ volume. The top panel shows the cosmic star formation rate density (SFRD), the middle panel shows the median energy per CC SN (in units of $10^{51}$ erg) as a function of redshift, and the bottom panel shows the median stellar birth pressure versus redshift. The \colibrefixedagndtcolor{} shaded regions in the middle and bottom panels denote the 16$^{\rm th}$ to 84$^{\rm th}$ percentile scatter in the \colibrefixedagntemp{} model. In the middle panel, the thin horizontal \colibrefixedagndtcolor{} dotted lines indicate the minimum and maximum allowed SN energies in the \colibrefixedagntemp{} model, while the horizontal long-dashed line marks the average SN energy across the entire simulation, all in units of $10^{51}$ erg. In the bottom panel, the \colibrefixedagndtcolor{} dash-dotted line indicates the value of the subgrid parameter $P_{\rm E,pivot}$ used in the \colibrefixedagntemp{} model.

We find that in the \colibrefixedagntemp{} model, the median CC SN energy increases monotonically with redshift due to the corresponding rise in stellar birth pressure, as dictated by equation (\ref{eq: stellar_birth_pressure_vs_SN_energy}). At $z > 2$, the median SN energy in \colibrefixedagntemp{} significantly exceeds the constant CC SN energy in \VardT{}, resulting in stronger SN feedback and causing \colibrefixedagntemp{} to agree better with the observed SFRD at these redshifts.

\bsp	
\label{lastpage}
\end{document}